\documentclass[english,aps,manuscript]{revtex4}
\usepackage[T1]{fontenc}
\usepackage[latin9]{inputenc}
\usepackage{textcomp}
\usepackage{amsmath}
\usepackage{amssymb}
\usepackage{graphicx}
\usepackage{esint}

\makeatletter

\newcommand{\lyxmathsym}[1]{\ifmmode\begingroup\def\b@ld{bold}
  \text{\ifx\math@version\b@ld\bfseries\fi#1}\endgroup\else#1\fi}

\@ifundefined{textcolor}{}
{%
 \definecolor{BLACK}{gray}{0}
 \definecolor{WHITE}{gray}{1}
 \definecolor{RED}{rgb}{1,0,0}
 \definecolor{GREEN}{rgb}{0,1,0}
 \definecolor{BLUE}{rgb}{0,0,1}
 \definecolor{CYAN}{cmyk}{1,0,0,0}
 \definecolor{MAGENTA}{cmyk}{0,1,0,0}
 \definecolor{YELLOW}{cmyk}{0,0,1,0}
 }
\newenvironment{lyxlist}[1]
{\begin{list}{}
{\settowidth{\labelwidth}{#1}
 \setlength{\leftmargin}{\labelwidth}
 \addtolength{\leftmargin}{\labelsep}
 }}
{\end{list}}

\makeatother

\usepackage{babel}
\begin{document}

\title{Phases of Gauge Theories}

\author{Michael C. Ogilvie}
\email[] {mco@physics.wustl.edu}

\affiliation{Dept. of Physics, Washington University, St. Louis, MO 63130 USA}
\begin{abstract}
One of the most fundamental questions we can ask about a given gauge
theory is its phase diagram. In the standard model, we observe three
fundamentally different types of behavior: QCD is in a confined phase
at zero temperature, while the electroweak sector of the standard
model combines Coulomb and Higgs phases. Our current understanding
of the phase structure of gauge theories owes much to the modern theory
of phase transitions and critical phenomena, but has developed into
a subject of extensive study. After reviewing some fundamental concepts
of phase transitions and finite-temperature gauge theories, we discuss
some recent work  that broadly
extends our knowledge of the mechanisms that determine
the phase structure of gauge theories.
A new class of models with a rich phase structure
has been discovered, generalizing our understanding of the confinement-deconfinement transition
in finite-temperature gauge theories. 
Models in this class have space-time
topologies with one or more compact directions. On $R^{3}\times S^{1}$,
the addition of double-trace deformations or periodic adjoint fermions
to a gauge theory can yield a confined phase in the region where the
$S^{1}$ circumference $L$ is small, so that the coupling constant
is small, and semiclassical methods are applicable. In this region,
Euclidean monopole solutions, which are constituents of finite-temperature
instantons, play a crucial role in the calculation of a non-perturbative
string tension. We review the techniques use to analyze this new class
of models and the results obtained so far, as well as their application
to finite-temperature phase structure, conformal phases of gauge theories
and the large-$N$ limit.
\end{abstract}
\maketitle

\section{Introduction: gauge theories}

Gauge theories are central to the modern understanding of the fundamental
forces of nature, and form the foundation of the standard model of
particle physics. Gauge theories promote global symmetries, such as
the $U(1)$ invariance associated with conservation of electric charge,
to local symmetries. In the case of quantum electrodynamics (QED),
the the quantum field theory of electromagnetic interactions, local
symmetry requires the introduction of a new vector field, associated
with the photon. The standard model of particle physics is based on
a local symmetry group of the form $SU(2)\times U(1)\times SU(3)$.
The standard model contains the Weinberg-Salam-Glashow model of the
electroweak interactions, with gauge group $SU(2)\times U(1)$ , which
unifies the electromagnetic interaction with the weak interaction.
The number of gauge pariticles equals the rank of the gauge group,
in this case four: the massless photon $\gamma$ and the massive vector
boson $W^{\pm}$ and $Z^{0}$. The other part of the standard model
is quantum chromodynamics (QCD), with gauge group $SU(3)$, describes
the strong interactions of quarks. There are eight gauge fields in
QCD, describing the gluons that bind quarks inside of hadrons. It
is widely believed that the standard model is not a complete description
of particle physics, and there are many different proposals for what
lies beyond the standard model. Gauge theories are integral to all
such proposals, even for those where four-dimensional gauge theories
are low-energy effective theories, as in string theory.

One of the most fundamental questions we can ask about a given gauge
theory is its phase structure. In the standard model, we observe three
fundamentally different types of behavior: the familiar Coulomb behavior
associated with the massless photon; the Higgs mechanism, responsible
for the masses of the $W^{\pm}$ and $Z^{0}$; and the confinement
of quarks and gluons by the gluon fields. These properties are characteristics
of different phases: QCD is in a confined phase at zero temperature,
while the electroweak sector of the standard model combines Coulomb
and Higgs phases. The phase structure of gauge theories has been extensively
studied. A large part of the interest in this subject centers on QCD
at non-zero temperature and density, but there is now great interest
in the phase structure of gauge theories associated with physics beyond
the standard model.

It is often convenient theoretically to view QCD in a simplified way,
in a form without dynamical quarks; this is often referred
to as pure $SU(3)$ gauge theory. 
As a consequence of dimensional transmutation, the
dimensionless gauge coupling $g^{2}$ can be replaced by a parameter
$\Lambda$ with dimensions of energy. Key observables with units of
mass such as hadron masses become pure numbers times $\Lambda$. In
other words, the pure gauge theory is a theory with no adjustable
dimensionless parameters, making it difficult to carry out analytical
approximations. Including dynamical quarks with realistic masses does
not make analytical work easier. However, there are many parameters
that can be varied within the context of gauge theories, shedding
light on important aspects of gauge theory behavior. Aside from their
great experimental interest, finite-temperature gauge theories offer
a dimensionless parameter $T/\Lambda$ which can be used, for example,
to change QCD from a confining phase at low temperatures to a plasma
phase at high temperatures. This additional parameter tells us something
about the physics of both phases. It is also natural to study the
phase structure of gauge as other parameters, such as the number of
colors $N$ or the number of quark flavors $N_{f}$, are varied. 

Recent work has shown the existence of a new class of gauge theory
models with important and desirable properties. All of these models have one
or more compact directions, and the most developed case is the geometry
$R^{3}\times S^{1}$, which is the geometry of Euclidean gauge theories
at finite temperature when the circumference $L$ of $S^{1}$ is identified
with the inverse temperature $\beta=1/T$. In addition to $L$, there
are many parameters that can be used to examine the phase structure.
Unlike conventional finite-temperature gauge theories, this new class
can be put into a confined phase when $L\ll\Lambda$ and $g^{2}\left(L\right)\ll1$
\cite{Unsal:2007vu}. Euclidean monopoles, the constituents of finite-temperature
instantons, are essential to a semiclassical calculation of the string
tension in this region. Moreover, this small-$L$ phase is smoothly
connected to the conventional, large-$L$ confining phase \cite{Myers:2007vc}.
There is a price to be paid for this. In the case of models with adjoint
fermions, the use of periodic boundary conditions in the compact direction
removes the spectral positivity of the transfer matrix in the compact
direction, so $L$ cannot be identified with $\beta$ and must be
regarded as a spatial direction. Models with double-trace deformations
can be motivated as the heavy-quark limit of periodic adjoint fermions,
and have a similar problem. The gains, however, are great: an analytic
understanding of confinement in a class of four-dimensional models,
the discovery of many new phases, and new approaches to conformality
and to the large-$N$ limit.

The analysis of this new class of models uses techniques and ideas
from many areas of theoretical physics, and the results are broadly
interesting as well. In section \ref{sec:Modern-theory-of}, we review
the modern theory of phase transitions and critical phenomena, concentrating
on key concepts and their appearance in a field-theoretic context.
Section \ref{sec:Symmetries-of-gauge-theories} introduces the basics
of finite-temperature gauge theories, including  important symmetries
of gauge theories associated with confinement and chiral symmetry
breaking. Sections \ref{sec:Phases-on-R3xS1} through \ref{sec:Large-N}
discuss recent developments. Section \ref{sec:Phases-on-R3xS1} discussed
the phase structure of gauge theories on $R^{3}\times S^{1}$, an
arena where it has proven possible to demonstrate confinement in certain
models using semiclassical methods. Section \ref{sec:Conformality-vs-Confinement}
focuses on the general issue of gauge theory phase structure in the
context of conformality and duality in gauge theories. In section
\ref{sec:Large-N}, $SU(N)$ gauge theories in the large-$N$ limit
are considered, and promising new approaches are compared with older
formulations. A final section concludes and summarizes.

The notations used throughout are as follows: All field theories are
taken to be in Euclidean space unless noted otherwise. Lower-case
Greek indices are used for space-time and the metric in Euclidean
space is
\begin{equation}
g_{\mu\nu}=g^{\mu\nu}=\delta_{\mu\nu}.
\end{equation}
Roman indices in the range $j\cdots n$ generally denote {}``spatial''
directions on $R^{3}\times S^{1}$, i.e., the three directions orthogonal
to the compact direction. Roman indices in the range $a\cdots d$
generally label group generators, while capital letters are used to
denote group representations: $F,Adj,S,A,R$. $S^{k}$ is the $k$-dimensional
surface of a $\left(k+1\right)$-dimensional hypersphere, so $S^{1}$
is the unit circle. $T^{k}$ is the $k$-dimensional hypertorus, so
$T^{1}$ is also $S^{1}$.

\section{\label{sec:Modern-theory-of}Modern theory of Phase transitions and
Critical Phenomena}

\subsection{\label{sub:Order-parameters-and}Order parameters and symmetries}

\begin{figure}
\includegraphics[scale=0.7]{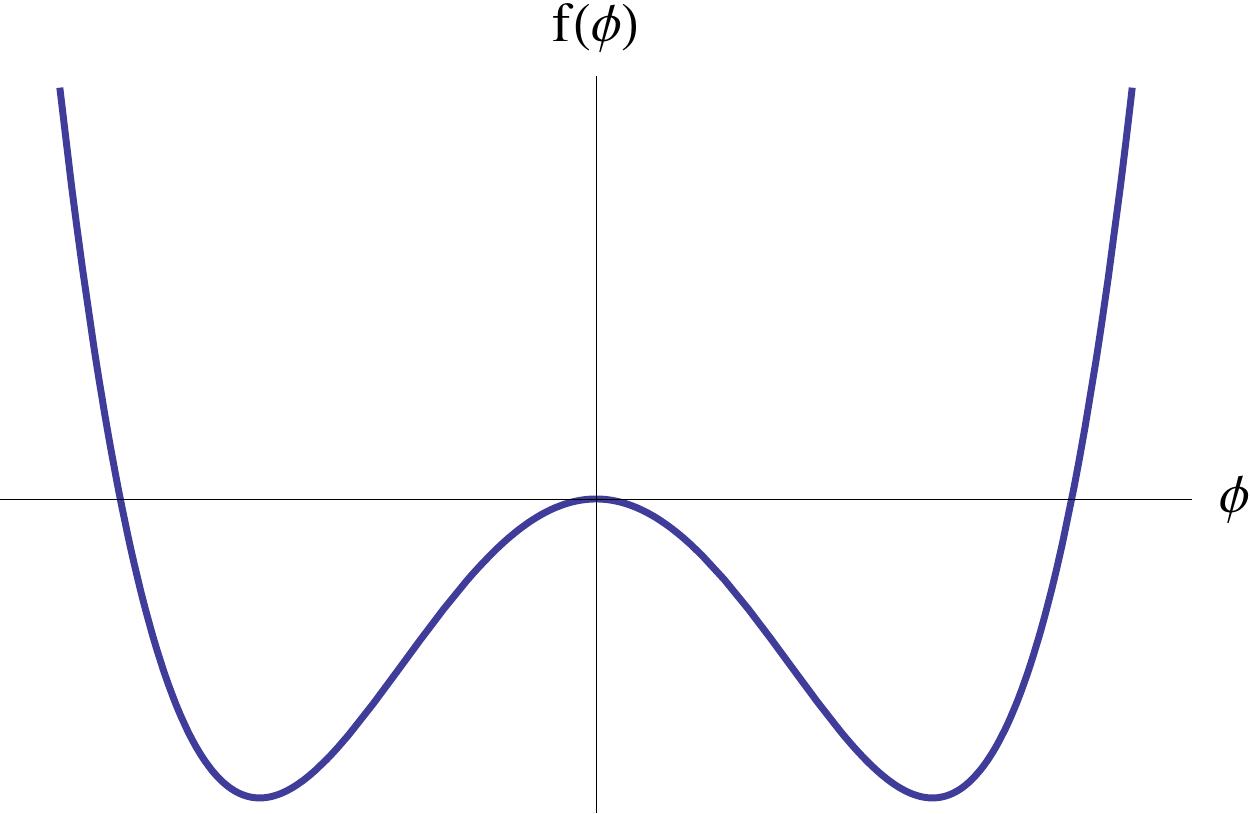}\caption{\label{fig:DoubleWell}The Landau-Ginsburg free energy density $f(\phi)$
for the case $r<0$ where the symmetry $\phi\rightarrow-\phi$ is
spontaneously broken.}
\end{figure}

The modern theory of phase transitions begins with Landau-Ginsburg
theory \cite{cha95}, and the key concepts of order parameters and
free energies as functions of those order parameters. The simplest
examples of order parameters come from ferromagnetic spin systems,
where the magnetization $m_{j}\left(\vec{x}\right)$ is the order
parameter, with $j$ specifying the number of components. In order
for the magnetization to be treated as a continuous variable, it is
necessary to imagine that some smoothing procedure be applied to average
the microscopic spins over many unit cells. The magnetization is naturally
a three-component vector, but underlying microscopic physics may cause
it to be effectively one- or two-dimensional. Assuming macroscopic
isotropy, the magnetization has a natural internal symmetry group
$O(3),$ $O(2)$ or $Z(2)$ depending on whether the order parameter
is three, two or one dimensional. Thus the dimensionality of the order
parameter need not be the same as the dimensionality of the system.
Suppose for a moment we are interested in the free energy of a spin
system in three spatial dimensions with a uniform magnetization $\phi$
which we take to be a scalar. This is the case where the magnetization
has an easy axis; then $\phi$ is the projection of the magnetization
onto the easy axis. The key assumption of Landau-Ginsburg theory is
that the free energy may be written as an integral over a local free
energy density $f\left(\phi\right)$ which is an analytic function
of $\phi$ as well as any underlying parameters, including the temperature.
Thus the free energy is given by
\begin{equation}
F\left[\phi\right]=\int d^{3}x\, f(\phi)=\int d^{3}x\left[\frac{r}{2}\phi^{2}+\frac{\lambda}{4}\phi^{4}+...\right]
\end{equation}
where only terms even in $\phi$ are allowed by the $Z(2)$ symmetry.
If higher-order terms in the polynomial expansion are to be ignored,
then we must have $\lambda>0$ for stability. If $r>0$, the free
energy is minimized when $\phi=0$. As shown in figure \ref{fig:DoubleWell}, if $r<0$,
the free energy will be minimized when
\begin{equation}
\phi=\pm\sqrt{\frac{-r}{\lambda}}
\end{equation}
giving the most common example of the idea of spontaneously broken
symmetry. The free energy concept can be generalized to that of a
function of a spatially-dependent magnetization $\phi\left(\vec{x}\right)$.
Including gradient terms in the expansion of the free energy, we obtain
\begin{equation}
F\left[\phi\right]=\int d^{3}x\left[\frac{a}{2}\left(\nabla\phi\right)^{2}+\frac{r}{2}\phi^{2}+\frac{\lambda}{4}\phi^{4}+...\right]
\end{equation}
where higher-order terms in $\nabla\phi$ are suppressed. It is now
possible to imagine calculating the free energy in an ensemble with
an external field, via the functional integral
\begin{equation}
Z\left[h\right]=\int\left[d\phi\right]\exp\left[-\frac{1}{T}F\left[\phi\right]+\int d^{3}x\, h\left(\vec{x}\right)\phi\left(\vec{x}\right)\right]
\end{equation}
This is a three-dimensional Euclidean scalar field theory, the simplest
example of the confluence of  quantum field theory (at zero temperature)
and classical statistical mechanics. Within Landau-Ginsburg theory,
the point $r=0$ marks the location of a second-order phase transition,
where the correlation length $\xi\propto r^{-1/2}$ becomes infinite.
From a field-theoretic point of view, the inverse of the correlation
length is a mass. From the modern point of view, a second-order phase
transition is intrinsically associated with a massless excitation. 

\subsection{\label{sub:Universality}Universality}

Landau-Ginsburg theory explains how non-analyticities arise in the
vicinity of second-order critical point. For example, the critical
point can be determined as a function of temperature: $r\left(T_{C}\right)=0$
locates the critical point. Analyticity of $r$ in $T$ gives $r\propto\left(T-T_{c}\right)$
near the critical point and hence $\phi\left(T\right)\propto\sqrt{T_{c}-T}$
in the low-temperature, symmetry-broken phase. The non-analytic behavior
is captured in a set of critical indices, which characterize the critical
behavior. For example, $\phi\left(T\right)\propto\left(T_{c}-T\right)^{\beta}$
so the critical index $\beta$ is $1/2$ in Landau-Ginsburg theory.
Landau-Ginsburg theory leads to a simple universality: models with
the same Landau-Ginsburg theory have the same phase diagram near the
critical point, when parametrized in terms of the Landau-Ginsburg
parameters, and the critical exponents at second-order phase transitions
are idential. This universality was supported by the known exact equivalence
between the Ising model and a lattice binary allow model, and led
to the identification of the liquid-gas critical end point with the
similar critical point in spin systems. However, both theoretical
and experimental determinations of critical indices led to the conclusion
that that critical indices, while falling into universal classes,
were not given by the Landau-Ginsburg values \cite{KADANOFF:1967zz}.
Furthermore, it was clear that the spatial dimensionality $d$ of
the system must play a role. For example, classical spin systems with
$d=1$ with short-range interactions do not have phase transitions
except at $T=0$ , in contradiction to the predictions of Landau-Ginsburg
theory. It was clear that fluctuations, ignored in Landau-Ginsburg
theory, were crucial. This led to the development of the modern renormalization
group \cite{Wilson:1973jj}.

\subsection{\label{sub:Renormalization-group}Renormalization group}

The renormalization group takes many forms, but common to all approaches
is the idea of a fixed point in the space of couplings. Consider an
Ising model of spatial dimension $d$. It has two parameters: the
lattice spacing $a$ and a dimensionless coupling $J$ which is the
energy of a ferromagnetic bond divided by $k_{B}T$. The Hamiltonian
is given by a sum over nearest-neighbor interactions:
\begin{equation}
-\frac{\mathcal{H}}{k_{B}T}=\sum_{nn}J\sigma_{j}\sigma_{k}
\end{equation}
where the spins $\sigma_{j}$ are associated with lattice sites $j$
and take on the values $\pm1$. On general grounds, the correlation
length $\xi$ between spins must have the form $\xi=aF\left(J\right)$.
At the critical point $J_{c}$, the correlation length is infinite:
$F\left(J_{c}\right)=\infty$. If an exact or approximate relation
between $\xi$ and $J$ is known, we can define a simple renormalization
group transformation as a change in the lattice spacing by a scale
factor $s>1$ such that $a\rightarrow a'=sa$ and $J\rightarrow J'(J)$
with the correlation length $\xi$ fixed: $\xi'=\xi$. In other words,
we must have $sF\left(J'\right)=F\left(J\right)$ or

\begin{equation}
J'=F^{-1}\left(s^{-1}F\left(J\right)\right)
\end{equation}
At the critical point $J_{c}$, we must have $J'=J$, which is a fixed
point of this renormalization group transformation. There are two
equivalent points of view: From a physical point of view the correlation
length is constant, but from a lattice point of view the dimensionless
correlation length $F(J)$, measured in units of lattice spacing,
decreases away from $J_{c}$. In the Ising model with $d\ge2$, there
are actually three fixed points. There are two fixed points with $\xi=0$,
at $J=0$ and at $J=\infty$, corresponding to $T=0$ and $T=\infty$
respectively. The second-order phase transition is associated with
the non-trivial fixed point between them. 

It is generally necessary to introduce many parameters into the effective
Hamiltonian describing the system at a length scale $\lambda a$,
which can be denoted as the set $\left\{ J_{a}\right\} $. In Ising
models, such additional couplings include a next-nearest-neighbor
coupling and a four-spins around a square coupling. It is here that
the renormalization group has its explanatory power. In the vicinity
of a fixed point $J^{*}=\left\{ J_{a}^{*}\right\} $ , the renormalization
group transform which changes the scale by a factor $\lambda$ can
be linearized as
\begin{equation}
J_{a}'-J_{a}^{*}\simeq\sum_{b}R_{ab}\left(J_{b}-J_{b}^{*}\right).
\end{equation}
The matrix $R$ can be diagonalized and the corresponding linear combinations
of spin interactions identified; the latter are usually referred to
as operators and denoted $\mathcal{O}_{c}$. In the new basis for
interactions the Hamiltonian can be written as $\mathcal{H}=\sum_{c}J_{c}\mathcal{O}_{c}$
and a renormalization group transformation has the form $J'_{c}-J_{c}^{*}=s^{p_{c}}\left(J_{c}-J_{c}^{*}\right)$.
If $p_{c}>0$, then the renormalization group flow generated by $J\rightarrow J'$
will take $J_{c}$ away from its fixed point value. In this case,
the coupling and its corresponding operators are termed relevant.
On the other hand, if $p_{c}<0$, the renormalization group flow will
naturally take $J_{c}$ towards its fixed point value, at least within
some basin of attraction where the linearization of the renormalization
group is valid. Such couplings and operators are called irrelevant.
There is also the case $p_{c}=0$, the marginal case, where higher-order
effects in the renormalization group transformation must be considered.
In the simplest case, a second-order phase transition is associated
with the existence of one relevant coupling $J_{r}$, accompanied
by a potentially infinite set of irrelevant couplings. 
This situation occurs, for example, in the Ising model in the
absence of symmetry-breaking fields.
Within the
linear approximation, the hyper-plane in coupling space defined by
$J_{r}=J_{r}^{*}$, is the critical surface where the correlation
length is infinite. More generally, the set of renormalization group
trajectories that flow into the fixed point forms a critical surface
of codimension one in parameter space. The existence of irrelevant
variables explains universality: many different models lie in the
same universality class, and their critical behavior is determined
not by the particular microscopic model, but by the physics near the
fixed point. 

In the field-theoretic approach to critical phenomena, regularization
of ultraviolet divergences typically requires the introduction of
a parameter $\mu$ with dimensions of mass. Typical regularization
schemes include a simple ultraviolet cutoff, Pauli-Villars regularization
and dimensional regularization. The observed coupling constants are
defined in terms of the arbitrary scale $\mu$, and are therefore
called running coupling constants. A change in $\mu$ does not change
the underlying physics, but implies a corresponding change in the
coupling constants, because a change in conventions does not change
the physics of the system. In the case of a pure gauge theory, there
is only one coupling $g\left(\mu\right)$, and its running is described
by
\begin{equation}
\mu\frac{dg}{d\mu}=\beta\left(g\right)
\end{equation}
where $\beta\left(g\right)$ is the renormalization group beta function.
In pure $SU(N)$ gauge theories, the perturbative $\beta$ function
is negative, indicating that the renormalization group flow is towards
$g=0$ at large $\mu$. We say that $g=0$ is an ultraviolet fixed
point, or that the theory is asymptotically free. There is an important
connection here with the construction of the continuum limit of lattice
gauge theories, where the inverse of lattice spacing $a$ plays the
role of an ultraviolet regulator. Physical quantities, such as a glueball
mass $M$, cannot depend on the lattice spacing, but can be written
parametrically as
\begin{equation}
M=a^{-1}F\left(g\left(a\right)\right)
\end{equation}
 where $g(a)$ is the coupling constant defined in the lattice action.
The continuum limit, where $M$ is to remain finite as $a\rightarrow0$,
can only be taken by simultaneously adjusting $g$ such that $F\left(g\right)\rightarrow0$.
From the perspective of fixed lattice spacing, the continuum limit
can only be taken at a second-order critical point, and physical masses
become small in lattice units. Although the most common gauge theories
have a simple fixed point structure in four dimensions, non-trivial
fixed points play an important role in some scenarios for physics
beyond the standard model, as will be discussed in section \ref{sec:Conformality-vs-Confinement}.

The close connection between quantum field theories at zero temperature
in $d$ Euclidean dimensions with classical statistical mechanics
models was crucial in developing the field-theoretic approach to critical
phenomena. The simplest example of this connection is between the
simple model of ferromagnetism described above and the $\phi^{4}$
field theory. There is a simple rule of thumb which is helpful in
matching up field-theoretic properties with renormalization group
properties: a super-renormalizable interaction is relevant, a renormalizable
interaction is marginal, and a non-renormalizable interaction is irrelevant.
For example, in the of the the $\lambda$$\phi^{4}$ field theory
in $d=4-\epsilon$ dimensions \cite{Wilson:1971bg,Wilson:1971dh,Wilson:1971dc},
the $\phi^{4}$ interaction is marginal in four dimensions, with an
infrared fixed point at $\lambda=0$. For $\epsilon>0$, a non-trivial
infrared fixed point $\lambda^{*}>0$ is present, and the $\phi^{2}$
term is the only relevant operator near the critical point. From this
point of view, Landau-Ginsburg theory fails to give the correct critical
indices for $d<4$ because it is essentially an expansion around $\lambda=0$
rather than $\lambda^{*}$. Above four dimensions, the IR fixed point
is at $\lambda=0$, and the critical indices are correctly predicted
by Landau-Ginsuburg theory. The field theoretic approach allows a
precise statement of universality: starting from the most general
renormalizable field theory using the order parameter as the field,
the second-order phase transitions and their properties follow from
the fixed points of the theory. Thus, the universality class of a
second-order phase transition depends on the dimensionality of space
and the symmetry group of the order parameter.

Although the renormalization group has led to tremendous progress
in our understanding of second-order phase transitions, the vast majoritiy
of phase transitions observed in nature are first-order. Such transitions
are generally accompanied by a discontinuity in one or more order
parameter, and there is a latent heat associated with the transition.
Within the framework of the effective potential, first-order transitions
occur when there are two or more disconnected global minima of the
free energy. A simple model that illustrates the general concept has
a Landau-Ginsburg free energy density of the form
\begin{equation}
f\left(\phi\right)=\frac{r}{2}\phi^{2}+\frac{\lambda}{4}\phi^{4}+\frac{g}{6}\phi^{6}
\end{equation}
Unlike the simpler $\phi^{4}$ theory, with $g>0$ to ensure stability
we need not have $\lambda>0$. For $\lambda>0$, there is a second-order
phase transition at $r=0$. However, if $r<0$, there is a first-order
phase transition at $\lambda=-4\sqrt{rg/3}$ where two non-trivial
minima have the same free energy as the local minimum at $\phi=0$.

This model contains an important lesson: Although the behavior of
systems at second-order transitions is determined by the local structure
of the the free energy density $f(\phi)$, the overall phase structure
is determined by the location of the global minimum of the free energy.
This is usually more difficult to obtain in systems with several order
parameters and complicated interactions.

\subsection{\label{sub:effective-potential}The effective potential}

The close correspondence between classical statistical mechanics and
quantum field theory extends beyond critical phenomena. In thermodynamics
and statistical mechanics, different independent variables, free energies
and statistical ensembles are often used. These ideas, so useful in
statistical physics, have direct analogs in quantum field theory.
We begin with the case of quantum field theories at zero temperature,
and later extend our results to finite temperature. Consider the following
generalization of the partition function
\begin{equation}
Z\left[J\right]=\int\left[d\phi\right]\exp\left[-S\left[\phi\right]+\int d^{d+1}x\, J\left(x\right)\phi\left(x\right)\right]
\end{equation}
This is a generating functional for all $n$-point correlation functions.
\begin{equation}
\frac{Z\left[J\right]}{Z\left[0\right]}=1+\sum_{n=1}^{\infty}\frac{1}{n!}\int d^{d+1}x_{1}..d^{d+1}x_{n}J\left(x_{1}\right)..J\left(x_{n}\right)\left\langle \phi\left(x_{1}\right)..\phi\left(x_{n}\right)\right\rangle .
\end{equation}
The logarithm of $Z\left[J\right]$, $-W\left[J\right]=\log Z\left[J\right]$
is the generating functional of connected correlation functions
\begin{equation}
-W\left[J\right]=-W\left[0\right]+\sum_{n=1}^{\infty}\frac{1}{n!}\int d^{d+1}x_{1}..d^{d+1}x_{n}J\left(x_{1}\right)..J\left(x_{n}\right)\left\langle \phi\left(x_{1}\right)..\phi\left(x_{n}\right)\right\rangle _{c}
\end{equation}
 where the connected correlation functions are, for example,
\begin{equation}
\left\langle \phi\left(x_{1}\right)\phi\left(x_{2}\right)\right\rangle _{c}=\left\langle \phi\left(x_{1}\right)\phi\left(x_{2}\right)\right\rangle -\left\langle \phi\left(x_{1}\right)\right\rangle \left\langle \phi\left(x_{2}\right)\right\rangle 
\end{equation}
While $W\left[0\right]$ is typically not physically meaningful, the
difference $W\left[J\right]-W\left[0\right]$ can be interpreted in
the case of constant $J$ as $\mathcal{VTE}\left(J\right)$, where
$\mathcal{VT}$ is the volume of Euclidean space-time, and $\mathcal{E}\left(J\right)$
is the vacuum energy density in the presence of the constant source
$J$. In this formalism, $J\left(x\right)$ plays a role analogous
to that of a spatially dependent external magnetic field in a ferromagnet,
and $\left\langle \phi\left(x\right)\right\rangle $ is the analog
of the local magnetization at $x$. Spontaneous symmetry breaking
is typically determined from the solution of the relation
\begin{equation}
\left\langle \phi\left(x\right)\right\rangle =-\frac{\delta W}{\delta J\left(x\right)}_{J=0}.
\end{equation}
This is generally not a closed equation for $\left\langle \phi\left(x\right)\right\rangle $,
and approximation solutions from, \emph{e.g.}, perturbation theory
must be used. The functional $W\left[J\right]$ is the field-theoretic
analog of the free energy of a spin system in an external magnetic
field. It will be convenient in what follows to work with $\phi$
as the independent variable rather than $J$. The effective action
$\Gamma\left[\phi_{c}\right]$ is a functional of the classical field
$\phi_{c}\left(x\right)$ defined by
\begin{equation}
\phi_{c}\left(x\right)=-\frac{\delta W}{\delta J\left(x\right)}
\end{equation}
 for arbitrary $J\left(x\right)$. Given that relation between $J(x)$
and $\phi_{c}(x)$, we define $\Gamma$ via a functional Legendre
transform:
\begin{equation}
\Gamma\left[\phi_{c}\right]=W\left[J\right]+\int d^{d+1}x\, J\left(x\right)\phi_{c}\left(x\right)
\end{equation}
 which satisfies
\begin{equation}
\frac{\delta\Gamma}{\delta\phi(x)}=J(x).
\end{equation}
This construction is precisely analogous to exchanging the role of
independent variable between the external field and the magnetization
in a ferromagnet.
For constant fields, $W[J]$ is the free energy of the system with
constant $J$, and $/Gamma[\phi]$ is the free energy of the system
with the conjugate variable $\phi_c$ held constant.
For the case of a single scalar field, the effective
potential is given at one loop as
\begin{equation}
\Gamma\left[\phi_{c}\right]=S\left[\phi_{c}\right]+\frac{\hbar}{2}Tr\left[\log\left(-\nabla^{2}+V''\left(\phi_{c}\right)\right)\right]
\end{equation}
$\Gamma$ also has a graphical interpretation as the generator of
one-particle-irreducible (1PI) Feynman graphs. 

In many applications, it is convenient to consider the effective potential
$V_{eff}(\phi_{c})$. It can be obtained from a derivative expansion
of $\Gamma[\phi_{c}]$ for slowly-varying fields:
\begin{equation}
\Gamma[\phi_{c}]=\int d^{d+1}x\,\left[V_{eff}\left(\phi_{c}\right)+\frac{1}{2}Z\left(\phi_{c}\right)\left(\partial_{\mu}\phi_{c}\right)^{2}+...\right]
\end{equation}
where quartic-derivative terms and higher are indicated by ellipses.
The effective potential should be understood at zero temperature as
the vacuum energy density of the system given that $\left\langle \phi\left(x\right)\right\rangle =\phi_{c}$.
We have for constant fields
\begin{equation}
\frac{\partial V_{eff}}{\partial\phi_{c}}=J
\end{equation}
and $V_{eff}(\phi_{c})$ is related to $\mathcal{E}(J)$ by a Legendre
transform
\begin{equation}
V_{eff}(\phi_{c})=\mathcal{E}(J)+J\phi_{c}.
\end{equation}
It is important to note that this construction is ambiguous when there
are multiple solutions for $\phi_{c}$. In general, $V_{eff}(\phi_{c})$
is given by the convex hull of the differential construction, a generalization
of Maxwell's equal-area construction.

$V_{eff}$ is an extension of the classical potential $V$ including
quantum effects. From the point of view of perturbation theory, $V_{eff}=V+O(\hbar)$,
and the $O(\hbar)$ correction term has ....
\begin{equation}
V_{eff}\left(\phi_{c}\right)=V\left(\phi_{c}\right)+\frac{\hbar}{2}\int\frac{d^{4}k}{\left(2\pi\right)^{4}}\log\left(k^{2}+V''\left(\phi_{c}\right)\right)
\end{equation}
Note that this expression requires renormalization. It is easy to
show that this expression may also be written as
\begin{equation}
V_{eff}\left(\phi_{c}\right)=V\left(\phi_{c}\right)+\hbar\int\frac{d^{3}k}{\left(2\pi\right)^{3}}\frac{1}{2}\sqrt{k^{2}+V''\left(\phi_{c}\right)}
\end{equation}
which clearly exposes the one-loop correction as a sum over the zero-point
energies of all the field modes. The background field method offers
a very powerful technique for evaluating $\Gamma\left[\phi\right]$.

\subsection{\label{sub:QFT-at-T-and-mu}Field theory at finite temperature}

In quantum statistical mechanics, the partition function is given
by $Z=Tr[\exp\left(-\beta H\right)]$ where $\beta$ is the inverse
of the temperature $T$. The propagation in Euclidean time is over
a finite extent, given by $\beta$, and the trace can be implemented
by requiring periodic boundary conditions on the fields. The partition
function is just the generating function of the field theory
\begin{equation}
Z=\int\left[d\phi\right]e^{-S}
\end{equation}
but the functional integral is now over all field configurations satisfying
$\phi\left(\vec{x},0\right)=\phi\left(\vec{x},\beta\right)$. Thus
the effects of non-zero temperature are included by changing the geometry
of the system from $R^{d+1}$ to $R^{d}\times S^{1}$ with periodic
boundary conditions for bosons; fermions require antiperiodic boundary
conditions.

The generalization of the one-loop effective potential to finite temperature
is illuminating. Due to the periodic boundary conditions in the Euclidean
time direction, the integral over $k_{d+1}$ is replaced by a sum
over Matsubara frequencies
\begin{equation}
V_{eff}\left(\phi_{c}\right)=V\left(\phi_{c}\right)+\frac{\hbar}{2}\frac{1}{\beta}\sum_{n\in Z}\int\frac{d^{3}k}{\left(2\pi\right)^{3}}\log\left(\left(\frac{2\pi n}{\beta}\right)^{2}+\vec{k}^{2}+V''\left(\phi_{c}\right)\right)
\end{equation}
which can be converted into the form
\begin{equation}
V_{eff}\left(\phi_{c}\right)=V\left(\phi_{c}\right)+\hbar\int\frac{d^{3}k}{\left(2\pi\right)^{3}}\left[\frac{1}{2}\sqrt{\vec{k}^{2}+V''\left(\phi_{c}\right)}+\log\left(1-e^{-\beta\sqrt{\vec{k}^{2}+V''\left(\phi_{c}\right)}}\right)\right]
\end{equation}
The additional term is instantly recognizable as $-p$, where $p$ is the pressure
of a relativistic Bose gas with mass $\sqrt{V''\left(\phi_{c}\right)}$.
This is of course consistent with the identification of $V_{eff}\left(\phi_{c}\right)$
with the free energy of the system, when evaluated at its global minimum.

\subsection{\label{sub:Symmetry-restoration}Symmetry restoration at high temperatures}

Generally speaking, broken symmetries are associated with low temperatures,
as seen, for example, in spin systems such as the Ising model. This
can be explicitly seen in scalar field theories at finite temperature
using the effective potential. The finite temperature part of the
one-loop effective potential has a natural expansion for high temperatures.
A more general form for such expressions will be given in section
\ref{sub:Role-of-quarks}, so for now we simply note that that for
a single real scalar field the first two terms in a high-temperature
expansion are given by \cite{Dolan:1973qd,Weinberg:1974hy}
\begin{equation}
V_{1T}\simeq-\frac{\pi^{2}T^{4}}{90}+\frac{1}{24}T^{2}V''\left(\phi_{c}\right).
\end{equation}
The first term is the usual blackbody term with no effect on the value
of $\phi_{c}.$ However, the sub-leading $T^{2}$ term acts to restore
symmetry. If the potential $V(\phi)$ has the form
\begin{equation}
V\left(\phi\right)=-\frac{1}{2}\mu^{2}\phi^{2}+\frac{1}{4!}\lambda\phi^{4},
\end{equation}
then $\mu^{2}>0$ will lead to spontaneous symmetry breaking at zero
temperature. However $V_{1T}$ supplies a temperature-dependent mass
term such that the potential for $\phi$ becomes
\begin{equation}
V\left(\phi\right)+V_{1T}\left(\phi\right)\simeq-\frac{\pi^{2}T^{4}}{90}+\frac{1}{2}\left(\frac{\lambda}{24}T^{2}-\mu^{2}\right)\phi^{2}+\frac{1}{4!}\lambda\phi^{4}
\end{equation}
and the symmetry $\phi\rightarrow-\phi$ will be restored at temperatures
above $T_{c}\simeq\sqrt{24/\lambda}\mu$. A notable exception to the
general rule of symmetry restoration at high temperature is the deconfinement
transition in pure gauge theories, where the broken phase lies above
the unbroken phase in temperature. This will be explain in detail
in Section \ref{sub:Center-symmetry}.

The temperature dependence of the effective potential is very similar
to the general form of the free energy as a function of order parameter
postulated by Landau theory. A slightly different construction, often
referred to as dimensional reduction, also leads to a free energy
of Landau's form. In dimensional reduction, the field modes with non-zero
Matsubara frequencies are integrated out, so that only modes which
are constant in Euclidean time remain. This effectively reduces a
$\left(d+1\right)$-dimensional theory to a $d$-dimensional theory
with temperature-dependent parameters in the action.

\subsection{\label{sub:Spatially-modulated-phases}Spatially modulated phases}

A further complication in the determination of the phase diagram is
the possible existence of spatially modulated phases. In the most
familiar cases, symmetry breaking occurs when some order parameter
$\phi\left(x\right)$ has a constant, non-zero expectation value independent
of $x$. However, there are phases in nature where order parameters
are not translationally and/or rotationally invariant. The most familiar
example is the formation of crystalline solid phases in ordinary materials,
where the order parameter is the density. In a liquid or vapor phase,
the density is uniform, but in a crystalline phase it is periodic.
Magnetic systems can also show spatially modulated phases, as exhibited
by simple models such as the anisotropic next-nearest-neighbor Ising
(ANNNI) model \cite{Selke:1988ss} and the chiral potts moder \cite{Ostlund:1981zz,Howes:1983mk}.
In field theories, spatially modulated phases are generally associated
with systems at finite density, and the sign problem is intimately
involved in this behavior \cite{Ogilvie:2009me,Ogilvie:2011mw}. In
QCD, some color superconducting phases are crystalline in nature \cite{Alford:2000ze};
these are analogs of the Larkin-Ovchinnikov-Fulde-Ferrell (LOFF) phases
of ordinary superconductors \cite{larkin:1964zz,Fulde:1964zz}.

There is a simple approach to spatially modulated phases due to Lifshitz
\cite{cha95}. As a simple model, consider a Landau-Ginsburg model
with a free energy density of the form
\begin{equation}
f\left(\phi\right)=\frac{a}{2}\left(\nabla\phi\right)^{2}+\frac{b}{2}\left(\nabla^{2}\phi\right)^{2}+\frac{r}{2}\phi^{2}+\frac{\lambda}{4}\phi^{4}
\end{equation}
where $b$ and $\lambda$ must be positive but $a$ and $r$ can be
of either sign. Taking the case $a<0$ and $r>0$, it is easy to see
that the homogeneous equilibrium state will be unstable to spatially-varying
perturbations of wavenumber $k$ for any values of $k$ satisfying
$bk^{4}+ak^{2}+r<0$. This model has three phases: an unbroken phase,
a broken phase, and a spatially modulated phase. The point $a=0,\, r=0$
is the Lifshitz point, where the three phases coexist.

\subsection{\label{sub:Non-equilibrium-information-from}Non-equilibrium behavior:
nucleation, spinodal decomposition and relaxation}

The perturbative effective potential can yield information about non-equilibrium
behavior. Recall that the perturbative effective potential does not
satisfy the convexity properties required of $V_{eff}$ by its Legendre
transform construction. However, the non-convex region of $V_{eff}$
contains information about non-equilibrium behavior. Metastable states
can be understood as local minima of the effective potential. Consider
a field theory with a double well potential of the form $V\left(\phi\right)=\lambda\left(\phi^{2}-v^{2}\right)^{2}$.
The presence of a small additional linear coupling $-J\phi$ will
bias the system towards one of the two minima at $\pm v$. This leads
to a discontinuity in the behavior of $\phi_{c}$ as a function of
$J$. However, the perturbative effective potential shows the persistence
of metastable states across $J=0$. The intuitive picture is simple
in the thin wall approximation, which has roots in classical nucleation
theory. Consider an initial, homogeneous field configuration where
$\phi=\phi_{local}$, the value of the local minimum of $V_{eff}$.
Coherent fluctuations in space will produce small droplets, also called
bubbles, of $\phi_{global}$, the global minimum. The difference between
the free energy densities $\Delta f=f(\phi_{local})-f(\phi_{global})$
drives the expansion of such bubbles. However, there is also a surface
tension $\sigma$ residing in the interface between the two phases,
which opposes the growth of bubbles. In the thin-wall approximation,
the free energy of a three-dimensional droplet of radius $R$ is approximately
\begin{equation}
F\left(R\right)=-\Delta f\cdot\frac{4\pi}{3}R^{3}+\sigma\cdot4\pi R^{2}
\end{equation}
where $\sigma$ is the interfacial surface tension at equilibrium.
The surface tension may be calculated from the one-dimensional kink
solution that interpolates between the two phases A droplet which
is smaller than the critical size $R_{c}$ will shrink due to surface
tension effects, while a droplet with $R>R_{c}$ will expand to convert
all space to the the equilibrium phase. The decay rate of metastable
states can be calculated within the functional integral formalism
using instanton techniques \cite{Langer:1967ax,Coleman:1977py,Callan:1977pt}. 

\begin{equation}
\Gamma=A\, e^{-S_{b}}
\end{equation}
where $S_{b}$ is the action of the so-called bounce solution. Within
the thin-wall approximation, it is equal to $F\left(R_{c}\right)$.
The pre-factor $A$ is given in terms of a functional determinant
representing fluctuations around the bounce solution.

Within the saddle-point approximation considered here, there is a
clear point at which the metastable phase ceases to exist, \emph{i.e.},
where the metastable states cease to exist as local minima. The subspace
of parameter space where this occurs is referred to as the spinodal,
which may be a point, line, \emph{et cetera. }The spinodal can be
located by the condition $V_{eff}'\left(\phi_{spinodal}\right)=V_{eff}''\left(\phi_{spinodal}\right)=0$,
and more generally by the vanishing of a mass associated with a metastable
state. In physical systems, fluctuations become important near the
spinodal point, and the distinction between metastable versus unstable
states becomes blurred.

Unstable states are of interest in their own right, and something
about their behavior can be learned using the effective action. Consider
a $\phi^{4}$ field theory in $d$ spatial dimensions at finite temperature.
We apply dimensional reduction so we have a three-dimensional field
theory with temperature-dependent coefficients in the potential. Suppose
the system begins in a high-temperature equilibrium state with unbroken
symmetry so $\phi=0$, but has a stable broken symmetry phase at low
temperatures. A rapid lowering of the temperature below the critical
point, known as a rapid quench, will place the system in a state of
unstable equilibrium, with $V_{eff}''\left(\phi=0\right)<0$. As in
the case of a metastable state, such an unstable state will lead to
a free energy
\begin{equation}
f\left(\phi=0\right)=V_{eff}\left(\phi=0\right)=V\left(\phi=0\right)+\frac{1}{2}\int\frac{d^{d}k}{\left(2\pi\right)^{d}}\log\left(\vec{k}^{2}+V''\left(\phi=0\right)\right)
\end{equation}
Because $V_{eff}''\left(\phi=0\right)<0$ , the modes near $\vec{k}=0$
in the $ $functional determinant give rise to an imaginary component
of the free energy. This imaginary part can be associated with the
decay rate of the unstable phase in a manner similar to the decay
of metastable states \cite{Weinberg:1987vp}. In the early stages
of equilibration, the unstable modes in the region $\vec{k}^{2}+V''\left(\phi=0\right)<0$
lead to exponentially growth of long-wavelength oscillations in the
order parameter. This process is known as spinodal decomposition.
When the order parameter is not conserved, \emph{e.g.} in the case
of magnetization, the fastest growth occurs around $\vec{k}=0$; however,
in the case of a conserved order parameters, such as a charge density,
the fastest growth occurs at a non-zero value of $\left|\vec{k}\right|$
\cite{cha95}.

In addition to nucleation and spinodal decomposition, there is a third
equilibration process, one which is always present in non-equilibrium
situations. Relaxation is the process whereby a system returns to
its equilibrium value after a small perturbation away from equilibrium;
the term may also be applied to individual field modes. An introductory
discussion of relaxation requires a treatment of damping, \emph{i.e.}
frictional or dissipative terms in the effective action. All three
processes have been extensively discussed in the context of the early
universe \cite{Kolb:1990vq,Kolb:1988aj}, particularly with respect
to inflation.

\section{\label{sec:Symmetries-of-gauge-theories}Phases of finite temperature
gauge theories}

Gauge theories with $T\ne0$ ({}``finite temperature'') have a rich
phase structure which has been extensively explored using a combination
of analytic methods and lattice simulations. Non-Abelian gauge theories
have global symmetries and associated order parameters which are analogous
to magnetization in spin systems, and much of the modern formalism
of critical phenomena is directly applicable. The symmetries and order
parameters associated with quark confinement and chiral symmetry breaking
are of particular interest as principal determinants of gauge theory
phase structure.

\subsection{Confinement}

The most striking feature of QCD is confinement: the force between
widely separated quark-antiquark pairs is a constant $\sigma$, known
as the string tension. This constant force implies a potential energy
between a quark-antiquark pair that grows as $\sigma r$ for large
distances r. The numerical value of $\sigma$, determined from phenomenology,
is approximately $0.18\, GeV^{2}\approx0.9\, GeV/fm$. The string
tension can be extracted from the asymptotic behavior of the quark-antiquark
potential. In turn, the static potential between two heavy quarks
can be determined in a gauge-invariant way from the expectation value
of the Wilson loop. This is a non-local operator associated with a
closed curve $\mathcal{C}$ in space-time parametrized as $x_{\mu}\left(\tau\right)$
where $\tau$ can be taken to be the unit interval and $x_{\mu}\left(1\right)=x_{\mu}\left(0\right)$.
The Wilson loop is defined as 
\begin{equation}
W\left[\mathcal{C}\right]=\mathcal{P}\exp\left[i\oint_{\mathcal{C}}dx_{\mu}A_{\mu}\left(x\right)\right]
\end{equation}
where $\mathcal{P}$ indicates path-ordering of the gauge fields $A_{\mu}\left(x\right)$
along the path $\mathcal{C}$ \cite{Wilson:1974sk}. Taking $W\left[\mathcal{C}\right]$
to be an element of the abstract gauge group $G$, the basic observables
are $Tr_{R}W\left[\mathcal{C}\right]$, where the trace is taken over
an irreducible representation $R$ of $G$. $Tr_{R}W\left[\mathcal{C}\right]$
has a physical interpretation as the non-Abelian phase factor associated
with a heavy particle in the representation $R$ moving adiabatically
around the closed loop $\mathcal{C}$. Typically, we are interested
in rectangular Wilson loops of width sides $L$ and $T$. The loop
can be associated with a process in which a particle-antiparticle
pair are created at one time, move to a separation $L$, propagate
forward in time for an interval $T$, and then annihilate. For $T\gg L$,
we have 
\begin{equation}
\left<Tr_{R}W\left[\mathcal{C}\right]\right>\simeq\exp\left[-V_{R}\left(L\right)T\right]
\end{equation}
where $V_{R}\left(L\right)$ is the heavy quark-antiquark potential
for particles in the representation $R$. For large distances, the
potential can grow no faster than linearly with $L$, and that linear
growth defines the string tension $\sigma_{R}$ for the representation
$R$: 
\begin{equation}
V\left(L\right)\rightarrow\sigma_{R}L+\mathcal{O}\left(1\right)
\end{equation}
 as $L\rightarrow\infty$. The $\mathcal{O}\left(1\right)$ correction
term represents a perimeter-law contribution, which must be renormalized;
the effect is closely related to the mass renormalization of a heavy particle
interacting with the gauge field. 

Beyond the perimeter law contribution,
there  are corrections which are  $\mathcal{O}\left(1/L\right)$ and higher
that are of physical interest.
Using an effective string description, Luscher showed that the
$\mathcal{O}\left(1/L\right)$ has a universal coefficient that depends
only on the dimensionality of space-time  \cite{Luscher:1980ac}; 
this prediction was confirmed in lattice simulations \cite{Luscher:2002qv}.
Subsequent analytical work has explored the effects of higher-order
terms \cite{Meyer:2006qx,Aharony:2009gg,Aharony:2010cx}. 
There have also been results on the broadening
of the flux tube as a function of $L$ \cite{Gliozzi:2010zv,Gliozzi:2010jh} . For
a lattice-oriented review of string effects in gauge theories,
see the review by Pepe \cite{Pepe:2010na}.

The statement that quarks are confined is the statement that the string
tension $\sigma_{F}$ of the fundamental representation of $SU(3)$
gauge theory is non-zero. However, each representation has its own
string tension. In lattice simulations of $SU(3),$ the fundamental
string tension is the smallest, and representations of larger dimension
generally have a larger string tension. A characteristic feature of
gauge theories in four dimensions is dimensional transmutation: the
classical action is scale-invariant, but the introduction of a mass
scale $\mu$ is necessary in order to define the running coupling
constant $g(\mu)$. It follows directly from the renormalization group
that any physical mass in four-dimensional pure gauge theories must
be proportional to a renormalization group invariant mass, usually
written simply as $\Lambda$. This implies that $\sigma_{R}=c_{R}\Lambda^{2}$,
where $c_{R}$ is a pure number. Put another way, the ratio $\sigma_{R}/\sigma_{F}$
is a pure number characterizing the representation $R$. In $(1+1)$-dimensional
gauge theories, the string tension $\sigma_{R}$ is proportional to
the quadratic Casimir invariant, leading to a behavior known as Casimir
scaling. It is generally believed that a satisfactory description
of confinement in four dimensions would include an understanding of
the string-tension scaling law. 

There are two subtleties associated with the string tension. The first
subtlety is that string tensions are generally determined in so-called
pure gauge theories, where only the gauge fields are dynamical; quarks
and other particles exist only as static classical charges. This is
done because of string breaking , also known as charge screening.
Consider a theory with dynamical particles of mass $m$ in the fundamental
representation $F$ of the gauge group. Then energy obtained by separating
a pair of static sources in the fundamental representation is $\sigma_{F}L$;
when that energy becomes on the order of $2m$, it will become energetically
favorable to produce a pair of dynamical particles at a cost in energy
of $2m$, and no string tension will be seen for $L\gtrsim2m/\sigma_{F}$. 

The second subtlety is related: in $SU(N)$ gauge theories, a state
with $N+1$ fermions in the fundamental representation, \emph{e.g.
}quarks in $SU(3)$, will form a baryon, a neutral state of $N$ fermions,
plus one additional fermion. Thus the string tension between $N+1$
fermions and $N+1$ antifermions observed at large distances should
be the same string tension observed between one fermion and one antifermion.
Thus for purposes of confinement, the relevant {}``charge'' is an
integer $k$ which can be taken between $1$ and $N$. Furthermore,
the string tension is the same for states of $N$-ality $k$ and $N-k$,
so that there are $\left[\frac{N}{2}\right]$ independent string tensions.
This behavior is associated with the $Z(N)$ symmetry underlying confinement
in $SU(N)$, discussed below. 

It is generally believed that there is some simple regular behavior
described by string tension scaling laws \cite{Greensite:2003bk}.
For example, in $(1+1)$-dimensions, Casimir scaling gives
\begin{equation}
\frac{\sigma_{k}}{\sigma_{1}}=\frac{k\left(N-k\right)}{N-1}
\end{equation}
This behavior is also found in the strong-coupling limit of lattice
gauge theories in all dimensions, and is consistent with the behavior
seen in lattice simulations. An alternative to Casimir scaling is
sine-law scaling 
\begin{equation}
\frac{\sigma_{k}}{\sigma_{1}}=\frac{\sin\left(\frac{\pi k}{N}\right)}{\sin\left(\frac{\pi}{N}\right)}
\end{equation}
a behavior predicted in certain supersymmetric gauge theories \cite{Douglas:1995nw},
in MQCD \cite{Hanany:1997hr}. This behavior is also roughly consistent
with lattice results. Note that both formulas give $\sigma_{N}=0$
and are invariant under $k\leftrightarrow N-k$. In the limit $N\rightarrow\infty$
at fixed $k$, both formulas give $\sigma_{k}\rightarrow k\sigma_{1}$.
Corrections to the large-$N$ limit start at $1/N$ for Casimir scaling
and $1/N^{2}$ for sine-law scaling, and it has been argued that the
large-$N$ limit requires $1/N^{2}$ corrections \cite{Armoni:2003nz}.
However, lattice simulations of $SU(N)$ gauge theories in $\left(2+1\right)$
dimensions indicate that $1/N$ corrections are strongly indicated
\cite{Bringoltz:2008nd}, and a careful analysis shows that Casimir
scaling is compatible with the large-$N$ limit \cite{Greensite:2011gg}.

\subsection{\label{sub:Center-symmetry}Pure gauge theories at finite temperature}

\begin{figure}
\includegraphics[scale=0.4]{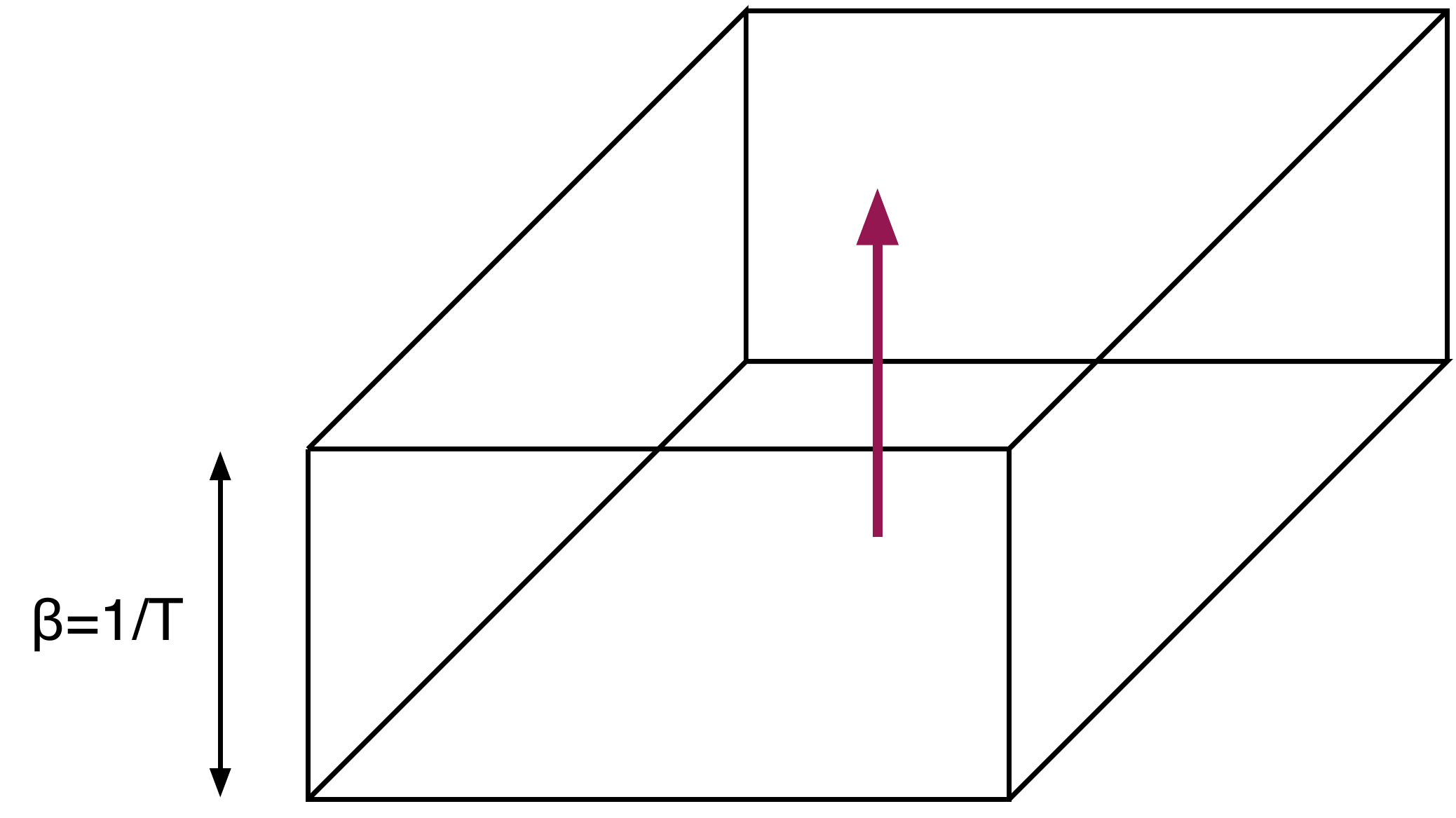}\caption{\label{fig:The-Polyakov-loop}The Polyakov loop is associated with
the worldline of a heavy particle.}
\end{figure}

\begin{figure}

\includegraphics[scale=0.4]{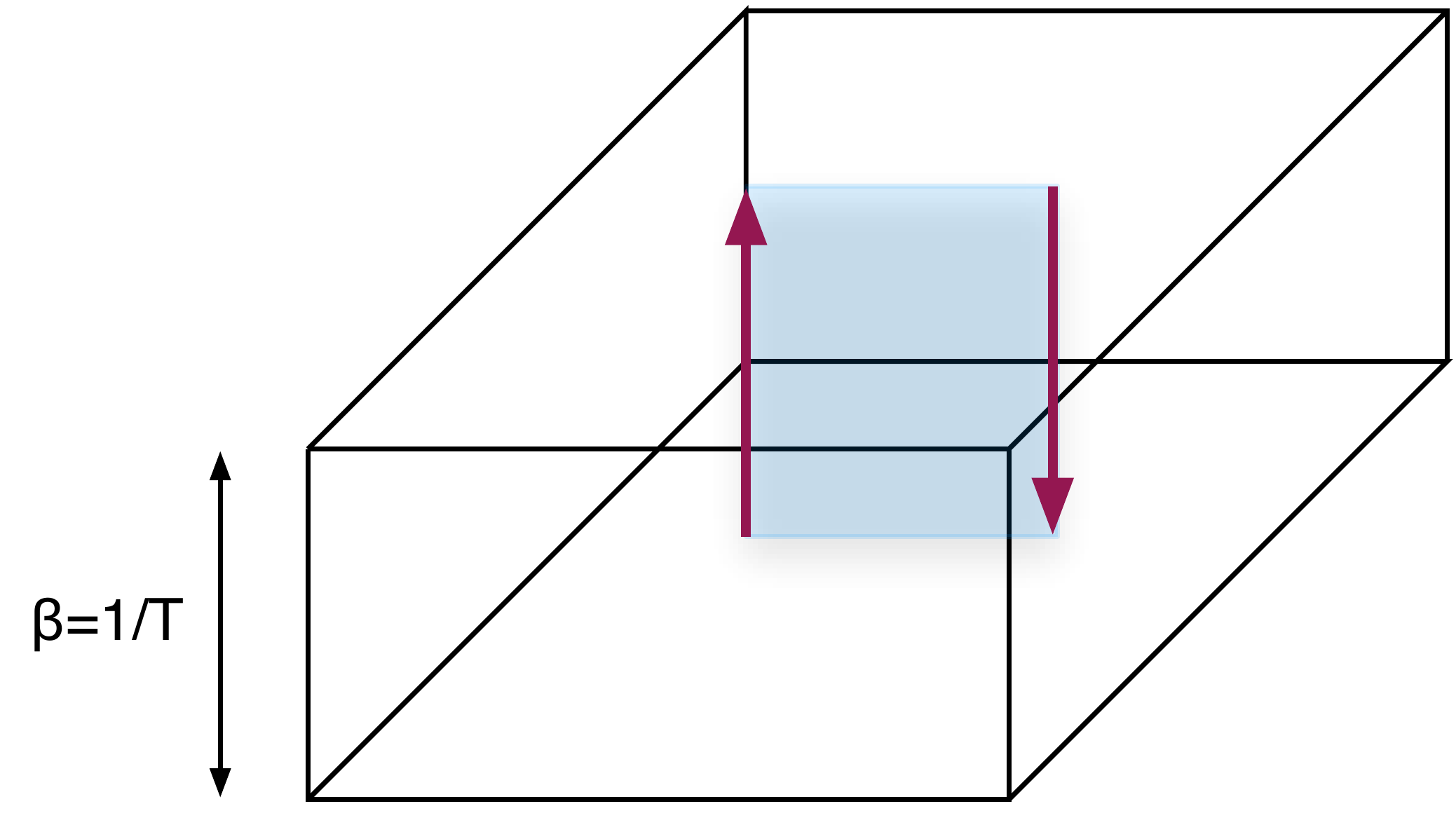}\caption{\label{fig:Polyakov-loop-2pt}The Polyakov loop two-point function
determines the electric string tension.}
\end{figure}

If one or more directions in space-time are compact, the string tension
may also be measured using the Polyakov loop $P$, also known as the
Wilson line. The Polyakov loop is essentially a Wilson loop that uses
a compact direction in space-time to close the curve using a topologically
non-trivial path in space time, as shown in figure \ref{fig:The-Polyakov-loop}. 
The typical use for the Polyakov loop
is for gauge theories at finite temperature, where space-time is $R^{3}\times S^{1}$.
The partition function being given by $Z=Tr\left[e^{-\beta H}\right]$,
the circumference of $S^{1}$ is given by the inverse temperature
$\beta=1/T$. In this case, we write
\begin{equation}
P\left(\vec{x}\right)=\mathcal{P}\exp\left[i\int_{0}^{\beta}dx_{4}A_{4}\left(x\right)\right]
\end{equation}
 and string tensions may be determined from a two-point function 
\begin{equation}
\left\langle Tr_{R}P\left(\vec{x}\right)Tr_{R}P^{\dagger}\left(\vec{y}\right)\right\rangle \sim e^{-\beta\sigma_{R}\left|\vec{x}-\vec{y}\right|}
\end{equation}
 a behavior that assumes that the one-point function $\left\langle Tr_{R}P\left(\vec{x}\right)\right\rangle =0$.
 This behavior is shown in figure \ref{fig:Polyakov-loop-2pt}.
The Polyakov loop one-point function $\left\langle Tr_{R}P\left(\vec{x}\right)\right\rangle $
can be interpreted as a Boltzmann factor $\exp\left(-\beta F_{R}\right)$,
where $F_{R}$ is the free energy required to add a static particle
in the representation $R$ to the system. Of course, $\left\langle Tr_{R}P\left(\vec{x}\right)\right\rangle =0$
implies that $F_{R}=\infty$ , which is thus a fundamental criterion
determining whether particles in the representation $R$ are confined.
Note that the introduction of a compact direction breaks the four-dimensional
symmetry of the theory, and the string tension measured by Polyakov
loops is not the same as the string tension measured by Wilson loops
lying in non-compact planes. In the case of finite temperature, it
is natural to use the terminology electric and magnetic string tension,
respectively. In the limit where the compactification radius becomes
large, \emph{e.g.} $\beta\rightarrow\infty$, the two string tensions
must coincide.

\subsubsection{Center symmetry}

One of the most important concepts in our understanding of confinement
is the role of center symmetry. The center of a Lie group is the set
of all elements that commute with every other element. For $SU(N)$,
this is $Z(N)$. Although the $Z(N)$ symmetry of $SU(N)$ gauge theories
can be understood from the continuum theory, it is easier to understand
from a lattice point of view. A lattice gauge theory associates link
variable $U_{\mu}\left(x\right)$ with each lattice site $x$ and
direction $\mu$. The link variable is considered to be the path-ordered
exponential of the gauge field from $x$ to $x+\hat{\mu}$: $U_{\mu}\left(x\right)=\exp\left[iaA_{\mu}\left(x\right)\right].$
Consider a center symmetry transformation on all the links in a given
direction on a fixed hyperplane perpendicular to the direction. The
standard example from $SU(N)$ gauge theories at finite temperature
is $U_{4}\left(\vec{x},t\right)\rightarrow zU_{4}\left(\vec{x},t\right)$
for all $\vec{x}$ and fixed $t$, with $z\in Z(N)$. Because lattice
actions such as the Wilson action consist of sums of small Wilson
loops, they are invariant under this global symmetry. However, the
Polyakov loop transforms as $P\left(\vec{x}\right)\rightarrow zP\left(\vec{x}\right)$,
and more generally 
\begin{equation}
Tr_{R}P\left(\vec{x}\right)\rightarrow z^{k_{R}}Tr_{R}P\left(\vec{x}\right)
\end{equation}
 where $k_{R}$ is an integer in the set $\left\{ 0,1,...,N-1\right\} $
and is known as the $N$-ality of the representation $R$. If $k_{R}\ne0$,
then unbroken global $Z(N)$ symmetry implies $\left\langle Tr_{R}P\left(\vec{x}\right)\right\rangle =0$.
Thus global $Z(N)$ symmetry defines the confining phase of a gauge
theory. For pure gauge theories at non-zero temperature, the deconfinement
phase transition is associated with the loss of $Z(N)$ symmetry at
the critical point $T_{d}$. Below that point $\left\langle Tr_{F}P\left(\vec{x}\right)\right\rangle =0$
but above $T_{d}$, $\left\langle Tr_{R}P\left(\vec{x}\right)\right\rangle \ne0$.

Notice that the case of zero $N$-ality representations is special
within this framework: there is no requirement from $Z(N)$ symmetry
that these representations are confined. This includes the adjoint
representation, the representation of the gauge particles. However,
lattice simulation indicate that $\left\langle Tr_{R}P\left(\vec{x}\right)\right\rangle $
is very small for these representations in the confined phase. Although
screening by gauge particles must dominate at large distances, these
zero $N$-ality representations have well-defined string tensions
at intermediate distances scales, \emph{e.g.}, on the order of a few
fermi for $SU(3)$, behaving in a manner very similar to representations
with non-zero $N$-ality \cite{Deldar:1999vi,Bali:2000un}.

\subsubsection{The nature of the confined phase and partially confined phases}

The requirements for confinement are simple: $Z(N)$ symmetry and
a mass gap. However, gauge theories with $SU(N)$ symmetry are not
simply $Z(N)$ systems. In models where the order parameter is an
element of $Z(N)$, the order parameter must average to zero in the
disordered phase, and the equilibrium disordered state must consist
of many domains, each with its own $Z(N)$ orientation. This need
not be the case with $SU(N)$ systems. There is a unique set of Polyakov
loop eigenvalues that are $Z(N)$ symmetric, and can form the basis
for a semiclassical understanding of confinement.

Although $Tr_{F}P$ is the order parameter for confinement of particles
in the fundamental representation, it does not by itself characterize
the confined phase, and the most general form of the free energy does
not depend on $P\,$ solely through $Tr_{F}P$ and $Tr_{F}P^{\dagger}$
\cite{Meisinger:2001cq}. The high-temperature effective potential
$V_{eff}\left(P\right)$ illustrates this point. It is not a function
solely of $Tr_{F}P$ and its conjugate, and cannot be written as an
infinite series in $Tr_{F}P$ and $Tr_{F}P^{\dagger}$. A simple example
will illustrate this point. Consider two diagonal matrices lying in
$SU(4)$, defined by
\begin{equation}
P_{1}=\left(\begin{array}{llll}
e^{i\pi/4}\\
 & e^{i3\pi/4}\\
 &  & e^{i5\pi/4}\\
 &  &  & e^{i7\pi/4}
\end{array}\right)
\end{equation}
and

\begin{equation}
P_{2}=\left(\begin{array}{llll}
e^{i\pi/2}\\
 & e^{i\pi/2}\\
 &  & e^{-i\pi/2}\\
 &  &  & e^{-i\pi/2}
\end{array}\right)
\end{equation}
 Both $P_{1}$ and $P_{2}$ have zero trace in the fundamental representation,
yet the traces of their square are differenct: $Tr_{F}P_{1}^{2}=0$
and $Tr_{F}P_{2}^{2}=-4$, establishing that $Tr_{F}\, P^{2}$ cannot
be a function solely of $Tr_{F}P$. 

At first sight, this may seem to contradict two standard results:
a) the characters form a complete, in fact orthogonal, basis on class
functions; b) all characters may be obtained from the fundamental
representation by repeated multiplication and the application of 
\begin{equation}
\chi_{a}(P)\chi_{b}(P)=\sum_{c}n\left(a,b;c\right)\chi_{c}(P)
\end{equation}
where all $n$'s are non-negative integers. Taken together, these
results might suggest that all group characters are polynomials in
$Tr_{F}U$ and its complex conjugate. Consider, however, the product
representation $N\otimes N$. It is reducible into $N(N+1)/2\oplus N(N-1)/2$,
which are symmetric and antisymmetric representations, respectively.
Their characters are respectively 
\begin{eqnarray}
\chi_{S}\left(P\right) & = & \frac{1}{2}\left[\left(Tr_{F}P\right)^{2}+Tr_{F}\left(P^{2}\right)\right]\nonumber \\
\chi_{A}\left(P\right) & = & \frac{1}{2}\left[\left(Tr_{F}P\right)^{2}-Tr_{F}\left(P^{2}\right)\right]\text{.}
\end{eqnarray}
 Note that the sum $\chi_{S}+\chi_{A}$ is a polynomial in $Tr_{F}P$,
but $\chi_{S}$ and $\chi_{A}$ are in general not. For example, the
restriction on the eigenvalues of $P$ imposed by $\det\left(P\right)=1$
allows us to prove that for $SU(3)$ 
\begin{equation}
\frac{1}{2}\left[\left(Tr_{F}P\right)^{2}-Tr_{F}\left(P^{2}\right)\right]=Tr_{F}\left(P^{\dagger}\right)
\end{equation}
in accord with the $SU(3)$ result $3\otimes3=6\oplus\overline{3}$.
In this case, it is true that $Tr_{F}\left(P^{2}\right)$ can be written
as a polynomial in $Tr_{F}P$ and $Tr_{F}P^{\dagger}$. However, for
$SU(4)$, unitarity of $P$ gives instead 
\begin{equation}
\frac{1}{2}\left[\left(Tr_{F}P\right)^{2}-Tr_{F}\left(P^{2}\right)\right]=\frac{1}{2}\left[\left(Tr_{F}P\right)^{2}-Tr_{F}\left(P^{2}\right)\right]^{\ast}
\end{equation}
 which shows that the $6$ representation of $SU(4)$ is real. $SU(N)$
group characters can be written as polynomials in $Tr_{F}\left(P\right)$
and its complex conjugate only for $SU(2)$ and $SU(3)$.

An alternative statement is that the Polyakov loop in the fundamental
representation, $Tr_{F}U$, is not sufficient to determine the eigenvalues
of $P$, beginning with the case of $SU(4)$. Let us label the eigenvalues
for the Polyakov loop as $z_{i}$, with $i=1$ to $N$. For $SU(3)$,
the characteristic polynomial for the Polyakov loop is 
\begin{equation}
\prod_{i=1}^{3}(z-z_{i})=z^{3}-(z_{1}+z_{2}+z_{3})\, z^{2}+(z_{1}z_{2}+z_{2}z_{3}+z_{3}z_{1})\, z-z_{1}z_{2}z_{3}\text{.}
\end{equation}
 Since the determinant of a special unitary matrix is $1$, we have
$z_{1}z_{2}z_{3}=1$, and the characteristic polynomial is 
\begin{equation}
z^{3}-z^{2}Tr_{F}P\,+zTr_{F}P^{+}-1
\end{equation}
so for $SU(3)$, knowledge of $Tr_{F}P$ determines all eigenvalues,
and the free energy can be written as a function of $Tr_{F}P$ alone.
For $SU(4)$, similar considerations allow the characteristic polynomial
to be written as 
\begin{equation}
z^{4}-z^{3}Tr_{F}P+z^{2}\frac{1}{2}\left[\left(Tr_{F}P\right)^{2}-Tr_{F}\left(P^{2}\right)\right]-zTr_{F}P^{+}+1
\end{equation}
and knowledge of $Tr_{F}P$ must be supplemented by the value of $Tr_{F}\left(P^{2}\right)$.
As $N$ increases, more information must be supplied to reconstruct
the eigenvalues.

The characteristic equation for Polyakov loops in $SU(N)$ may be
written as
\begin{equation}
\det\left[P-\lambda I\right]=\sum_{j=0}^{N}\left(-\lambda\right)^{j}\chi_{j}\left(P\right)
\end{equation}
where $\chi_{j}$ is the character for the representation formed from
the antisymmetrized product of the fundamental $j$ times; the corresponding
Young tableaux is $j$ vertical boxes. Of course, $\chi_{j}\left(P\right)$
has $N$-ality $ $$j$, and $\chi_{0}\left(P\right)=\chi_{N}\left(P\right)=1$.
At the classical level, a Polyakov loop that represents the confining
phase must have $\chi_{j}\left(P\right)=0$ for $1\le j\le N-1$,
leading to a characteristic equation $1+\left(-\lambda\right)^{N}=0$.
This in turn implies the important result that any model in which
confinement is associated with a particular field configuration (up
to gauge transformations) must have Polyakov loop eigenvalues evenly
spaced on the unit circle in the confined phase. This is precisely
what must happen in the large-$N$ limit, and is known to occur in
analytically tractable lattice models of the large-$N$ limit \cite{Billo:1996pu}.
This behavior is not possible in simple $Z(N)$ models where symmetric
phases always represent some average over different $Z(N)$ domains.
However, it should be noted that there are $Z(N)$ spin systems with
features similar to models with variables in $SU(N)$. The best-known
is the Blume-Emory-Griffiths model, which was originally proposed
as a simple model of $^{3}He$-$^{4}He$ mixtures \cite{BlumePhysRevA.4.1071}.
Essentially, an Ising spin model is extended such that each spin $\sigma$
can have the value $0$ as well as $\pm1$. The behavior of such extended
$Z(N)$ lattice models is very similar to treatments of $SU(N)$ lattice
models with fundamental and adjoint interactions, where the adjoint
interaction can be adjusted to give a $Z(N)$ limit \cite{Halliday:1981te,Halliday:1981tm};
see also \cite{Drouffe:1983fv}.

In the confining phase of a gauge theory, $Z(N)$ symmetry is unbroken,
and all representations with non-zero $N$-ality are confined. In
the deconfined phase, $Z(N)$ symmetry is completely lost, and particles
are no longer confined, independent of their representation. For $N\ge4$,
additional phases are possible where $Z(N)$ is broken down to a non-trivial
subgroup \cite{Meisinger:2001cq,Myers:2007vc,Myers:2009df,Meisinger:2009ne}.
In the case of $Z(4)$, there can be breaking of center symmetry down
to $Z(2)$. In this partially confined phase, states consisting of
two fundamental representation fermions are not confined, but single
fermions are. In the case of $SU(N)$, $Z(N)$ can break to $Z(k)$,
where $k$ is any divisor of $N$. States with $1$ to $k-1$ fermions
are confined, but states with $k$ fermions are not. It is often convenient
to include the confined and deconfined phases as $k=N$ and $k=1$,
respectively. As we will see in section \ref{sec:Phases-on-R3xS1},
such partially confined phases been found in gauge theories on $R^{3}\times S^{1}$,
using both lattice simulations and perturbation theory \cite{Myers:2007vc}.
It should be noted that not all gauge groups have non-trivial centers.
The gauge group $G(2)$ provides an interesting example of a gauge
theory without a center \cite{Holland:2003jy} that has received significant
attention \cite{Pepe:2005sz}.

\subsubsection{Origin of the deconfinement transition}

As we have seen in section \ref{sec:Modern-theory-of}, the critical
properties of a four-dimensional field theory at finite temperature
can often best be understood in terms of an effective three-dimensional
theory. From this point of view, the Polyakov loop can be considered
to be a scalar order parameter with $Z(N)$ invariance in some effective
three-dimensional theory. A complete treatment of the effective potential,
including non-perturbative effects, would presumably yield the phase
structure, including both the confined and deconfined phases. 

Perturbation theory is a reliable indicator of broken center symmetry
and thus deconfinement at high temperature, because the running coupling
constant $g(T)$ is small if $T\gg\Lambda$. The one-loop effective
potential for a pure gauge theory in the background of a static Polyakov
loop $P$ can be easily evaluated in a gauge where $A_{4}$ is time-independent
and diagonal \cite{Gross:1980br,Weiss:1980rj}. It is easy to see
that $V_{eff}^{1l}\left(P\right)$ is given by
\begin{equation}
V_{eff}^{1l}\left(P\right)=2T\, Tr_{Adj}\sum_{n\in Z}\int\frac{d^{3}k}{\left(2\pi\right)^{3}}\log\left[\left(2\pi nT-A_{4}\right)^{2}+\vec{k}^{2}\right]
\end{equation}
where the factor of $2$ represents the two helicity states of each
mode. Note that there is no classical contribution. Discarding the
zero-point energy term, we obtain the one-loop finite-temperature
effective potential for gauge bosons
\begin{equation}
V_{eff}^{1l}\left(P\right)=2T\, Tr_{Adj}\int\frac{d^{3}k}{\left(2\pi\right)^{3}}\log\left[1-P\,\exp\left(-\left|\vec{k}\right|/T\right)\right]
\end{equation}
which is the free energy density of the gauge bosons in the background
of the Polyakov loop $P$.

The logarithm in this expression for $V_{eff}^{1l}\left(P\right)$
can be expanded, leading to an interpretation of $V_{eff}^{1l}\left(P\right)$
as a sum of contributions from gluon worldlines wrapping around the
compact direction an arbitrary number of times. Explicitly, we have
the expression
\begin{equation}
V_{eff}^{1l}\left(P\right)=-\frac{2}{\pi^{2}}\sum_{n=1}^{\infty}\frac{1}{n^{4}}Tr_{Adj}P^{n}.
\label{eq:V_pure_glue}
\end{equation}
From this form, it is easy to see that $ $$V_{eff}^{1l}\left(P\right)$
is minimized when all the moments $Tr_{A}P^{n}$ are maximized. This
occurs when $P\in Z(N)$, which gives $Tr_{A}P^{n}=N^{2}-1$. This
indicates that the one-loop gluon effective potential favors the deconfined
phase. The pressure $p$ is the negative of the free energy density
at the minimum, 
\begin{equation}
p\left(T\right)=2\left(N^{2}-1\right)\frac{\pi^{2}T^{4}}{90},
\end{equation}
 which is exactly $p$ for a blackbody with $2\left(N^{2}-1\right)$
degrees of freedom. 

In the gauge where $A_{4}$ is diagonal and time-independent, we can
parametrize $A_{4}$ in the fundamental representation of $SU(N)$
as a diagonal, traceless $N\times N$ matrix
\begin{equation}
\left(A_{4}\right)_{jk}=T\theta_{j}\delta_{jk}
\end{equation}
so that
\begin{equation}
P_{jk}=e^{i\theta_{j}}\delta_{jk}
\end{equation}
with $\sum_{j=1}^{N}\theta_{j}=0$. Using the decomposition $F\otimes\bar{F}=1\oplus Adj$
and the corresponding decomposition
\begin{equation}
Tr_{F}P\, TrP^{+}=1+Tr_{Adj}P,
\end{equation}
one realizes that the $N^{2}$ eigenvalues of the product representation
$F\otimes\bar{F}$ have the form $\exp\left[i\Delta\theta_{jk}\right]$
, where we define $\Delta\theta_{jk}\equiv\theta_{j}-\theta_{k}$.
The effective potential is given by
\begin{equation}
V_{eff}^{1l}\left(P\right)=-\frac{2}{\pi^{2}}\sum_{j,,k=1}^{N}(1-\frac{1}{N}\delta_{jk})\sum_{n=1}^{\infty}\frac{1}{n^{4}}\exp\left[in\Delta\theta_{jk}\right].
\end{equation}
The infinite sum over $n$ may be carried out explicitly in terms
of the fourth Bernoulli polynomial. For our purposes, a convenient
explicit form is 
\begin{equation}
V_{eff}^{1l}\left(P\right)=-T^{4}\sum_{j,k=1}^{N}(1-\frac{1}{N}\delta_{jk})\left[\frac{\pi^{2}}{45}-\frac{1}{24\pi^{2}}\left|\Delta\theta_{jk}\right|_{2\pi}^{2}\left(\left|\Delta\theta_{jk}\right|_{2\pi}-2\pi\right)^{2}\right]
\end{equation}
where $\left|\Delta\theta_{jk}\right|_{2\pi}$ lies in the interval
between 0 and $2\pi$.

In many systems, broken symmetry phases are found at low temperatures
and symmetry is restored at high temperatures. The phase structure
of gauge theories as a function of temperature is unusual because
the broken-symmetry phase is the high-temperature phase. A lattice
construction of the effective action for Polyakov loops, valid for
strong-coupling, is instructive \cite{Polonyi:1982wz,Ogilvie:1983ss,Green:1983sd,Gross:1984wb}.
The spatial link variables may be integrated out exactly if spatial
plaquette interactions are neglected. Each spatial link variable then
appears only in two adjacent temporal plaquettes, and may be integrated
out exactly using the same techniques that are used in the Migdal-Kadanoff
real-space renormalization group \cite{Ogilvie:1983ss,Billo:1996pu}.
The resulting effective action has the form
\begin{equation}
S_{eff}=\sum_{\left\langle jk\right\rangle }K\,\left[Tr_{F}P_{j}Tr_{F}P_{k}^{\dagger}+Tr_{F}P_{k}Tr_{F}P_{j}^{\dagger}\right]
\end{equation}
where $K$ is a function of the lattice gauge coupling $g^{2}$ and
the extent of the lattice in the Euclidean time direction $n_{t}$,
which is related to the temperature by $n_{t}a=1/T$. In the strong-coupling
limit of the underlying gauge theory, the explicit form for $K$ is
$K\simeq\left(1/g^{2}N\right)^{n_{t}}$ to leading order. In the weak-coupling
limit, a Migdal-Kadanoff bond-moving argument gives $K\simeq2N/g^{2}n_{t}$.
This effective action represents a $Z(N)$-invariant nearest-neighbor
interaction of a spin system where the Polyakov loops are the spins.
It depends only on gauge-invariant quantities. Standard expansion
techniques show that the $Z(N)$ symmetry is unbroken for small $K$,
and broken for $K$ large. This model explains why the high-temperature
phase of gauge theries is the symmetry-breaking phase: the relation
between $K$ and the underlying gauge theory parameters is such that
$K$ is small at low temperatures, and large at high temperatures,
exactly the reverse of a classical spin system where the coupling
is proportional to $T^{-1}$. For small values of $n_{t}$, the deconfinement
transition can be easily extracted, but the phase transition is in
the strong-coupling region and far from the continuum limit. A systematic
treatment of strong-coupling corrections has recently been shown to
yield values for the critical lattice couplings $\beta_{c}\equiv2N/g^{2}$
for $SU(2)$ and $SU(3)$ that are within a few percent of simulation
results for $4\le N_{t}\le16$ \cite{Langelage:2010yr}. 
For Abelian lattice gauge theories in $(2+1)$-dimensions, a duality
transformation maps the high-temperature symmetry-breaking phase
of the gauge theory into the high-temperature unbroken phase of
the dual spin system, and the low-temperature symmetry-breaking
phase of the spin system into the unbroken phase of the gauge theory,
consistent with the general behavior seen using strong-coupling
arguments.

\subsubsection{Universality}

The lattice construction of the Polyakov loop effective action is
a concrete realization of Svetitsky-Yaffe universality \cite{Svetitsky:1982gs},
which states that a second-order deconfinement transition in a $(d+1)$-dimensional
gauge theory is in the universality class of classical spin systems
in $d$ dimensions with the same global symmetry. Lattice simulations
indicate that all pure $SU(N)$ gauge theories have a deconfining
phase transition at some temperature $T_{d}$, above which center
symmetry is broken. In accordance with predictions based on universality,
the deconfinement transition for an $SU(2)$ gauge theory in $3+1$
dimensions has been well-established as being in the universality
class of the three-dimensional Ising model, exhibiting a second-order
transition at $T_{d}$. The deconfinement transition for $SU(3)$
in $3+1$ dimensions is first-order. This is consistent with Landau-Ginsburg
predictions for a system with a $Z(3)$ symmetry. The transitions
for $N>3$ appear to be first-order in $3+1$ dimensions as well, with a smooth limit as
$N$ goes to infinity \cite{Lucini:2002ku,Lucini:2003zr}, so
the direct applicability of the universality argument in $3+1$ dimensions
is more limited than one might have expected.

\subsubsection{Equation of state}

Lattice simulations can be used to determine both the pressure $p\left(T\right)$
and the internal energy $\epsilon\left(T\right)$, along with the
deconfinement transition temperature $T_{d}$ and related properties
of the transition; see \emph{e.g. \cite{Karsch:2001cy}} and references
therein. For pure gauge theories below $T_{d}$, both $p$ and $\epsilon$
are extremely small. This is expected because the mass of the lightest
color singlet state, the scalar glueball, is substantially larger
than $T_{d}$. The pressure must be continuous at $T_{d}$, but the
first-order character of the deconfinement transition for $N\ge3$
leads to a latent heat, indicated by a discontinuity in $\epsilon$.
The thermodynamics quantities $p$ and $\epsilon$ as well as the order
parameter $\left\langle Tr_{F}P\right\rangle $, show a rapid rise
in the interval from $T_{d}$ to roughly $3T_{d}$. Both $p$ and
$\epsilon$ show a monotonic but slow rise towards their blackbody
values at higher temperatures.

The pressure $p\left(T\right)$ and related thermodynamic quantities
can be calculated in perturbation theory in terms of $g^{2}\left(T\right)$,
the running coupling at temperature $T$. However, there is a barrier
to perturbative calculations at $\mathcal{O}\left(g^{6}\right)$,
due to infrared divergences in the magnetic sector \cite{Linde:1980ts}.
The best that can be done in perturbation theory is an expression
for $p\left(T\right)$ that is valid up to $\mathcal{O}\left(g^{6}\log g\right)$
\cite{Kajantie:2002wa}. Comparison of different order of perturbation
theory up to $\mathcal{O}\left(g^{6}\log g\right)$ shows a reasonable
convergence only at temperatures greater than about $10\Lambda_{\overline{MS}}$.

\subsubsection{Phenomenology of the deconfinement transition}

The availability of high-quality lattice data for the $SU(3)$ pressure
\cite{Boyd:1996bx,Papa:1996an,Beinlich:1997ia,Okamoto:1999hi} has
led to a variety of attempts to model it phenomenologically. Deconfinement
can be characterized broadly as a change in the number of degrees
of freedom as the temperature is raised, and quark and gluon degrees
of freedom manifest in thermodynamic behavior. The key to building
a successful model is incorporating a mechanism for characterizing
the change in degrees of freedom with temperature. Phenomenological
models built around minimizing the free energy as a function of the
order parameter are appealing and simple \cite{Meisinger:2001cq,Dumitru:2010mj,Dumitru:2012fw}. 

One simple model \cite{Meisinger:2001cq,Nishimura:2009me} follows
from noting that the one-loop free energy for massive particles in
a Polyakov loop background lead to expansions in which the first two
terms have simple forms when $M/T\ll1$. Thus we arrive at a potential
of the form
\begin{equation}
V_{G}\left(P\right)=-\frac{2T^{4}}{\pi^{2}}\sum_{n=1}^{\infty}\frac{Tr_{Adj}P^{n}}{n^{4}}+\frac{M^{2}T^{2}}{2\pi^{2}}\sum_{n=1}^{\infty}\frac{Tr_{Adj}P^{n}}{n^{2}}.\label{eq:V_G}
\end{equation}
It is important to note that the mass parameter $M$ should not be
interpreted as a gauge boson mass, nor do we limit ourselves to $ML\ll1$:
the additional term in $ $$V_{g}$ is purely phenomenological. The
crucial feature of this potential is that for sufficiently large values
of the dimensionless parameter $M/T$, the second term dominates and
the potential leads to a $Z(N)$-symmetric, confining minimum for
$P$. On the other hand, for small values of $M/T$, the first term
dominates and the pure gauge theory will be in the deconfined phase.
This model has the nice property that $V_{G}$ is a good representation
of the gauge boson contribution for high temperatures. 

\begin{figure}
\includegraphics[scale=0.8]{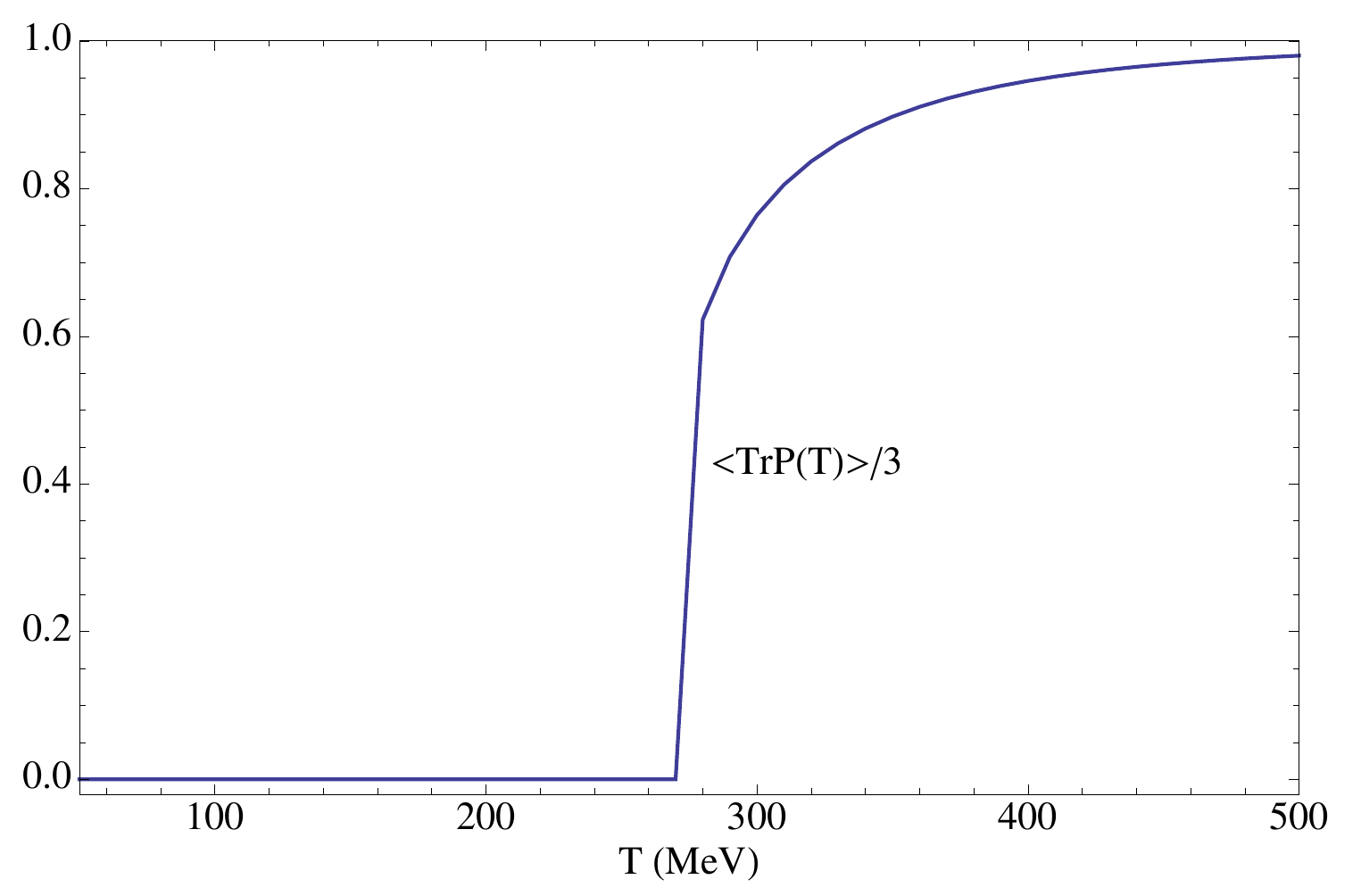}

\caption{\label{fig:TrP-vs-T}The Polyakov loop $Tr_{F}P$ as a function of
$T$ in $SU(3)$ using a phenomenological model \cite{Meisinger:2001cq,Nishimura:2009me}.}
\end{figure}

Minimizing $V_{G}$ as a function of the eigenvalues of $P$ leads
to expressions for the pressure and associated thermodynamic quantities.
In the case of $SU(3)$, we can set the mass scale $M$ by requiring
that $V_{g}$ yields the correct deconfinement temperature for the
pure gauge theory, with a value of $T_{d}=270\, MeV$, giving $M=596\, MeV$.
The pressure $p$ is given by the value of $-V_{G}$ at the minimum
of the potential. The behavior obtained for the pressure, the energy
density $\epsilon$, and the interaction measure $\Delta\equiv\left(\epsilon-3p\right)/T^{4}$
are all roughly consistent with lattice simulations for $T>T_{d}$
and can be improved with the use of one or two additional free parameters
\cite{Dumitru:2010mj,Dumitru:2012fw}. As shown in Fig. \ref{fig:TrP-vs-T},
the order parameter $TrP$ for the deconfinement transition jumps
at $T_{d}$, indicating a first-order deconfinement transition for
$SU(3)$. 

\subsection{\label{sub:Role-of-quarks}Role of quarks and other particles}

\subsubsection{Chiral symmetry breaking}

If confinement is the most striking feature of QCD, chiral symmetry
breaking is certainly a close second. If we imagine for a moment a
variant of QCD in which all $N_{f}$ quarks have the same mass, there
will be a global $SU(N_{F})$ symmetry which rotates the different
flavors into each other. In this theory, the left- and right-handed
components of the Dirac fields, defined by
\begin{equation}
q_{L/R}=\frac{1}{2}(1\pm\gamma_{5})q
\end{equation}
 are connected by the Lagrangian mass term
\begin{equation}
m\bar{qq}=m\left(\bar{q}_{L}q_{R}+\bar{q}_{R}q_{L}\right)
\end{equation}
which can be thought of as an interaction that flips the helicity
of a quark. If the masses of the quarks are set to zero, then the
left- and right-handed components are no longer coupled, and the symmetry
group becomes $SU(N_{f})_{L}\otimes SU(N_{f})_{R}$. At the classical
level, there is also a $U(1)_{A}$ axial symmetry generated by $\gamma_{5}$,
but it is anomalous.

In our universe, the two lightest quarks, the $u$ and $d$ quarks,
have masses $m_{u}$ and $m_{d}$ which are more than an order of
magnitude smaller than $\Lambda_{QCD}$, so we are close to having
an $SU(2)_{L}\otimes SU(2)_{R}$ symmetry. On the basis of models
and, more recently, lattice simulations, we believe that in the massless
limit this symmetry would be spontaneously broken to $SU(2)$, the
symmetry group of the massive theory, by the existence of a condensate
$\left\langle 0\left|\bar{u}u+\bar{d}d\right|0\right\rangle $. The
breaking of this continuous global symmetry leads to the existence
of three Goldstone bosons, the pions.

Unfortunately, our ability to understand chiral symmetry breaking
analytically has been limited. We assume that the basic mechanism
is similar to the mechanism of superconductivity. In the case of superconductivity,
electron-hole interactions mediated by phonons lead to an expectation
value for a composite field, while in gauge theories, the gauge boson
interactions somehow give rise to some effective four-fermion interaction
which leads to the formation of a $\bar{q}q$ condensate. However,
there is no convincing derivation of this effective interaction from
the gauge theory. In the absence of a more fundamental understanding,
 Nambu-Jona Lasinio (NJL) models are often used to describe the chiral
symmetry effects. In NJL models, a four-fermion interaction induces
chiral symmetry breaking; the four-fermion interaction is non-renormalizable,
and NJL models cannot be regarded as fundamental. There has been a
great deal of work on NJL models, both as phenomenological models
for hadrons and as effective theories of QCD \cite{Klevansky:1992qe,Hatsuda:1994pi}.

NJL models use purely fermionic interactions as a proxy for the gauge
theory interactions that give rise to chiral symmetry breaking. Typically,
the relation between the gauge theory and an associated NJL model
is fixed by matching important hadronic parameters such as $f_{\pi}$.
In the case of NJL models at finite temperature, and particularly
PNJL models, it is common to assume that the NJL model parameters
are fixed by $T=0$ hadronic parameters and remain constant as $T$
is increased, at least up to the deconfinement temperature. 

As an example of an NJL model consider the representative model \cite{Nishimura:2009me}
\begin{equation}
L_{NJL}=\bar{\psi}\left(i\gamma\cdot\partial-m_{0}\right)\psi+\frac{g_{S}}{2}\left[\left(\bar{\psi}\lambda^{a}\psi\right)^{2}+\left(\bar{\psi}i\gamma_{5}\lambda^{a}\psi\right)^{2}\right]+g_{D}\left[\det\bar{\psi}\left(1-\gamma_{5}\right)\psi+h.c.\right]
\end{equation}
where $\psi$ is associated with $N_{f}$ flavors of Dirac fermions
in the fundamental or adjoint representation of the gauge group $SU(N)$.
The $\lambda^{a}$'s are the generators of the flavor symmetry group
$U(N_{f})$; $g_{S}$ represents the strength of the four-fermion
scalar-pseudoscalar coupling and $g_{D}$ fixes the strength of an
anomaly induced term. For simplicity, we can take the mass matrix
$m_{0}$ to be diagonal: $\left(m_{0}\right)_{jk}=m_{0j}\delta_{jk}$.
If $m_{0}$ and $g_{D}$ are taken to be zero, $L_{NJL}$ is invariant
under the global symmetry $U(N_{f})_{L}\times U(N_{f})_{R}$. With
a nonzero, flavor-independent mass $m_{0}$ and $g_{D}\neq0$, the
symmetry is reduced to $SU(N_{f})_{V}\times U(1)_{V}$. This model
can be solved approximately in a self-consistent manner. The effective
potential $V_{F0}$ can be calculated at zero temperature as a function
of the chiral condensates $\sigma_{j}\equiv\bar{\psi}_{j}\psi_{j}$
: 
\begin{eqnarray}
V_{F0}\left(m_{j}\right) & = & \sum_{j}g_{S}\sigma_{j}^{2}+2g_{D}\left(N_{f}-1\right)\prod_{j}\sigma_{j}-2N\sum_{j=1}^{N_{f}}\int\frac{d^{3}k}{(2\pi)^{3}}\omega_{k}^{\left(j\right)}\label{eq:V_F0}
\end{eqnarray}
where the last term representing the sum of fermionic zero-point energies
from the functional determinant. The zero-point energy of each mode
$\omega_{k}^{\left(j\right)}=\sqrt{k^{2}+m_{j}^{2}}$, is written
in terms of the constituent mass $m_{j}$, given in this model as
\begin{equation}
m_{j}\equiv m_{j}^{0}-2g_{S}\sigma_{j}-2g_{D}\left(N_{f}-1\right)\prod_{k\ne j}\sigma_{k}.
\end{equation}
The sum over zero-point energies is divergent, and requires regularization.
Differentiation with respect to $\sigma_{j}$ yields a set of gap
equations which can be solved numerically for the chiral condesate.

\subsubsection{Quarks and deconfinement}

The addition of quarks or other particles to a gauge theory can completely
change the theory's finite temperature behavior. Such particles can
directly affect deconfinement, because they can alter or even destroy
center symmetry. These effects can be seen directly in the effective
potential.

The calculation of the one-loop effective potential for gauge bosons
can be extended to other particles such as quarks in a very general
way \cite{Meisinger:2001fi}. Because the effective potential is central
for subsequent discussion, we treat the calculation in detail. All
of the required expressions can be obtained from the one-loop effective
potential for a charged scalar boson of mass $M$ moving in a background
$U(1)$ Polyakov loop in $d$ spatial dimensions. This is given by
\begin{equation}
V_{B}\left(\theta\right)=\frac{1}{\beta}\int\frac{d^{d}k}{\left(2\pi\right)^{d}}\ln\left[1-e^{-\beta\omega_{k}+i\theta}\right]+\frac{1}{\beta}\int\frac{d^{d}k}{\left(2\pi\right)^{d}}\ln\left[1-e^{-\beta\omega_{k}-i\theta}\right]
\end{equation}
where $\omega_{k}$ is given by $\sqrt{k^{2}+M^{2}}$. Now consider
a scalar boson in the fundamental representation of an $SU(N)$ gauge
group moving in a uniform Polyakov loop background. The one-loop contribution
to the finite-temperature effective potential is
\begin{equation}
V=\frac{1}{\beta}\int\frac{d^{d}k}{\left(2\pi\right)^{d}}Tr_{F}\left[\ln\left(1-Pe^{-\beta\omega_{k}}\right)+\ln\left(1-P^{\dagger}e^{-\beta\omega_{k}}\right)\right]
\end{equation}
where the first term is due to particles and the second term is due
to antiparticles. A global unitary transformation puts $P$ into the
diagonal form 
\begin{equation}
P_{jk}=\delta_{jk}\exp\left(i\theta_{j}\right)
\end{equation}
 and the partition function can be written as 
\begin{equation}
V=\sum_{j}V_{B}\left(\theta_{j}\right)\text{.}
\end{equation}
As a second example, consider the case of the gauge bosons themselves,
which lie in the adjoint representation of the gauge group. The Polyakov
loop in the adjoint representation is an $\left(N^{2}-1\right)\times\left(N^{2}-1\right)$
matrix. The partition function for the $N^{2}-1$ particles is 
\begin{equation}
s\frac{1}{2}\sum_{j,k=1}^{N}(1-\frac{1}{N}\delta_{jk})V_{B}\left(\theta_{j}-\theta_{k}\right)
\end{equation}
 where the $\delta_{jk}$ removes a singlet contribution, and the
factor of $1/2$ corrects for over-counting since $V_{B}$ has both
a particle and antiparticle contribution. The factor $s$ accounts
for spin degeneracy; in $3+1$ dimensions $s=2$, a consequence of
the two possible polarization states of gauge bosons.

For our third and final example, consider the evaluation of fermionic
partition functions, which can be reduced to the general bosonic problem.
A typical fermionic contribution of particle and antiparticle has
the form 
\begin{equation}
V_{F}\left(\theta\right)=-\frac{1}{\beta}\int\frac{d^{d}k}{\left(2\pi\right)^{d}}\ln\left[1+e^{-\beta\omega_{k}+i\theta}\right]-\frac{1}{\beta}\int\frac{d^{d}k}{\left(2\pi\right)^{d}}\ln\left[1+e^{-\beta\omega_{k}-i\theta}\right]
\end{equation}
which is easily written as 
\begin{equation}
V_{F}\left(\theta\right)=-V_{B}\left(\pi+\theta\right)\text{.}
\end{equation}
For fermions in the fundamental representation of $SU(N)$, the partition
function is 
\begin{equation}
s\sum_{j}V_{F}\left(\theta_{j}\right)=-s\sum_{j}V_{B}\left(\pi+\theta_{j}\right)
\end{equation}
 where the factor $s$ again accounts for spin degeneracy.

A low-temperature expansion for $V_{B}\left(\theta\right)$ can be
generated for arbitrary spatial dimension $d$ by expanding the logarithm
and integrating term by term:
\begin{equation}
V_{B}\left(\theta\right)=-\frac{M^{d/2+1/2}}{2^{d/2-3/2}\pi^{d/2+1/2}\beta^{d/2+1/2}}%
\sum_{n=1}^{\infty}\frac{1}{n^{d/2+1/2}}K_{\left(d+1\right)/2}\left(n\beta M\right)\cos\left(n\theta\right)
\end{equation}
which gives for scalar bosons in the fundamental representation
\begin{equation}
V_{B}\left(\theta\right)=-\frac{M^{d/2+1/2}}{2^{d/2-1/2}\pi^{d/2+1/2}\beta^{d/2+1/2}}%
\sum_{n=1}^{\infty}\frac{1}{n^{d/2+1/2}}K_{\left(d+1\right)/2}\left(n\beta M\right)Tr_{F}\left(P^{n}+P^{\dagger n}\right)
\end{equation}
with similar results for bosons in other representations. For fermions,
we have 
\begin{equation}
V_{F}\left(\theta\right)=\frac{M^{d/2+1/2}}{2^{d/2-3/2}\pi^{d/2+1/2}\beta^{d/2+1/2}}\sum_{n=1}^{\infty}\frac{\left(-1\right)^{n}}{n^{d/2+1/2}}K_{\left(d+1\right)/2}\left(n\beta M\right)\cos\left(n\theta\right)\text{.}
\end{equation}
In a path integral representation, the factors of $\left(-1\right)^{n}$
are a consequence of fermionic antiperiodic boundary conditions.

A high-temperature expansion for $V_{B}\left(\theta\right)$ may also
be obtained in $3+1$ dimensions
\begin{eqnarray}
V_{B}\left(\theta\right) & = & -\frac{2}{\pi^{2}\beta^{4}}\left[\frac{\pi^{4}}{90}-\frac{1}{48}\theta_{+}^{4}+\frac{\pi}{12}\theta_{+}^{3}-\frac{\pi^{2}}{12}\theta_{+}^{2}\right]+\frac{M^{2}}{2\pi^{2}\beta^{2}}\left[\frac{1}{4}\theta_{+}^{2}-\frac{\pi}{2}\theta_{+}+\frac{\pi^{2}}{6}\right]\label{eq:high-T-expansion}\\
 &  & -\frac{1}{2\pi\beta^{4}}\sum_{l \in Z}R\left(\beta M,\theta,l\right)-\frac{M^{4}}{16\pi^{2}}\left[\ln\left(\frac{\beta M}{4\pi}\right)+\gamma-\frac{3}{4}\right]\nonumber 
\end{eqnarray}
where $\theta_{+}$ is $\theta$ modulo $2\pi$ such that $0\le\theta<2\pi$.
The sum $\sum_{l}$ over $l$ is over all integers with
\begin{equation}
R\left(\beta M,\theta,l\right)\equiv\frac{1}{3}\left[\left(\beta M\right)^{2}+\left(\theta-2\pi l\right)^{2}\right]^{3/2}-\frac{1}{3}\left|\theta-2\pi l\right|^{3}-\frac{1}{2}\left|\theta-2l\pi\right|\beta^{2}M^{2}-\frac{\left(\beta M\right)^{4}}{16\pi\left|l\right|}
\end{equation}
 except that the divergent term is omitted when $l=0$.

The first term in eqn.\ \ref{eq:high-T-expansion} is the blackbody
free energy for two degrees of freedom, and depends only on the temperature
and the angle $\theta$. The second term, which is the leading correction
due to the mass $M$ at high temperatures, often appears in discussions
of symmetry restoration at high temperatures with $\theta=0$. The
third term is closely associated with the $n=0$ Matsubara mode, which
is the most infrared singular contribution to a finite temperature
functional determinant. This term is responsible for non-analytic
behavior in finite temperature perturbation theory via the summation
of ring diagrams. For example, in a scalar theory it gives rise to
the $\lambda^{3/2}$ contribution to the free energy; in QED, the
contribution is $e^{3}$ \cite{Kapusta:2006pm}. The last term is
logarithmic in the dimensionless combination $\beta M$ and independent
of $\theta$. In calculations of effective potentials, it typically
combines with zero-temperature logarithms in such a way that the temperature
$T$ sets the scale of running coupling constants at high $T$.

\subsubsection{Lattice results}

Physical QCD has neither center symmetry, because there are quarks,
not chiral symmetry, because there is a quark mass term in the Lagrangian.
Nevertheless, simulations of $SU(3)$ gauge theories with quarks have
revealed a complicated phase structure, where the number of flavors
and the quark masses are crucial parameters. Quark masses span several
orders of magnitude, from the light $u$ and $d$ quarks, with $u,d\ll\Lambda_{QCD}$
to the very heavy $t$ quark with $m_{t}\gg\Lambda_{QCD}$. Because
$m_{s}$ is roughly on the order of $\Lambda_{QCD}$ , lattice simulations
aimed at realistic descriptions of hadronic physics are generally
carried out with $2+1$ flavors for sea quarks, that is, two light
flavors and one intermediate relative to $\Lambda_{QCD}$.

For gauge theories at finite temperature, the static approximation
is the next logical step beyond pure gauge theories, and represents
very well the effect of very heavy particles. In the case where quarks
or other particles in the fundamental representation satisfy $\beta M\gg1$,
we need keep only the leading term in the low-temperature expansion
\begin{equation}
V_{H}=-h_{F}\left[Tr_{F}P+Tr_{F}P^{\dagger}\right].
\end{equation}
This generalizes immediately to particles in a representation $R$.
Recalling the spin model interpretation, we see that a heavy particle
behaves like an external field $h_{R}$ coupled to the Polyakov loop
$Tr_{R}P$ in the representation $R$. For fermions in the fundamental
representation, this breaks $Z(N)$ symmetry explicitly. In the case
of $SU(3)$ and other theories with first-order deconfinement transitions,
the addition of a small $Z(N)$ symmetry-breaking field does not remove
the transition, as it would in the case of a second-order transition.
Instead, a small symmetry-breaking field with $h_{F}>0$ makes the
deconfined phase more favorable at the deconfinement temperature,
and there is a new deconfinement temperature, giving rise to a critical
line in the $T-h_{F}$ plane. On both sides of the line, $Tr_{F}P$
is non-zero, but there is a discontinuity that grows smaller with
increasing $h_{F}$. This behavior persists until a critical value
of $h_{F}$ is reached, where the first-order line terminates in a
critical end point, believed to be in the usual $\phi^{4}$ universality
class \cite{Hasenfratz:1993az,Meisinger:1995qr,Kashiwa:2012wa}.

\begin{figure}

\includegraphics[scale=0.4]{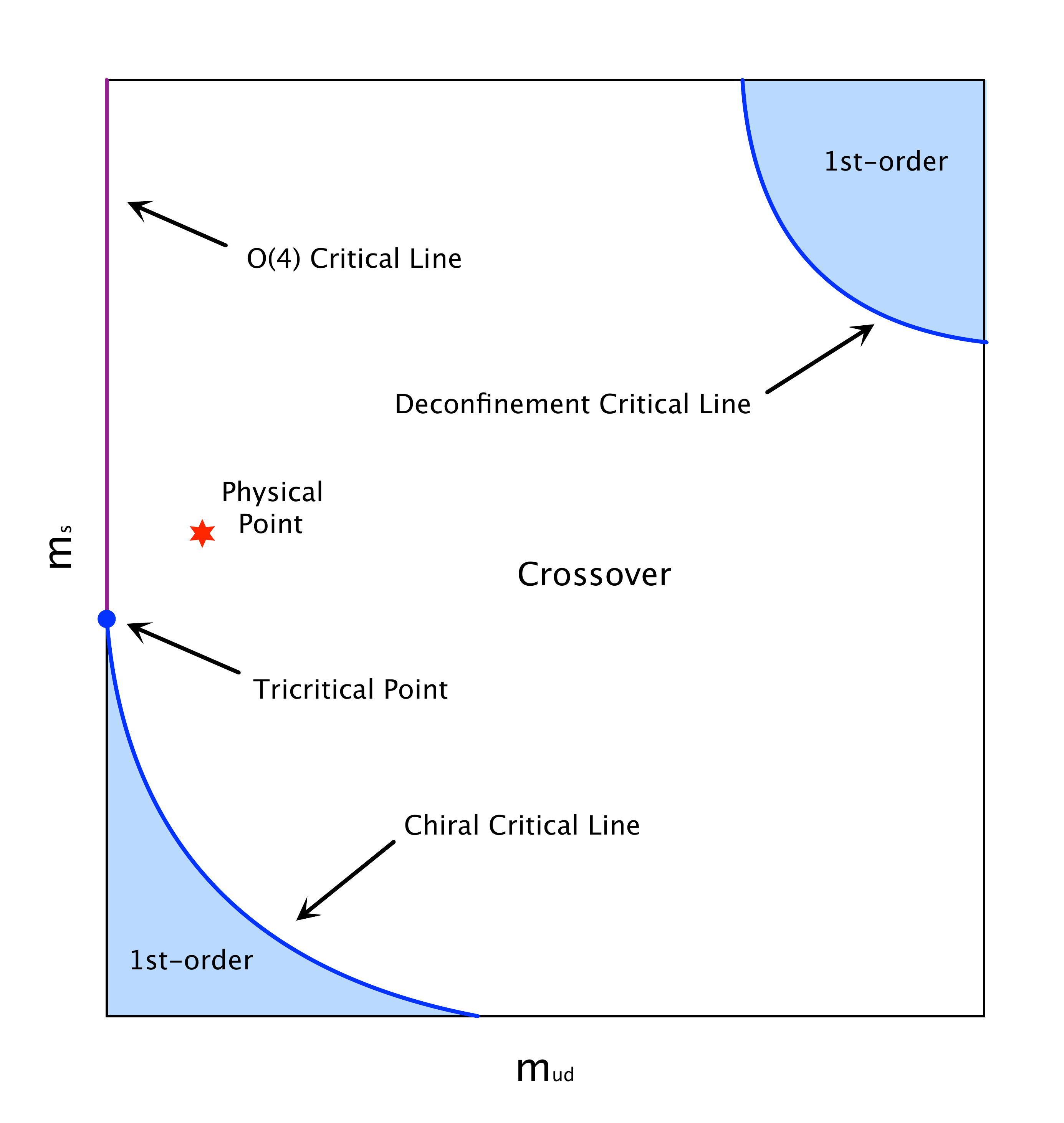}

\caption{\label{fig:Columbia plot}Schematic version of the Columbia plot showing
phase structure as a function of the $m_{ud}$ and $m_{s}$ quark
masses}

\end{figure}

As quark masses are lowered, chiral symmetry effects become important.
Phase transitions, if present, lie at temperatures below the deconfinement
transition temperature of the pure gauge theory, and the phase structure
is very sensitive to quark masses. The results of the extensive lattice
studies that have been done is usually displayed in a so-called Columbia
plot, which shows the phase structure of three-flavor QCD, with up,
down and strange quarks, by showing lines of phase transitions as
functions of a common up and down quark mass $m_{ud}$ and the strange
quark mass $m_{s}$. A version of this plot is shown in Figure \ref{fig:Columbia plot}.
There are several interesting limiting cases contained in this diagram:
\begin{itemize}
\item $m_{ud}=m_{s}=\infty$: Because the quarks are completely removed
from the dynamics, this limit is a pure gauge theory. For $SU(3)$,
the phase transition is first-order and associated with the spontaneous
breaking of center symmetry, as described above. If the quark masses
are decreased from infinity, center symmetry is explicitly broken,
but the phase transition persists for large values of the quark masses,
manifesting as a jump in $\left\langle Tr_{F}P\right\rangle $ from
one non-zero value to another. As explained above, this region of
first-order transitions terminates in what appears in the figure as
a second order line. 
\item $m_{ud}=0$ ; $m_{s}=\infty$ This is the two-flavor chiral limit
of QCD, with a second-order transition in the $\mathcal{O}\left(4\right)$
universality class that extends down from $m_{s}=\infty$ until it
meets another critical line at a tricritical point.
\item $m_{ud}=m_{s}=0$ This is the three-flavor chiral limit of QCD. Of
course, it is less realistic than the two-flavor limit because the
strange quark is sufficiently heavy to be on the order $\Lambda$.
This is a first-order transition that extends into the $m_{ud}-m_{s}$
plane before terminating in a second-order line.
\end{itemize}
Physical QCD does not appear to have a finite temperature phase transition.
Nevertheless, remnants of both the deconfinement transition of the
pure gauge theory and the chiral transition of massless quarks remain.
Near the critical temperature of the chiral transition in the massless
theory, there is a crossover region. This is a relatively narrow range
of temperatures over which $p$, $\epsilon$, $\bar{\psi}\psi$ and
$Tr_{F}P$ change rapidly. In physical QCD, this crossover marks the
change from low-temperature hadronic behavior to high-temperature
quark-gluon plasma behavior.

\subsubsection{connection of chiral symmetry breaking and confinement }

If a system has two or more order parameters, their behavior is typically
correlated. This can be understood as arising from terms in the effective
potential allowed by symmetry that couple the order parameters. In
general, a first-order transition observed as a discontinuity in one
order parameter will also give rise to a discontinuity in others.
In the case of QCD and related theories, lattice simulations with
fundamental representation fermions show that deconfinement and chiral
symmetry restoration are strongly correlated.

The connection of chiral symmetry breaking and deconfinement can be
seen in strong-coupling lattice models \cite{Gocksch:1984yk}. Integration
over the spatial link variables does two things: generates the nearest-neighbor
Polyakov loop coupling that drives the deconfinement transition in
strong coupling, and produces a four-fermion coupling that leads to
spontaneous chiral symmetry breaking at low $T$. In order to explore
the interrelationship of confinement and chiral symmetry breaking
in a continuum model, we can use a generalization of Nambu-Jona Lasinio
models known as Polyakov-Nambu-Jona Lasinio (PNJL) models \cite{Fukushima:2003fw};
see \cite{Mocsy:2003qw} for an alternative approach. NJL models have
been used to study hadronic physics at finite temperature, but they
include only chiral symmetry restoration, and do not model deconfinement.
This omission is rectified by the PNJL models, which include both
chiral restoration and deconfinement. In PNJL models, fermions with
NJL couplings move in a nontrivial Polyakov loop background, and the
effects of gluons at finite temperature is modeled in a semi-phenomenological
way.

\begin{figure}
\includegraphics[scale=0.7]{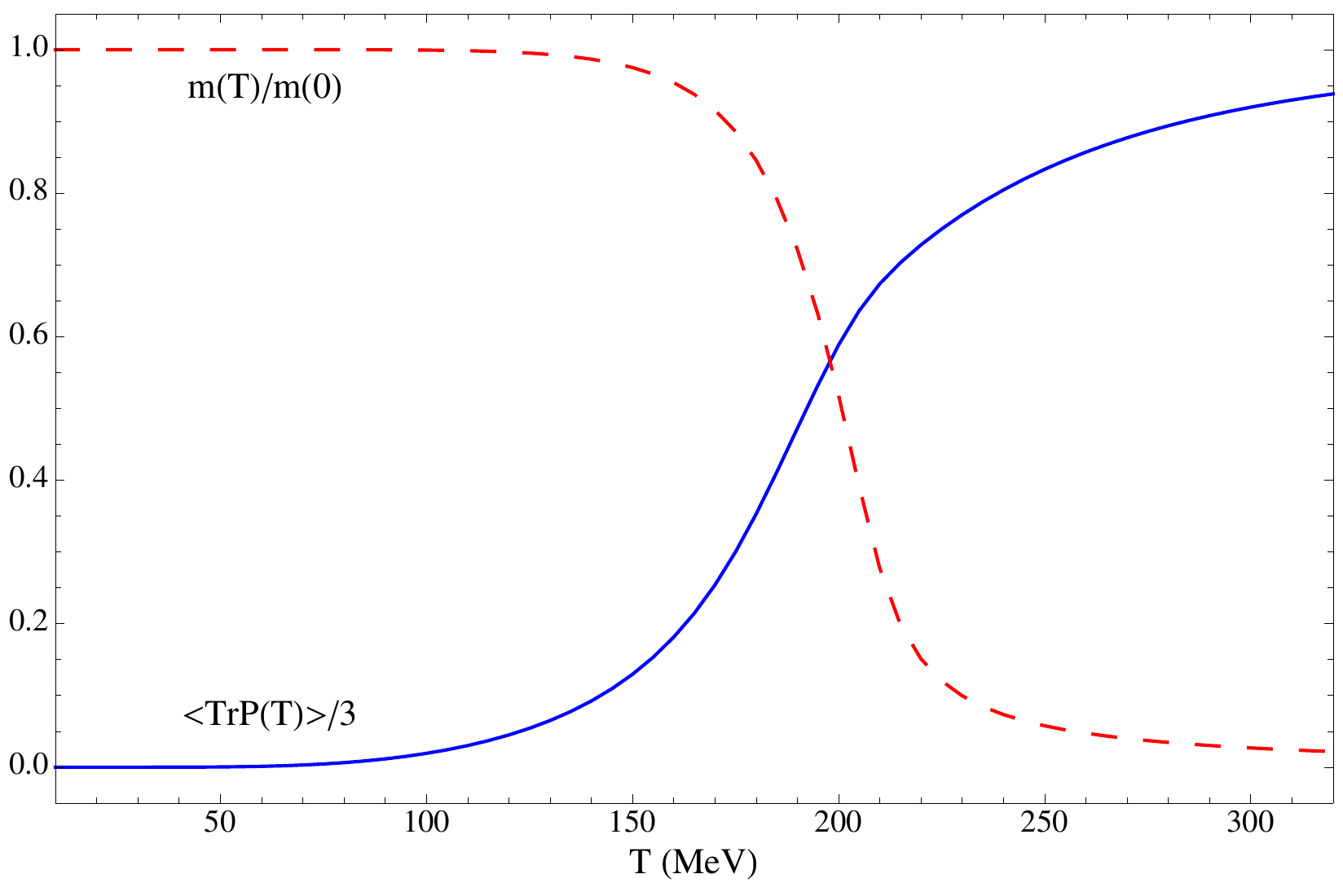}

\caption{\label{fig:PNJL}Order parameters in a PNJL model of QCD with two
light flavors as a function of temperature \cite{Nishimura:2009me}.}

\end{figure}

The key idea of PNJL models is familiar: all particles move in a constant
Polyakov loop background, and the effective potential will be a function
of both the chiral order parameter $\bar{\psi}\psi$ and $P$. The
coupling between the chiral condensate and the Polyakov loop arises
because the covariant derivative $D_{\mu}$ replaces the conventional
derivative $\partial_{\mu}$ in the fermion kinetic term. However,
$D_{\mu}$ contains only a constant $A_{4}$ gauge field background
that gives rise to a non-trivial Polyakov loop. The complete PNJL
effective potential consists of three terms 
\begin{equation}
V_{PNJL}=V_{G}\left(P\right)+V_{F0}\left(\sigma_{j}\right)+V_{FT}\left(P,\sigma_{j}\right)
\end{equation}
where $V_{G}\left(P\right)$ reproduces the behavior of the pure gauge
theory, $V_{F0}\left(\sigma_{j}\right)$ is the NJL effective potential
at $T=0$, and $V_{FT}\left(P,\sigma_{j}\right)$ is the $T\ne0$
part of the fermion effective potential, given by 
\begin{equation}
V_{FL}\left(P,m\right)=-2\sum_{j}Tr_{F}[T\int\frac{d^{3}k}{(2\pi)^{3}}ln(1+Pe^{-\beta\omega_{k}^{\left(j\right)}})+h.c.]
\end{equation}
where the $\omega_{k}^{\left(j\right)}$'s are defined in terms of
the constituent masses as they were for the NJL model in Section \ref{sub:Role-of-quarks}.
If $V_{G}$ is given by eqn.\ \ref{eq:V_G} and $V_{F0}$ by \ref{eq:V_F0},
then minimizing $V_{PNJL}$ as a function of $\sigma$ for two flavors
with $m_{0}=5.5\, MeV$ gives the behavior shown in Fig. \ref{fig:PNJL}
\cite{Nishimura:2009me}. This shows the explanatory power of PNJL
models. The constituent mass $m$ is heavy at low temperatures, due
to chiral symmetry breaking. The larger the constituent mass, the
smaller the $Z(3)$ breaking effect of the fermions, reflected in
the small value of $\lyxmathsym{\textlangle}Tr_{F}P\lyxmathsym{\textrangle}$
at low temperatures. On the other hand, a small value for $\lyxmathsym{\textlangle}Tr_{F}P\lyxmathsym{\textrangle}$
reduces the effectiveness of finite-temperature effects in restoring
chiral symmetry. These synergistic effects combine in the case of
fundamental representation fermions to give a single crossover temperature
at which both order parameters are changing rapidly, consistent with
the rapid crossover see in lattice simulations.

Chiral and deconfinement behavior are not always as closely correlated
as they are with quarks in the fundamental representation. In the
case of adjoint fermions, lattice simulations have shown that chiral
symmetry is restored at a substantially higher temperature than the
deconfinement temperature, with the ratio of critical temperatures
$T_{c}/T_{d}$ around $7.8$ \cite{Karsch:1998qj,Engels:2005te}.
This behavior can also be reproduced in adjoint PNJL models \cite{Nishimura:2009me}.

\subsection{finite density QCD}

The phase diagram of QCD in the temperature-chemical potential plane
is of great interest. Unfortunately, lattice gauge theory is not directly
applicable because of the sign problem. The nature of the problem
can be seen from the one-loop effective potential. The effective potential
$V_{FT\mu}$ for a single flavor quarks at temperature $T=\beta^{-1}$
and chemical potential $\mu$ in a static Polyakov loop background
$P$ is 
\begin{equation}
V_{FT\mu}=-2T\int\frac{d^{3}k}{\left(2\pi\right)^{3}}Tr_{F}\left[\ln\left(1+Pe^{\beta\mu-\beta\omega_{k}}\right)+\ln\left(1+P^{\dagger}e^{-\beta\mu-\beta\omega_{k}}\right)\right]
\end{equation}
reflecting the fact that a non-zero chemical potential behaves as
an imaginary $U(1)$ Polyakov loop. This expression is complex for
$\mu\ne0$ when $P$ is non-trivial. This can be seen very clearly
for heavy quarks, where $V_{FT\mu}$ can be written as
\begin{equation}
V_{FT\mu}\approx f_{q}\approx-h_{F}\left[e^{\beta\mu}Tr_{F}P+e^{-\beta\mu}Tr_{F}P^{+}\right].
\end{equation}
Because $Tr_{F}P$ is complex for $N\ge3$, the effective action for
the gauge fields is complex. This is one version of the sign problem
for gauge theories at finite density: the Euclidean path integral
involve complex weights. This problem is a fundamental barrier to
lattice simulations of QCD at finite density \cite{deForcrand:2005xr,Philipsen:2005mj}
and a significant problem in other fields \cite{Loh:1990zz}.

A large part of the interest in finite-density QCD lies in the possibility
of a color superconducting phase of hadronic matter in the interior
of neutron stars \cite{Alford:2000ze}. At non-zero density, there
 are additional order parameters reflecting the possibility of Cooper
pair formation whenever an attractive channel exists between fermions.
As we have seen, chiral symmetry breaking can be viewed as the formation
of a condensate of $\bar{q}q$ pairs in the vacuum. Other bifermion
condensates are possible when the chemical potential is non-zero,
giving rise to what is known as color superconductivity. This is an
extension of the BCS mechanism from the formation of Cooper pairs
of electrons to $qq$ condensates. The mechanism for BCS superconductivity
is an attractive force between Cooper pairs mediated by phonons. Early
work on color superconductivity considered both interactions mediated
by gluon exchange and induced by instantons via the axial $U(1)$
anomaly. In practice, it is common to assume an NJL model with a set
of four-fermion interactions, perhaps with an anomaly-induced term
as well that is a $2N_{f}$ fermion coupling.

Because quarks carry both color and flavor, it is necessary to consider
both when discussing appropriate order parameters for color superconductivity.
Quarks carry flavor, color and spinor indices, so the quark condensate
in a given channel can be characterized by
\begin{equation}
\left\langle \psi_{ia}^{\alpha}\psi_{jb}^{\beta}\right\rangle =P_{ij\: ab}^{\alpha\beta}\Delta
\end{equation}
where $\alpha$ and $\beta$ are color indices, $j$ and $k$ are
flavor indices, and $a$ and $b$ are spinor indices. The color-flavor-spin
matrix characterizes a particular pairing, and the gap parameter $\Delta$
gives the magnitude of the condensate in this channel. The condensate
may be momentum-dependent. The familiar BCS condensate is a special
case where the condensate is independent of position, so the pairing
is between fermions of equal and opposite momentum, and a spin singlet.
Unlike the fermion-antifermion condensates, the fermion-fermion condensates
are not gauge invariant, and thus are similar to the Higgs scalar
expectation value, which similarly depends on the choice of gauge.
As with the Higgs mechanism, physical observables like the spectrum
are gauge-invariant. There are strong arguments why the color-flavor
locking phase (CFL) phase is favored for high-density QCD, combining
antisymmetry in color and flavor with a Lorentz scalar behavior, giving
\begin{equation}
\left\langle \psi_{i}^{\alpha}C\gamma_{5}\psi_{j}^{\beta}\right\rangle \propto\epsilon^{\alpha\beta A}\epsilon_{ijA}
\end{equation}
where $C$ is the charge-conjugation matrix acting on Dirac spinors.
For a recent review of our current understanding of the phase diagram
of dense QCD, see \cite{Fukushima:2010bq}.

\section{\label{sec:Phases-on-R3xS1}Phases of gauge theories on $R^{3}\times S^{1}$}

In the last few years, it has proven possible to construct four-dimensional
gauge theories for which confinement may be reliably demonstrated
using semiclassical methods \cite{Myers:2007vc,Unsal:2007vu}. These
models combine $Z(N)$ symmetry, the effective potential for $P$,
instantons, and monopoles into a satisfying picture of confinement
for a special class of models. All of the models in this class have
one or more small compact directions. Models with an $R^{3}\times S^{1}$
topology have been most investigated, and discussion here will focus
on this class. The use of one or more compact directions will cause
the running coupling constant of an asymptotically free gauge theory
to be small, so that semiclassical methods are reliable. For example,
if the circumference $L$ of $S^{1}$ is small, \emph{i.e.}, $L\ll\Lambda^{-1}$
, then $g(L)\ll1$. However, this leads to an immediate problem: generally
speaking, one or more small compact directions lead to breaking of
$Z(N)$ symmetry in those directions. As we have seen in Section \ref{sub:Center-symmetry},
for the case of finite temperature gauge theories, where $L=\beta=1/T$,
the effective potential for the Polyakov loop is easily calculated
to lowest order in perturbation theory; for a pure gauge theory it
is given by eq.\ (\ref{eq:V_pure_glue}):
\begin{equation}
V_{gauge}\left(P,\beta\right)=
\frac{-2}{\pi^{2}\beta^{4}}\sum_{n=1}^{\infty}\frac{Tr_{Adj}P^{n}}{n^{4}}
\end{equation}
where the trace of $P$ in the adjoint representation is given by
$Tr_{A}P=\left|Tr_{F}P\right|^{2}-1$. This effective potential is
minimized when $P=zI$ where $z\in Z(N)$, indicating that $Z(N)$
symmetry is spontaneously broken at high temperatures where the one-loop
expression is valid. However, it is possible to maintain $Z(N)$ symmetry
even when $L$ is small by modifying the action. This leads to a perturbative
calculation of possible phase structures, which turns out to be very
rich, as well as a perturbative understanding of Polyakov loop physics
in the confined phase. Furthermore, the restoration of $Z(N)$ symmetry
for small $L$ leads to a non-perturbative mechanism for confinement,
as measured by Wilson loops orthogonal to the compact direction. In
this confinement mechanism, a key role is played by finite-temperature
instantons, also known as calorons, and their monopole constituents.
Thus we obtain a realization of a long-held scenario for quark confinement,
based on ideas originally proposed by Mandelstam \cite{Mandelstam:1974vf,Mandelstam:1974pi}
and 't Hooft \cite{'tHooft:1977hy,'tHooft:1979uj}.

\subsection{Restoring center symmetry via deformations or periodic adjoint fermions }

\begin{figure}

\includegraphics[scale=0.6]{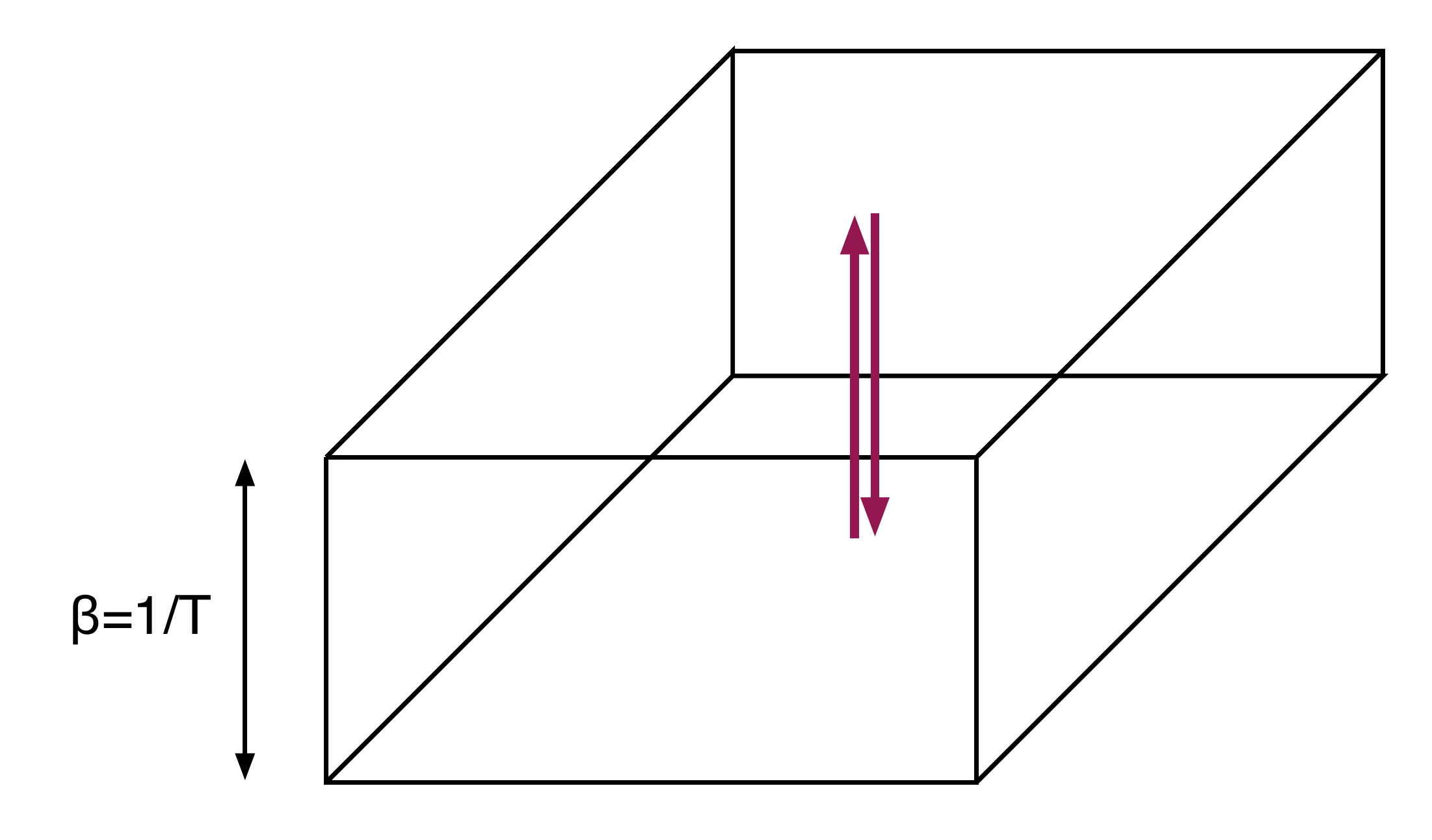}

\caption{\label{fig:double-trace}A double-trace Polyakov loop on $R^{3}\times S^{1}$.}

\end{figure}

There are two broad approaches to maintaining $Z(N)$ symmetry for
small $L$. The first approach deforms the pure gauge theory by adding
additional terms to the gauge action \cite{Myers:2007vc,Ogilvie:2007tj,Unsal:2008ch}.
The general form for such a deformation is
\begin{equation}
S\rightarrow S+\beta\int d^{3}x\,\sum_{k=1}^{\infty}a_{k}Tr_{A}P\left(\vec{x},x_{4}\right)^{k}
\end{equation}
where the value of $x_{4}$ is arbitrary and can be taken to be $0$.
Such terms are often referred to as double-trace deformations; see
Fig. \ref{fig:double-trace}. If the coefficients $a_{k}$ are sufficiently
large, they will counteract the effects of the one-loop effective
potential, and $Z(N)$ symmetry will hold for small $L$. Strictly
speaking, only the first $\left[N/2\right]$ terms are necessary to
ensure confinement. As discussed in Section \ref{sub:Center-symmetry},
it is easy to prove that for a classical Polyakov loop $P$, the conditions
$Tr_{F}P^{k}=0$ with $1\le k\le\left[N/2\right]$ determine the unique
set of Polyakov loop eigenvalues that constitute a confining solution,
\emph{i.e.}, one for which $Tr_{R}P=0$ for all representations with
$k_{R}\ne0$ \cite{Meisinger:2001cq}. The explicit solution is simple:
up to a factor necessary to ensure $\det P=1$, the eigenvalues of
$P$ are given by the set of $N$'th roots of unity, which are permuted
by a global $Z(N)$ symmetry transformation. The effective potential
associated with $S$ is given approximately by 
\begin{equation}
V_{eff}\left(P,\beta\right)=\frac{-2}{\pi^{2}\beta^{4}}\sum_{n=1}^{\infty}\frac{Tr_{Adj}P^{n}}{n^{4}}+\sum_{k=1}^{\left[\frac{N}{2}\right]}a_{k}Tr_{Adj}P^{k}.
\end{equation}
For sufficiently large and positive values of the $a_{k}$'s, the
confined phase yields the lowest value of $V_{eff}$. However, a rich
phase structure emerges from the minimization of $V_{eff}$ for intermediate
values of the coefficients $a_{k}$. For $N\ge3$, the effective potential
predicts that one or more phases may separate the deconfined phase
from the confined phase. In the case of $SU(3)$, a single new phase
is predicted, and has been observed in lattice simulations \cite{Myers:2007vc}.
For larger values of $N$, there is a rich set of possible phases,
including some where $Z(N)$ breaks down to a proper subgroup $Z(p)$.
In such phases, particles in the fundamental representation are confined,
but bound states of $p$ particles are not \cite{Ogilvie:2007tj}.

Lattice simulations of $SU(3)$ and $SU(4)$ agree for small $L$
with the theoretical predictions based on effective potential arguments
\cite{Myers:2007vc}. The phase diagram of $SU(3)$ as a function
of $T=L^{-1}$ and $a_{1}$ has three phases: the confined phase,
the deconfined phase, and a new phase, the skewed phase. In general,
the three phases of the eigenvalues of the Polykov loop may be taken
to be the set $\left\{ \theta_{1},\theta_{2},\theta_{3}\right\} $
where $\theta_{1}+\theta_{2}+\theta_{3}$=0. For all three phases,
it is possible to use $Z(3)$ symmetry to make $Tr_{F}P$ real, and
reduce the phases to the set $\left\{ 0,\theta,-\theta\right\} $
such that $Tr_{F}P=1+2\cos\theta$. The deconfined phase is represented
by $\theta=0$, the confined phase is given by $\theta=2\pi/3$, and
the skewed phase by $\theta=\pi.$ An important result obtained from
the lattice simulation of $SU(3)$ is that the small-$L$ confining
region, where semiclassical methods yield confinement, are smoothly
connected to the conventional large-$L$ confining region. In the
case of $SU(4)$, a sufficiently large value of $a_{1}$ leads to
a partially-confining phase where $Z(4)$ is spontaneously broken
to $Z(2)$. Particles with $k=1$ are confined in this phase, \emph{i.e.},
$\left\langle Tr_{F}P\left(\vec{x}\right)\right\rangle =0$, but particles
with $k=2$ are not, as indicated by $\left\langle Tr_{F}P^{2}\left(\vec{x}\right)\right\rangle \ne0$.
As in the case of $SU(3)$, $Tr_{F}P$ can be made real. Perturbation
theory then predicts a deconfined phase where the phases of the eigenvalues
of $P$ are $\left\{ 0,,0,0,0\right\} ,$ a confined phase where they
are $\left\{ \pi/4,3\pi/4,5\pi/4,7\pi/4\right\} $, and a partially
confined phase where the phases are $\left\{ \pi/2,\pi/2,-\pi/2,-\pi/2\right\} $.
The confined phase corresponds to the matrix $P_{1}$ and the partially
confined phase to the matrix $P_{2}$ described in Section \ref{sub:Center-symmetry}.

Another approach to preserving $Z(N)$ symmetry for small $L$ uses
fermions in the adjoint representation with periodic boundary conditions
in the compact direction \cite{Unsal:2007vu}. In this case, it would
be somewhat misleading to use $\beta$ as a synonym for $L$, because
the transfer matrix for evolution in the compact direction is not
positive-definite. Periodic boundary conditions in the compact direction
imply that the generating function of the ensemble, \emph{i.e.}, the
partition function, is given by
\begin{equation}
Z=Tr\left[\left(-1\right)^{F}e^{-LH}\right]
\end{equation}
 where $F$ is the fermion number and $H$ is the Hamiltonian in the
compact direction. This graded ensemble, familiar from supersymmetry,
can be obtained from an ensemble $Tr\left[\exp\left(\beta\mu F-\beta H\right)\right]$
with chemical potential $\mu$ by the replacement $\beta\mu\rightarrow i\pi$.
This system can be viewed as a gauge theory with periodic boundary
conditions in one compact spatial direction of length $L=\beta$,
and the transfer matrix in the time direction is positive-definite,

The use of periodic boundary conditions for the adjoint fermions dramatically
changes their contribution to the Polyakov loop effective potential.
In perturbation theory, the replacement $\beta\mu\rightarrow i\pi$
shifts the Matsubara frequencies from $\beta\omega_{n}=\left(2n+1\right)\pi$
to $\beta\omega_{n}=2n\pi$. The one loop effective potential is now
essentially that of a bosonic field, but with an overall negative
sign due to fermi statistics \cite{Meisinger:2001fi}. The sum of
the effective potential for the fermions plus that of the gauge bosons
gives 
\begin{equation}
V_{eff}\left(P,\beta,m,N_{f}\right)=\frac{1}{\pi^{2}\beta^{4}}\sum_{n=1}^{\infty}\frac{Tr_{Adj}P^{n}}{n^{2}}\left[2N_{f}\beta^{2}m^{2}K_{2}\left(n\beta m\right)-\frac{2}{n^{2}}\right]\label{eq:Veff-adjoint-fermions}
\end{equation}
where $N_{f}$ is the number of adjoint Dirac fermions and $m$ is
their mass \cite{Meisinger:2009ne}. Note that the first term in brackets,
due to the fermions, is positive for every value of $n$, while the
second term, due to the gauge bosons, is negative. 

The largest contribution to the effective potential at high temperatures
is typically from the $n=1$ term, which can be written simply as
\begin{equation}
\frac{1}{\pi^{2}\beta^{4}}\left[2N_{f}\beta^{2}m^{2}K_{2}\left(\beta m\right)-2\right]\left[\left|Tr_{F}P\right|^{2}-1\right]
\end{equation}
 where the overall sign depends only on $N_{f}$ and $\beta m$. If
$N_{f}\ge1$ and $\beta m$ is sufficiently small, this term will
favor $Tr_{F}P=0$. On the other hand, if $\beta m$ is sufficiently
large, a value of $P$ from the center, $Z(N)$, is preferred. Note
that an $\mathcal{N}=1$ super Yang-Mills theory would correspond
to $N_{f}=1/2$ and $m=0$, giving a vanishing perturbative contribution
for all $n$ \cite{Davies:1999uw,Davies:2000nw}. In that case, non-perturbative
effects lead to a confining effective potential for all values of
$\beta$. In the case of $N_{f}\ge1$, each term in the effective
potential will change sign in succession as $m$ is lowered towards
zero. For larger values of $N$, this leads to a cascade of phases
separating the confined and deconfined phases \cite{Myers:2009df}.
Numerical investigation shows that the confined phase is obtained
if $N\beta m\lesssim4.00398$ \cite{Meisinger:2009ne}. As $m$ increases,
it becomes favorable that $Tr_{F}P^{n}\ne0$ for successive values
of $n$. If $N$ is even, the first phase after the confined phase
will be a phase with $Z(N/2)$ symmetry. As $m$ increases, the last
phase before reaching the deconfined phase will have $Z(2)$ symmetry,
in which $k=1$ states are confined, but all states with higher $k$
are not. Lattice simulations of $SU(3)$ with periodic adjoint fermions
are completely consistent with the picture \cite{Cossu:2009sq} predicted
by the effective potential, with a skewed phase separating the confined
phase and deconfined phase. For $N\ge3$, there are generally phases
intermediate between the confined and deconfined phases which are
not of the partially-confined type. Careful numerical analysis appears
to be necessary on a case-by-case basis to determine the phase structure
for each value of $N$ \cite{Myers:2009df}. There has been very interesting
work on lattice gauge theories in the large-$N$ limit that indicates
the same rich phase structure \cite{Bringoltz:2009mi,Bringoltz:2009kb};
this work will be discussed in Section \ref{sec:Large-N}.

There are some interesting additional issues arising when periodic
adjoint fermions are used to obtain $Z(N)$ symmetry for small $L$.
There are strong indications from strong-coupling lattice calculations
\cite{Gocksch:1984yk} and the closely related Polyakov-Nambu-Jona
Lasinio (PNJL) models \cite{Fukushima:2003fw} that the mass $m$
that appears in the effective potential $V(P)$ should be regarded
as a constituent mass that includes the substantial effects of chiral
symmetry breaking. In strong-coupling lattice calculations and PNJL
models, this effect is responsible for the coupling of $P$ and $\bar{\psi}\psi$
in simulations of QCD at finite temperature. Thus it is important
that simulations of $SU(3)$ with $N_{f}=2$ flavors of adjoint fermions
show explicitly that the $Z(N)$-invariant confined phase is regained
when the fermion mass is sufficiently small \cite{Cossu:2009sq}.
However, this raises another issue. For a simple double-trace deformation,
the small-$L$ and large-$L$ regions are smoothly connected. A semi-phenomenological
analysis based on a PNJL model \cite{Nishimura:2009me} suggests that
this is also the case with periodic adjoint fermions, but additional
modifications of the action may be necessary to realize the connection.

\subsection{Monopoles and instantons on $R^{3}\times S^{1}$}

The non-perturbative dynamics of confining gauge theories on $R^{3}\times S^{1}$
are based on Polyakov's analysis of the Georgi-Glashow model in three
dimensions \cite{Polyakov:1976fu}. This is an $SU(2)$ gauge model
coupled to an adjoint Higgs scalar. The models we are considering
thus differ by the addition of a fourth compact dimension and a change
to the action designed to maintain $Z(N)$ symmetry. The four-dimensional
Georgi-Glashow model is the standard example of a gauge theory with
classical monopole solutions when the Higgs expectation value is non-zero.
These monopoles make a non-perturbative contribution to the partition
function $Z$. In three dimensions, these monopoles are instantons.
Polyakov showed that a gas of such three-dimensional monopoles gives
rise to non-perturbative confinement in three dimensions, even though
the theory appears to be in a Higgs phase perturbatively. 

\begin{figure}
\includegraphics[scale=0.6]{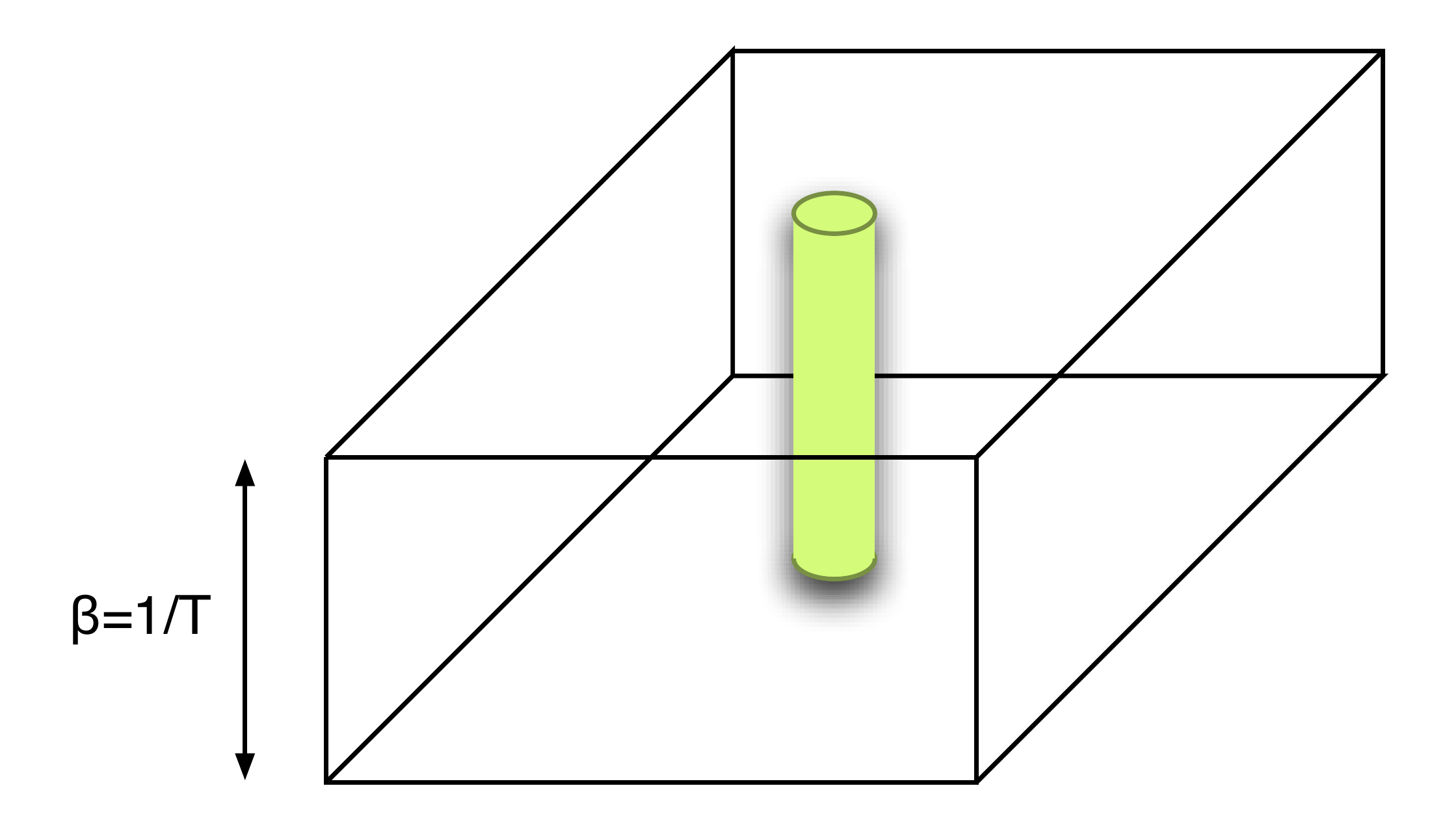}
\caption{\label{fig:Short-monopole-worldline}Short monopole worldline on $R^{3}\times S^{1}$.}
\end{figure}

Because $L$ is small in the $R^{3}\times S^{1}$ models we consider,
the three-dimensional effective theory describing the behavior of
Wilson loops in the non-compact directions will have many features
in common with the three-dimensional theory first discussed by Polyakov.
In the four-dimensional theory, monopole solutions with short worldline
trajectories in the compact direction exist, and behave as three-dimensional
instantons in the effective theory; see Fig. \ref{fig:Short-monopole-worldline}.
In models on $R^{3}\times S^{1}$, the role of the three-dimensional
scalar field is played by the fourth component of the gauge field
$A_{4}$. In a gauge where the Polyakov loop is diagonal and independent
of $x_{4}$, $P$ has a vacuum expected value induced by the perturbative
effective potential. However, there is another way to understand the
presence of monopoles in this phase, based on studies of instantons
in pure gauge theories at finite temperature and the properties of
the KvBLL caloron solution \cite{Lee:1998bb,Kraan:1998kp,Kraan:1998pm}.
If the Polyakov loop has a non-trivial expectation value, finite-temperature
instantons in $SU(N)$ may be decomposed into $N$ monopoles, and
the locations of the monopoles become parameters of the moduli space
of the instanton. In the case of $SU(2)$, an instanton may be decomposed
into a conventional BPS monopole and a so-called KK (Kaluza-Klein)
monopole. The presence of the KK monopole solution differentiates
the case of a gauge field at finite temperature from the case of an
adjoint scalar breaking $SU(N)$ to $U(1)^{N-1}$, in which case there
are $N-1$ fundamental monopoles. We will consider in detail the simplest
case of $N=2$.

The BPS monopole is found using the using the standard arguments \cite{Prasad:1975kr,Bogomolny:1975de}.
The Euclidean Lagrangian $\mathcal{L}$ can be taken to be 
\begin{equation}
\mathcal{L}=\frac{1}{4}\left(F_{\mu\nu}\right)^{2}++V_{eff}\left(P\right)
\end{equation}
where $V_{eff}$ includes both the one-loop gluonic effective potential
and the additional term that prevents $Z(N)$ symmetry breaking. This
can also be written as 
\begin{equation}
\mathcal{L}=\frac{1}{2}\left(D_{j}A_{4}\right)^{2}+\frac{1}{2}\left(B_{j}\right)^{2}+V_{eff}\left(P\right).
\end{equation}
We can associate with $\mathcal{L}$ an energy defined by 
\begin{equation}
E=\int d^{3}x\left[\frac{1}{2}\left(B_{j}\right)^{2}+\frac{1}{2}\left(D_{j}A_{4}\right)^{2}+V_{eff}\left(P\right)\right]
\end{equation}
as well as an action $S=LE$. We will concern ourselves for now with
the solutions in the BPS limit, in which the effective potential $V_{eff}$
is neglected, but the boundary condition on $P$ at infinity imposed
by the potential is retained. We can write the energy as 
\begin{eqnarray}
E & = & \int d^{3}x\left[\frac{1}{2}\left(B_{j}\pm D_{j}A_{4}\right)^{2}\mp B_{j}D_{j}A_{4}\right].
\end{eqnarray}
This expression is a sum of squares plus a term which can be converted
to a surface integral, giving rise to the BPS inequality 
\begin{equation}
E\geq\mp\int dS_{j}B_{j}A_{4}.
\end{equation}
The BPS inequality is saturated if the equality $B_{j}=\mp D_{j}A_{4}$
holds. For the case of a single monopole at the origin, we require
the fields at spatial infinity to behave as 
\begin{eqnarray}
\lim_{r\rightarrow\infty}A_{4}^{a} & = & w\frac{x^{a}}{r}\nonumber \\
\lim_{r\rightarrow\infty}A_{i}^{a} & = & \epsilon^{aij}\frac{x_{j}}{gr^{2}}.
\end{eqnarray}
Note that $w$ is related to the eigenvalues of $P$ at large distances
by $w=2\theta/gL$. Note that $A_{4}$ has the usual hedgehog form.
$A_{i}^{a}$ is chosen such that covariant terms vanish at infinity:
$\left(D_{i}A_{4}\right)^{a}=0$. With the 't Hooft-Polyakov ansatz,
the general expressions for the fields become 
\begin{eqnarray}
A_{4}^{a} & = & wh\left(r\right)\frac{x^{a}}{r}\nonumber \\
A_{i}^{a} & = & a\left(r\right)\epsilon^{aij}\frac{x_{j}}{gr^{2}}
\end{eqnarray}
where we define $w>0$ and require $h(\infty)=1$ or $-1,$ and $a(\infty)=1$
to obtain the correct asymptotic behavior. We must also have $h=a=0$
at $r=0$ to have well-defined functions at the origin. We identify
a magnetic flux 
\begin{equation}
\Phi=\pm\int dS_{j}B_{j}^{a}\frac{x^{a}}{r}=\mp\frac{4\pi}{g}
\end{equation}
where the $+$ sign corresponds to the case $h(\infty)=1$ and $-$
corresponds to $h(\infty)=-1$. The energy of the BPS monopole can
be written as 
\begin{eqnarray}
E_{BPS} & = & \mp\Phi w=\frac{4\pi w}{g}.
\end{eqnarray}

In addition to the BPS monopole, there is another, topologically distinct
monopole which occurs at finite temperature when $A_{4}$ is treated
as a Higgs field \cite{Davies:1999uw}. Starting from a static monopole
solution where $\left|A_{4}\right|=w$ at spatial infinity, we apply
a special gauge transformation 
\begin{equation}
U_{special}=\exp\left[-\frac{i\pi x_{4}}{L}\tau^{3}\right]
\end{equation}
 where $\tau^{i}$ is the Pauli matrix. $U_{special}$ transforms
$A_{\mu}$ in such a way that the value of $A_{4}$ at spatial infinity
is shifted: $w\rightarrow w-2\pi/gL$. If we instead start from a
static monopole solution such that $A_{4}=2\pi/gL-w$ at spatial infinity,
then the action of $U_{special}$ gives a monopole solution with $A_{4}=-w$
at spatial infinity. A final constant gauge transformation $U_{const}=\exp\left[i\pi\tau^{2}/2\right]$
yields a new monopole solution with $A_{4}=w$ at spatial infinity.
The distinction between the BPS solution, which is independent of
$x_{4}$, and the KK solution is made clear by consideration of the
topological charge. The action of $U_{special}$ followed by $U_{const}$
increases the topological charge by $1$ and changes the sign of the
monopole charge. Thus the KK solution is topologically distinct from
the BPS solution because it carries instanton number $1$. The $\overline{BPS}$
antimonopole has magnetic charge opposite to the BPS monopole, and
hence the same as that of the $KK$ monopole. The $\overline{KK}$
monopole has the same magnetic charge as the $BPS$ monopole, but
carries instanton number $-1$. This is all completely consistent
with the KvBLL decomposition of instantons in the pure gauge theory
with non-trivial Polyakov loop behavior, where $SU(2)$ instantons
can be decomposed into a BPS monopole and a KK monopole. Our picture
of the confined phase is one where instantons and anti-instantons
have ``melted'' into their constituent monopoles and anti-monopoles,
which effectively forms a three-dimensional gas of magnetic monopoles.
In the BPS limit, both the magnetic and scalar interactions are long-ranged;
this behavior appears prominently, for example, in the construction
of $N$-monopole solutions in the BPS limit. 

The BPS solution has action 
\begin{equation}
S_{BPS}=\frac{4\pi wL}{g}=\frac{8\pi\theta}{g^{2}}.
\end{equation}
 For the KK solution, we have instead 
\begin{equation}
S_{KK}=\frac{4\pi\left(2\pi-gLw\right)}{g^{2}}=\frac{4\pi\left(2\pi-2\theta\right)}{g^{2}}.
\end{equation}
The sum $S_{BPS}+S_{KK}$ is exactly $8\pi^{2}/g^{2}$, the action
of an instanton. For $\theta=\pi/2$, the $Z(2)$-symmetric value
for $SU(2),$ $S_{BPS}=S_{KK}$. This extends to $SU(N)$, where the
action of a monopole of any type is $8\pi^{2}/g^{2}N$. 

Although we used the BPS construction to exhibit the existence and
some properties of the monopole solutions of our system, we must move
away from the BPS limit to ensure that magnetic interaction dominate
at large distances, \emph{i.e.}, that the three-dimensional scalar
interactions associated with $A_{4}$ and are not long-ranged. This
behavior is natural in the confined, where the characteristic scale
of the Debye (electric) screening mass associated with $A_{4}$ is
large, on the order of $g/L$. It is well known that the BPS bound
for the monopole mass holds as an equality only when the scalar potential
is taken to zero. Numerical studies \cite{Kirkman:1981ck} have shown
that the monopole action is given in general for $SU(2)$ as 
\begin{equation}
LE_{BPS}C\left(\epsilon\right)
\end{equation}
 where $C$ a function of the quartic term in the potential that varies
from $C=1$ in the BPS limit to a maximum value $C\left(\infty\right)=1.787$.
Thus corrections to the BPS result for the monopole mass and action
due to the potential terms are less than a factor of two. We will
henceforth use the exact results for the actions in the BPS limit,
neglecting corrections from $V_{eff}$ for the sake of simplicity
of notation. 

The $SU(2)$ construction of BPS and KK monopoles extends to $SU(N)$
in the standard way, via the embedding of $SU(2)$ subgroups in $SU(N)$.
There are $N-1$ BPS monopoles and 1 KK monopole inside an instanton.
In the confined phase, each of the $N$ monopoles has action $8\pi^{2}/g^{2}N$.
It has long been thought that instanton effects must be suppressed
in the large-$N$ limit, because instanton effects would vanish as
$\exp\left(-cN\right)$ in the limit $N\rightarrow\infty$ with $\lambda\equiv g^{2}N$
fixed \cite{Witten:1978bc}. In contrast, we see that the effects
of monopole constituents of instantons are not suppressed by the large-$N$
limit.

\subsection{Monopoles and space-like string tensions}

It is important to understand that Wilson loops in planes orthogonal
to the compact direction should show area law behavior, even if the
$Z(N)$ symmetry associated with the compact direction is broken.
This is an old observation about the deconfined phase \cite{DeTar:1985kx,DeGrand:1986uf}
which is very clearly observed in lattice simulations of $SU(2)$
and $SU(3)$ at temperatures above the deconfinement transition \cite{Bali:1993tz,Karsch:1994af}.
At first sight, this seems to directly conflict with the association
of deconfinement with the loss of area-law behavior for Wilson loops.
However, the introduction of a compact direction, as in the case of
finite temperature, explicitly breaks space-time symmetry. In the
case of finite temperature, Wilson loops measuring electric flux have
perimeter behavior in the deconfined phase; Wilson loops measuring
magnetic flux still obey an area law. This asymmetry in behavior can
be understood on the basis of center symmetry. The full center symmetry
of an $SU(N)$ gauge theory on a $d$-dimensional hypertorus $T^{d}$
is $Z(N)^{d}$. While the $Z(N)$ symmetry may break spontaneously
in the short compact direction, the other $Z(N)$ symmetries are unbroken,
and thus the associated Wilson loops obey an area law.

\begin{figure}
\includegraphics{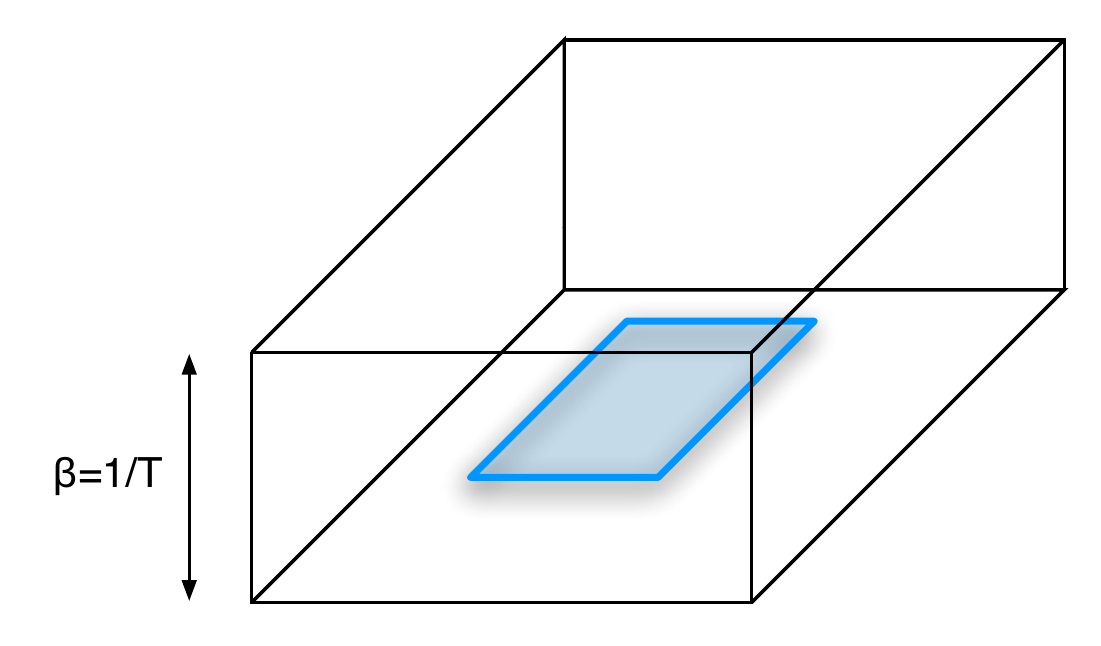}
\caption{\label{fig:Spatial-Wilson-loop}Spatial Wilson loop in $R^{3}\times S^{1}$
geometry.}

\end{figure}

In order to understand the effects of monopoles play in the confined
phase, we must analyze their interactions. We begin with a discussion
of quantum fluctuations around the monopole solutions. The contribution
to the partition function of a single BPS monopole at finite temperature
was considered by Zarembo \cite{Zarembo:1995am}. The measure factor
$d\mu^{a}$ associated with the collective coordinates (moduli) of
the monopole solution, including the Jacobians from the zero modes
is given by \cite{Davies:2000nw} 
\begin{equation}
\int d\mu^{a}=\mu^{4}\int\frac{d^{3}x}{\left(2\pi\right)^{3/2}}J_{x}\int_{0}^{2\pi}\frac{d\phi}{\left(2\pi\right)^{1/2}}J_{\phi}
\end{equation}
where $x$ is the position and $\phi$ the $U(1)$ phase of the monopole
and $\mu$ is a Pauli-Villars regulator. The label $a$ denotes the
type of monopole, $a=\left\{ BPS,KK,\overline{BPS},\overline{KK}\right\} $.
The Jacobians are 
\begin{equation}
J_{x}=S_{a}^{3/2},\,\, J_{\phi}=NLS_{a}^{1/2}.
\end{equation}
Each of the four zero modes contributes a factor of $\mu$. In the
BPS limit, each monopole carries an overall factor 
\begin{eqnarray}
Z{}_{a} & = & c\mu^{7/2}\left(NL\right)^{1/2}S_{a}^{2}\exp\left[-S_{a}+\mathcal{O}\left(1\right)\right]\int d^{3}x\nonumber \\
 & = & \xi_{a}\exp\left[-S_{a}\right]\int d^{3}x
\end{eqnarray}
in its contribution to $Z$ \cite{Zarembo:1995am}. The factor $\xi_{a}$
is $c\mu^{7/2}\left(NL\right)^{1/2}S_{a}^{2}$ where $c$ is a numerical
constant and the factor of $d^{3}x$ represents the integration over
the location of the monopole. From the construction of the KK monopole,
we see that we have $\xi_{KK}\left(\theta\right)=\xi_{BPS}\left(\pi-\theta\right)$.

The renormalization of the functional determinant arising from quantum
fluctuations around the monopole solution is particularly simple in
the confined phase, as first observed by Davies \emph{et al.} in the
corresponding supersymmetric model \cite{Davies:1999uw}. The dependence
on the Pauli-Villars regulator is removed, as usual, by coupling constant
renormalization. The relation at one loop of the bare coupling and
the regulator mass $\mu$ to a renormalization-group invariant scale
$\Lambda$ is 
\begin{equation}
\mbox{\ensuremath{\Lambda}}^{b_{0}}=\mu^{b_{0}}e^{-8\pi^{2}/g^{2}N}
\end{equation}
 where $b_{0}$ is the first coefficient of the $\beta$ function
divided by $N$: 
\begin{equation}
b_{0}=\frac{11}{3}-\frac{4}{3}\cdot\frac{n_{f}C(R_{f})}{N}-\frac{1}{6}\cdot\frac{n_{b}C(R_{b})}{N}
\end{equation}
 where $n_{f}$ is the number of flavors of Dirac fermions in a representation
$R_{f}$, $n_{b}$ is the number of flavors of real scalars in a representation
$R_{b}$, and $C(R)$ is obtained from $Tr_{R}\left(T^{a}T^{b}\right)=C(R)\delta^{ab}$.
For the case of a pure gauge theory with a deformation, there are
four collective coordinates and this gives a factor of $\mu^{4}$.
The functional integral over gauge degrees of freedom gives rise to
a factor $\det'\left[-D^{2}\right]^{-1}\propto\mu^{-1/3}$ and the
action contributes a factor $\exp\left(-8\pi^{2}/g^{2}N\right)$ in
the confined phase. Thus the contribution of a single monopole to
the partition function gives a factor 
\begin{equation}
\mu^{4-\frac{1}{3}}e^{-8\pi^{2}/g^{2}N}=\mu^{11/3}e^{-8\pi^{2}/g^{2}N}=\Lambda^{11/3}.
\end{equation}
Thus detailed calculation confirms what we might have guessed on dimensional
grounds: the contribution $\xi_{a}e^{-8\pi^{2}/g^{2}N}\propto L^{-3}\left(\Lambda L\right)^{11/3}$.
Note that the eliminations of renormalization-dependent quantities
by renormalization-independent quantities depends crucially on the
coefficient of $1/g^{2}$ in the action. 

The interaction of the monopoles is essentially the one described
by Polyakov in his original treatment of the Georgi-Glashow model
in three dimensions \cite{Polyakov:1976fu}, slightly generalized
to include both the BPS and KK monopoles. Let us consider, say, a
BPS-type monopole and KK-type monopole located at $\vec{x}_{1}$ and
$\vec{x}_{2}$ in the non-compact directions, with static worldlines
in the compact direction. The interaction energy due to magnetic charge
of such a pair is 
\begin{equation}
E_{BPS-KK}=-\left(\frac{4\pi}{g}\right)^{2}\frac{1}{4\pi\left|\vec{x}_{1}-\vec{x}_{2}\right|}
\end{equation}
 and the associated action is approximately $S_{BPS}+S_{KK}+LE_{BPS-KK}$.
As discussed above, this will be larger than the value obtained from
the Bogomolny bound, but of the same order of magnitude. There is
an elegant way to capture the dynamics of the monopole plasma, using
an Abelian scalar field $\sigma$ dual to the magnetic field. Assuming
that the Abelian magnetic gauge field is three-dimensional for small
$L$, we may write 
\begin{equation}
L\int d^{3}x\,\frac{1}{2}B_{k}^{2}=\int d^{3}x\frac{g^{2}}{32\pi^{2}L}\left(\partial_{k}\sigma\right)^{2}
\end{equation}
 where the normalization of $\sigma$ is chosen to simplify the form
of the interaction terms. The three-dimensional effective action is
given by

\begin{equation}
L_{eff}=\frac{g^{2}}{32\pi^{2}L}\left(\partial_{j}\sigma\right)^{2}-\sum_{a}\xi_{a}e^{-S_{a}+iq_{a}\sigma}
\end{equation}
where the sum is over the set $\left\{ BPS,KK,\overline{BPS},\overline{KK}\right\} $.
Each species of monopole has its own magnetic charge sign $q_{a}=\pm$
as well as its own action $S_{a}$. The coefficients $\xi_{a}$ represent
the functional determinant associated with each kind of monopole,
but the combination $\xi_{a}\exp\left(-S_{a}\right)$ may be usefully
regarded as a monopole activity in terms of the statistical mechanics
of a gas of magnetic charges. The generating functional 
\begin{equation}
Z_{\sigma}=\int\left[d\sigma\right]\exp\left[-\int d^{3}x\, L_{eff}\right]
\end{equation}
 is precisely equivalent to the generating function of the monopole
gas. This equivalence may be proved by expanding $Z_{\sigma}$ in
a power series in the $\xi_{a}$'s, and doing the functional integral
over $\sigma$ for each term of the expansion. 

The magnetic monopole plasma leads to confinement in three dimensions.
For our effective three-dimensional theory, any Wilson loop in a hyperplane
of fixed $x_{4}$, for example a Wilson loop in the $x_{1}-x_{2}$
plane, will show an area law; see figure \ref{fig:Spatial-Wilson-loop}.
The original procedure of Polyakov \cite{Polyakov:1976fu}
may be used to calculate the string tension, where the presence of
a large planar Wilson loop causes the dual field $\sigma$ to have
a discontinuity on the surface associated with the loop and a half-kink
profile on both sides. However, an alternative procedure is simpler
where the discontinuity in the gauge field strength induced by the
Wilson loop is moved to infinity so that the string tension is obtained
from the kink solution connecting the two vacua of the dual field
$\sigma$ \cite{Unsal:2008ch}.

In the confined phase, the action and functional determinant factors
for all four types of monopoles are the same, so we denote them by
$S_{M}$ and $\xi_{M}$. The potential term in the mixed and confined
phases then reduces to 
\begin{equation}
-\sum_{a}\xi_{a}e^{-S_{a}+iq_{a}\sigma}\rightarrow4\xi_{M}e^{-S_{M}}\left[1-\cos\left(\sigma\right)\right]
\end{equation}
 which has minima at $\sigma=0$ and $\sigma=2\pi$; we have added
a constant for convenience such that the potential is positive everywhere
and zero at the minima. A one-dimensional soliton solution $\sigma_{s}\left(z\right)$
connects the two vacua, and the string tension $\sigma_{3d}$ for Wilson
loops in the three non-compact directions is given by 
\begin{equation}
\sigma_{3d}=\int_{-\infty}^{+\infty}dz\, L_{eff}\left(\sigma_{z}(z)\right)
\end{equation}
 which can be calculated via a Bogomolny inequality to be 
\begin{equation}
\sigma_{3d}=\frac{4g}{\pi}\sqrt{\frac{\xi_{M}}{L}e^{-S_{M}}}.
\end{equation}
Of course, this result depends on $L$ and cannot be used outside
the region $\Lambda L\ll1$; nevertheless, this is a concrete realization
of confinement in a four-dimensional field theory via non-Abelian
monopoles.

\subsection{The special case of supersymmetry}

The earliest indication that confinement on $R^{3}\times S^{1}$ was
interesting and tractable came about in an unlikely way, in an effort
to resolve a controversy over adjoint fermion condensates in four-dimensional
$\mathcal{N}=1$ supersymmetric $SU(N)$ gauge theories. Davies \emph{et
al.} resolved the controversy by considering supersymmetric $SU(N)$
gauge theories on $R^{3}\times S^{1}$ \cite{Davies:1999uw,Davies:2000nw}.
In such theories, supersymmetry requires that the number of adjoint
fermion degrees of freedom ({}``gluinos'') equals the number of
gauge degrees of freedom, and that the adjoint fermions have the same
boundary conditions as the gauge fields. This model interpolates between
four-dimensional $\mathcal{N}=1$ supersymmetric $SU(N)$ gauge theories
as $L\rightarrow\infty$ and three-dimensional $\mathcal{N}=2$ supersymmetric
$SU(N)$ gauge theories in the limit $L\rightarrow0$ with no intervening
phase transition as $L$ is varied. In the three-dimensional limit,
the scalar field of the $d=3$ $\mathcal{N}=2$ theory is simply the
$n=0$ Matsubara mode of $A_{4}^{3}$. We follow closely the
original notation
and treatment of $SU(2)$ by Davies \emph{et al.} \cite{Davies:1999uw},
but see also \cite{Unsal:2007vu,Unsal:2007jx};  
some subtleties of the calculation that were previously unappreciated are discussed
in detail in \cite{Poppitz:2012sw}.
As before, a non-vanishing expectation value for $A_{4}^{3}$ breaks
the $SU(2)$ symmetry to $U(1)$, and the low-energy effective theory
will have only $U(1)$ symmetry. The superpotential of this model
is trivial in perturbation theory and the effective potential is completely
non-perturbative. The superpotential can be written in terms of chiral
and anti-chiral $\mathcal{N}=1$ superfields
\begin{eqnarray*}
\Phi & = & Z+\sqrt{2}\theta^{\alpha}\Psi_{\alpha}+\theta^{\alpha}\theta_{\alpha}\mathcal{F}\\
\Phi & = & \bar{Z}+\sqrt{2}\bar{\theta}^{\alpha}\bar{\Psi}_{\alpha}+\bar{\theta}^{\alpha}\bar{\theta}_{\alpha}\bar{\mathcal{F}}
\end{eqnarray*}
where $Z$ is a complex scalar $\phi+i\gamma$ that includes both
the three-dimensional scalar $\phi=A_{4}^{3}$ and a field $\gamma$
which is dual to the $U(1)$ magnetic field $B_{j}^{3}$ and thus
proportional to $\sigma$ in our notation. The effective action is

\begin{equation}
S_{eff}=S_{cl}+\frac{L}{g^{2}}\int d^{3}x\left[\int d^{2}\theta\mathcal{W}\left(\Phi\right)+\int d^{2}\bar{\theta}\bar{\mathcal{W}}\left(\Phi\right)\right]
\end{equation}
where $\mathcal{W}\left(\Phi\right)$ is given by
\begin{equation}
\mathcal{W}\left(\Phi\right)=\left(\frac{g^{2}}{4\pi L}\right)^{2}M\left[\exp\left(-\frac{4\pi L}{g^{2}}\Phi\right)+\exp\left(-\frac{8\pi^{2}}{g^{2}}+\frac{4\pi L}{g^{2}}\Phi\right)\right].
\end{equation}
The parameter $M$ is $M=16\pi^{2}L^{2}M_{PV}^{3}$ with $M_{PV}$
the Pauli-Villars regulator associated with the functional determinant.
The effective potential is given by

\begin{equation}
V_{eff}=\bar{\mathcal{F}}\mathcal{F}=\frac{\partial\mathcal{W}}{\partial Z}\frac{\partial\bar{\mathcal{W}}}{\partial\bar{Z}}
\end{equation}
 with

\begin{equation}
\mathcal{F}=\frac{\partial\mathcal{W}}{\partial Z}=-\frac{Mg^{2}}{4\pi L}\left[\exp\left(-\frac{4\pi L}{g^{2}}\Phi\right)-\exp\left(-\frac{8\pi^{2}}{g^{2}}+\frac{4\pi L}{g^{2}}\Phi\right)\right]
\end{equation}
The minimum of $V_{eff}$ preserves supersymmetry: $\mathcal{F}\left(\left\langle Z\right\rangle =0\right)$
implies $\left\langle \phi\right\rangle =\pi/\beta$ and $\left\langle \gamma\right\rangle =0$,
and leads to $Tr_{F}P=0$ . The effective potential for $\gamma$
is then given by
\begin{equation}
V_{eff}\left(\phi=\pi/\beta,\gamma\right)=2\left(\frac{Mg^{2}}{4\pi L}\right)^{2}\exp\left(-\frac{8\pi^{2}}{g^{2}}\right)\left[1-\cos\left(\frac{8\pi L}{g^{2}}\gamma\right)\right]
\end{equation}
The string tension is again obtained from the one-dimensional kink
solution of the equation of motion obtained from the effective action
for the dual field. This model represents a lower boundary of where
periodic adjoint fermions work to restore $Z(N)$ symmetry, with the
superpartner of the gauge boson corresponding to $N_{f}=1/2$ for
the number of Dirac fermions. For $N_{f}=1/2$, the one-loop effective
potential Eqn. \ref{eq:Veff-adjoint-fermions} favors the deconfined
phase for when the fermion mass is positive, and vanishes identically
when it is zero.

\subsection{Polyakov loop string tensions}

Polyakov loop string tensions are calculable perturbatively in the
small-$L$ confining region from small fluctuations about the confining
minimum of the effective potential \cite{Meisinger:2004pa,Meisinger:2009ne};
see also \cite{Unsal:2008ch}. The effective Lagrangian can be written
in terms of the phases of the Polyakov loop eigenvalues $\theta_{j}$
\begin{equation}
\frac{1}{g^{2}}\sum_{j=1}^{N}\left(\nabla\theta_{j}\right)^{2}+V_{eff}\left(\theta\right)
\end{equation}
where the kinetic term is obtained from the standard kinetic term
for gauge theories and the potential term is the one-loop effective
potential. In the high-$T$ (small-$L$) confining region, the minimum
of the effective potential will have $Z(N)$ symmetry. A convenient
form for such a solution is
\begin{equation}
\left(P_{0}\right)_{jk}=wz^{j}\delta_{jk}
\end{equation}
where $z=\exp\left(2\pi i/N\right)$ and $w$ is a phase factor that
ensures $\det P_{0}=1$. For small fluctuations, we write
\begin{equation}
P=P_{0}e^{i\delta\theta}
\end{equation}
 where $\delta\theta$ lies in the Cartan algebra of $SU(N)$, \emph{i.e.},
is a diagonal traceless matrix, and therefore has $N-1$ independent
components. The moments of the Polyakov loop are 
\begin{equation}
Tr_{F}P^{k}=w^{k}\sum_{j=1}^{N}z^{jk}e^{ik\delta\theta_{j}}
\end{equation}
 If $k$ is not divisible by $N$, we have approximately
\begin{equation}
Tr_{F}P^{k}\simeq ikw^{k}\sum_{j=1}^{N}z^{jk}\delta\theta_{j}=ikw^{k}\delta\tilde{\theta}_{k}
\end{equation}
 where $\delta\tilde{\mbox{\ensuremath{\theta}}}$ is the discrete
Fourier transform of $\delta\theta$, related by
\begin{equation}
\delta\tilde{\theta}_{k}=\sum_{j=1}^{N}z^{jk}\delta\theta_{j}
\end{equation}
\begin{equation}
\delta\mbox{\ensuremath{\theta}}_{j}=\frac{1}{N}\sum_{k=1}^{N}z^{-jk}\delta\tilde{\theta}_{k}.
\end{equation}
 Note that the reality of $\delta\theta$ implies that $\delta\tilde{\theta}_{k}^{*}=\delta\tilde{\theta}_{N-k}$;
the last Fourier component, $\delta\tilde{\phi}_{N}$, is identically
zero, due the tracelessness of $\delta\phi$. If $k$ is divisible
by $N$ we have instead
\begin{equation}
Tr_{F}P^{k}\simeq N-\frac{1}{2}k^{2}\sum_{j=1}^{N}\left(\delta\theta_{j}\right)^{2}=N-\frac{1}{2N}k^{2}\sum_{m=1}^{N}\left(\delta\tilde{\theta}_{m}\delta\tilde{\theta}_{N-m}\right).
\end{equation}
These formulae allow us to write the three-dimensional effective action
in terms of the $\delta\tilde{\theta}_{k}$ to quadratic order. For
each value of $k$, the terms in the potential which contribute have
$n\equiv k$ mod $N$, $n\equiv N-k$ mod $N$, or $n\equiv0$ mod
$N$. We obtain a different mass $\sigma_{k}^{P}/T$ for each Fourier
component $\delta\tilde{\theta}_{k}$ measured by using Polyakov loops
of $N$-ality $k$. There are $\left[N/2\right]$ different string
tensions because $\sigma_{k}^{P}=\sigma_{N-k}^{P}$. The string tensions
obtained, for example, in lattice simulations will depend on the operators
used. For example, the operators $(Tr_{F}P)^{k}\sim\left(\delta\tilde{\theta}_{1}\right)^{k}$
and $Tr_{F}(P^{k})\sim\delta\tilde{\theta}_{k}$ both have $N$-ality
$k$, but the two-point correlation function of the first operator
will decay as $\exp\left[-k\sigma_{1}^{P}r/T\right]$, while the second
operator decays as $\exp\left[-\sigma_{k}^{P}r/T\right]$. Group characters
will in general have a complicated pattern of mixing. Consider the
symmetric/anti-symmetric representations of dimension $N(N\pm1)/2$,
with characters given by 
\begin{equation}
\chi_{S/A}(P)=(1/2)\left[(Tr_{F}P)^{2}\pm Tr_{F}P^{2}\right].
\end{equation}
 These operators will have two different contributions, corresponding
to $2\sigma_{1}^{(t)}$ and $\sigma_{2}^{(t)}$. At higher orders
in perturbation theory, there will be mixing, and ultimately only
the lightest state for a given $N$-ality will be seen at large distances.
This phenomena, often referred to as string breaking,
 was first observed in lattice simulations in
$SU(2)$ Higgs models \cite{Philipsen:1998de,Knechtli:1998gf,Knechtli:2000df}
and later seen in pure $SU(2)$ gauge theory
in $(2+1)$ dimensions \cite{Pepe:2009in}.

The behavior seem in the models considered in
this section is very different from the behavior seen in exactly
solvable two-dimensional gauge theories or in lattice simulations
of four-dimensional gauge theories. This is not surprising. For the
case of double-trace deformations, there are at least $\left[N/2\right]$
independent parameters which can be adjusted, and which can vary in
turn the $\sigma_{k}^{P}$. For the case of periodic adjoint fermion,
the results are continuous functions of $\beta m$.  The string tensions
are of order $g$: 
\begin{eqnarray}
\left(\frac{\sigma_{k}^{P}}{T}\right)^{2} & = & g^{2}N\frac{N_{f}m^{2}}{\pi^{2}}\sum_{j=0}^{\infty}\left[K_{2}\left((k+jN)\beta m\right)+K_{2}\left((N-k+jN)\beta m\right)-2K_{2}\left((j+1)N\beta m\right)\right]\nonumber \\
 &  & -g^{2}N\frac{T^{2}}{3N^{2}}\left[3\csc^{2}\left(\frac{\pi k}{N}\right)-1\right]
\end{eqnarray}
where the gluon contribution has been summed in the last term. Note
that the symmetry $\sigma_{k}^{P}=\sigma_{N-k}^{P}$ is manifest in
this formula. The $m=0$ limit has the simple form 
\begin{equation}
\left(\frac{\sigma_{k}^{P}}{T}\right)^{2}=\frac{\left(2N_{f}-1\right)g^{2}T^{2}}{3N}\left[3\csc^{2}\left(\frac{\pi k}{N}\right)-1\right]\label{eq:spatialtension}
\end{equation}
and is a good approximation for $\beta m\ll1$. This scaling law is
not at all like either Casimir or sine-law scaling, because the usual
hierarchy $\sigma_{k+1}^{P}\ge\sigma_{k}^{P}$ is here reversed. Because
we expect on the basis of $SU(3)$ simulations that the high-temperature
confining region is continuously connected to the conventional low-temperature
region, there must be an inversion of the string tension hierarchy
between the two regions for all $N\ge4$. Other features of the string
tension behavior are in line with our expectations. For the case $N_{f}=1/2$
, corresponding to a single multiplet of adjoint Majorana fermions,
the perturbative string tension vanishes, and it is the non-perturbative
contribution to the effective potential induced by monopoles that
gives rise to the string tension in this case \cite{Davies:1999uw,Davies:2000nw}.
The large-$N$ limit of Eqn. \ref{eq:spatialtension} is smooth. For
fixed $k$ as $N\rightarrow\infty$, we have 
\begin{equation}
\left(\frac{\sigma_{k}^{P}}{T}\right)^{2}\sim\frac{\left(2N_{f}-1\right)\lambda T^{2}}{\pi^{2}k^{2}}
\end{equation}
where $\lambda$ is the 't Hooft coupling $g^{2}N$. The Polyakov
loop string tensions obtained in the small-$L$ confining region,
while conforming to general principles, do not appear to tell us much
about confinement at large $L$.

\subsection{Other geometries}

It is possible to apply the same methods used for gauge theories on
$R^{3}\times S^{1}$ to other geometries, such as $S^{1}\times S^{3}$
or $R^{2}\times T^{2}$. Meyers and Hollowood have performed a detailed
study of $SU(N)$ gauge theories on $S^{1}\times S^{3}$ with periodic
adjoint fermions \cite{Hollowood:2009sy}. In this geometry, $R^{3}$
is replaced by $S^{3}$, so there are two length scales introduced
by the geometry, the radius of the three-sphere $R=R_{S^{3}}$ and
$L=R_{S^{1}}$. We require $\min\left[R_{S^{1}},R_{S^{3}}\right]\ll\Lambda$
so that we are in the weak-coupling region. The projection onto gauge-invariant
states, manifested as integration over the eigenvalues of the Polyakov
loop, ensures non-trivial behavior. Because the spatial volume is
finite, there is no actual phase transition for finite $N$, only
a crossover as $R/L$ is varied. However, the large-$N$ limit does
give a phase transition whose behavior is closely approximated even
for moderate values of $N$.

As in the case of $R^{3}\times S^{1}$, the information about the
phase structure is contained in the effective potential. Each field
adds a factor like
\begin{equation}
\pm Tr_{R}\log\left[-D_{0}^{2}-\triangle\right]
\end{equation}
to the effective potential, where $\triangle$ is the appropriate
Laplacian on $S^{3}$ for each kind of field, $\pm$ for bosonic and
fermionic fields. With $N_{f}$ flavors of fermion, the effective
potential is given by

\begin{equation}
S(P)=\sum_{n=1}^{\infty}\frac{1}{n}\Big\{\big(1-z_{v}(nL/R)\big)Tr_{\text{Adj}}\big(P^{n}\big)+\sum_{f=1}^{N_{f}}z_{f}(nL/R,m_{f}R)Tr_{F}\big(P^{n}\big)\Big\}
\end{equation}
where the first term is due to Haar measure. The second term is from
the gauge boson, and as expected takes the form of a sum over paths
winding around $S^{1}$, with $z_{v}$ given by 
\begin{equation}
z_{v}(L/R)=2\sum_{\ell=1}^{\infty}\ell(\ell+2)e^{-L(\ell+1)/R}=\frac{6e^{-2L/R}-2e^{-3L/R}}{(1-e^{-L/R})^{3}}.
\end{equation}
The final term, due to the fermions, has a similar interpretation,
with
\begin{equation}
z_{f}(L/R,mR)=2\sum_{\ell=1}^{\infty}\ell(\ell+1)e^{-L\sqrt{(\ell+1/2)^{2}+m^{2}R^{2}}/R}.
\end{equation}
 Although the fermionic contribution cannot be put into a simple closed
form, Hollowood and Myers derive the useful form
\begin{equation}
z_{f}\left(\frac{L}{R},mR\right)=\frac{2m^{2}R^{3}}{L}K_{2}(Lm)-\frac{mR}{2}K_{1}(Lm)+4\int_{mR}^{\infty}dx\frac{x^{2}+\tfrac{1}{4}}{e^{2\pi x}+1}\sin(L\sqrt{x^{2}-m^{2}R^{2}}/R)
\end{equation}
 In the limit $m\rightarrow0$, this reduces to
\begin{equation}
z_{f}(L/R,0)=\frac{4e^{-3L/2R}}{(1-e^{-L/R})^{3}}\equiv\sum_{\ell=1}^{\infty}2\ell(\ell+1)e^{-L(\ell+1/2)/R}
\end{equation}
and in the limit $R\to\infty$ with fixed $m$ and $L$, we recover
\begin{equation}
z_{f}(L/R,mR)\longrightarrow\frac{2m^{2}R^{3}}{L}K_{2}(Lm)
\end{equation}
which is the expression that one obtains by working directly on $R^{3}\times S^{1}$
\cite{Myers:2009df}.

In the case of adjoint fermions, we have
\begin{equation}
S(P)=\sum_{n=1}^{\infty}\frac{1}{n}\Big(1-z_{v}(nL/R)+N_{f}z_{f}(nL/R,m_{f}R)\Big)\sum_{ij=1}^{N}\cos(n(\theta_{i}-\theta_{j}))\ .
\end{equation}
 In the large-$N$ limit, the effective potential becomes a functional
of the distribution $\rho\left(\theta\right)$ of Polyakov loop eigenvalues,
with
\begin{equation}
S[\rho(\theta)]=N^{2}\int d\theta\,\int d\theta'\,\rho(\theta)\rho(\theta')\sum_{n=1}^{\infty}\frac{f(nL/R,mR)}{n}\cos(n(\theta-\theta'))
\end{equation}
 where
\begin{equation}
f(L/R,mR)=1-z_{B}(L/R)+N_{f}z_{F}(L/R,mR)
\end{equation}
 and $\rho$ is normalized to 
\begin{equation}
\int_{0}^{2\pi}d\theta\,\rho(\theta)=1\ .
\end{equation}
This expression for $S[\rho(\theta)]$ can be made even simpler by
writing $\rho$ in terms of its Fourier components
\begin{equation}
\rho(\theta)=\frac{1}{2\pi}\,\sum_{n=-\infty}^{\infty}\rho_{n}e^{in\theta}
\end{equation}
 where $\rho_{n}=Tr_{F}P^{n}/N$. Then we may write the effective
potential as
\begin{equation}
S[\rho(\theta)]=\frac{N^{2}}{2}\sum_{n=1}^{\infty}\frac{f(nL/R,mR)}{n}|\rho_{n}|^{2}
\end{equation}
It is now obvious that the effective potential will become unstable
in the contribution of $N$-ality $n$ precisely when $f(nL/R,mR)=0$.
These techniques can also be applied to the study of gauge theories
at finite temperature and density on $S^{3}\times S^{1}$ \cite{Hands:2010zp,Hands:2010vw,Hollowood:2011ep}.
Another interesting geometry is $R^{2}\times T^{2}$, where $SU(N)$
gauge theories are in the universality class of $Z(N)\times Z(N)$
spin models because there are two compact directions \cite{Simic:2010sv,Anber:2011gn}.
Dimensional reduction leads to two-dimensional models where many powerful
techniques are available, as discussed in Section \ref{sec:Conformality-vs-Confinement}.

\subsection{Higgs theories on $R^{3}\times S^{1}$}

As shown by 't Hooft \cite{'tHooft:1977hy,'tHooft:1979uj}, there
is a fundamental conflict between the Higgs mechanism and confinement.
The dual superconductor picture of confinement gives a simple picture
of this conflict as being between an electric condensate in the Higgs
mechanism and a magnetic condensate in a confining model. With analytic
control of confinement in gauge theory on $R^{3}\times S^{1}$, an
adjoint Higgs field can be introduced to study the interplay of confinement
and the Higgs mechanism analytically \cite{Nishimura:2010xa,Nishimura:2011md}.
The simplest case is $SU(2)$, with a classical Euclidean action given
by 
\begin{equation}
S_{c}=\int d^{4}x\left[\frac{1}{4}\left(F_{\mu\nu}^{a}\right)^{2}+\frac{1}{2}\left(D_{\mu}\phi\right)^{T}\cdot D_{\mu}\phi+V\left(\phi\right)\right]
\end{equation}
where the Higgs potential is
\begin{equation}
V\left(\phi\right)=\frac{1}{2}m^{2}\phi^{2}+\frac{1}{4}\lambda\left(\phi^{2}\right)^{2}
\end{equation}
 with $\phi^{2}=\phi^{T}\phi$. The action has a $Z(2)_{H}$ global
symmetry given by $\phi\rightarrow-\phi$, in addition to $Z(2)_{C}$
center symmetry which transform $P$ to $-P$.

The scalar field $\phi$ is not gauge invariant, and cannot serve
as an order parameter for the breaking of the $Z(2)_{H}$ symmetry
associated with $\phi$ when gauge interactions are present. This
is an old problem, a consequence of Elitzur's theorem \cite{Elitzur:1975im}.
Higgs models with scalar fields in the fundamental and adjoint representations
behave differently. For Higgs models with scalar fields in the fundamental
representation, the confined and Higgs phases are connected \cite{Fradkin:1978dv},
in a manner similar to the connection between liquid and gas phases.
In this case, center symmetry is explicitly broken, and large Wilson
loops do not have area-law behavior due to screening by the scalars.
In the adjoint case, center symmetry is preserved by the action, and
there is a distinct phase transition between the confined and Higgs
phases. On $R^{3}\times S^{1}$, there are three distinct gauge-invariant
order parameters associated with the $Z(2)_{C}\times Z(2)_{H}$ symmetry.
The first of these is the trace in the fundamental representation
of the Polyakov loop $P$ itself, $\left\langle Tr_{F}P\right\rangle $.
It transforms non-trivially under $Z(2)_{C}$ but is invariant under
$Z(2)_{H}$. The second is $\left\langle Tr_{F}\left[P^{2}\left(x\right)\phi(x)\right]\right\rangle $
which is invariant under $Z(2)_{C}$, but transforms non-trivially
under $Z(2)_{H}$. Finally, there is $\left\langle Tr_{F}\left[P\left(x\right)\phi(x)\right]\right\rangle $,
which transforms non-trivially under both groups.

Using the three order parameters, one can show that there are four
distinct phases: a deconfined phase, a confined phase, a Higgs phase,
and a mixed confined phase. The mixed confined phase occurs where
one might expect a phase in which there is both confinement and the
Higgs mechanism, but the behavior of the order parameters distinguishes
the two phases. In the mixed confined phase, the $Z(2)_{C}\times Z(2)_{H}$
global symmetry breaks spontaneously to a $Z(2)$ subgroup that acts
non-trivially on both the scalar field and the Polyakov loop. In the
mixed confined phase, the role of the scalar field is played a linear
combination of the Higgs field $\phi$ and $A_{4}$ in the construction
of BPS and KK monopole solutions. In all four phases, Wilson loops
orthogonal to the compact direction are expected to show area-law
behavior due to unbroken center symmetry in the non-compact directions.
This confining behavior can be attributed to a dilute monopole gas
in a broad region that includes portions of all four phases. 

The supersymmetric analog of this model is the Seiberg-Witten model
\cite{Seiberg:1994rs}, which is an $\mathcal{N}=2$ supersymmetric
gauge theory with gauge group $SU(2)$. Seiberg and Witten found that
in this model the addition of an $\mathcal{N}=1$ mass perturbation
leads to confinement by magnetic monopoles. Recently, Poppitz and
Unsal have examined the behavior of this model on $R^{3}\times S^{1}$,
and concluded that the confined phase seen for small compactification
circumference on $R^{3}\times S^{1}$ is connected to the confining
phase at infinite compactification circumference \cite{Poppitz:2011wy}.
In their work, Euclidean monopoles in which a linear combination of
$A_{4}$ and $\phi$ plays the role of the scalar field appear in
a very similar fashion to the non-supersymmetric model.

\section{\label{sec:Conformality-vs-Confinement} Conformality and duality}

\subsection{Non-trivial fixed points in $d=3+1$ gauge theories}

Up until this point, we have implicitly restricted ourselves to theories
like QCD which have a simple renormalization group structure: they
have an ultraviolet fixed point at $g=0$, and hence are asymptotically
free, and an infrared fixed point at $g=\infty$ when the theory is
defined on $R^{4}$. Gauge theories with non-trivial fixed point structure
are known to occur if the number of particle representations included
in the theory is sufficiently large. The first two coefficients of
the perturbative contribution to the renormalization-group $\beta$
function of a gauge theory are independent of the renormalization
scheme, and given by
\begin{equation}
\beta(g^{2})=\frac{dg^{2}}{d\log q^{2}}=-\frac{b_{1}}{16\pi^{2}}g^{4}-\frac{b_{2}}{(16\pi^{2})^{2}}g^{6}+\cdots
\end{equation}
where
\begin{eqnarray*}
b_{1} & = & \frac{11}{3}\, C_{2}(G)-\frac{4}{3}\, N_{f}T(R)\\
b_{2} & = & \frac{34}{3}\,[C_{2}(G)]^{2}-N_{f}T(R)\left[\frac{20}{3}\, C_{2}(G)+4C_{2}(R)\right]
\end{eqnarray*}
The coefficients $b_{1}$ and $b_{2}$ depend on the group and the
group representations of the particles via $C_{2}(R)$, the value
of the quadratic Casimir operator in representation $R$ ($G$ denotes
the adjoint representation, so $C_{2}(G)=N_{c}$), while $T(R)$ is
the normalization of the group generators in $R$: $Tr\left(T_{R}^{a}T_{R}^{b}\right)=T(R)\delta^{ab}$.

\begin{figure}
\includegraphics[scale=0.26]{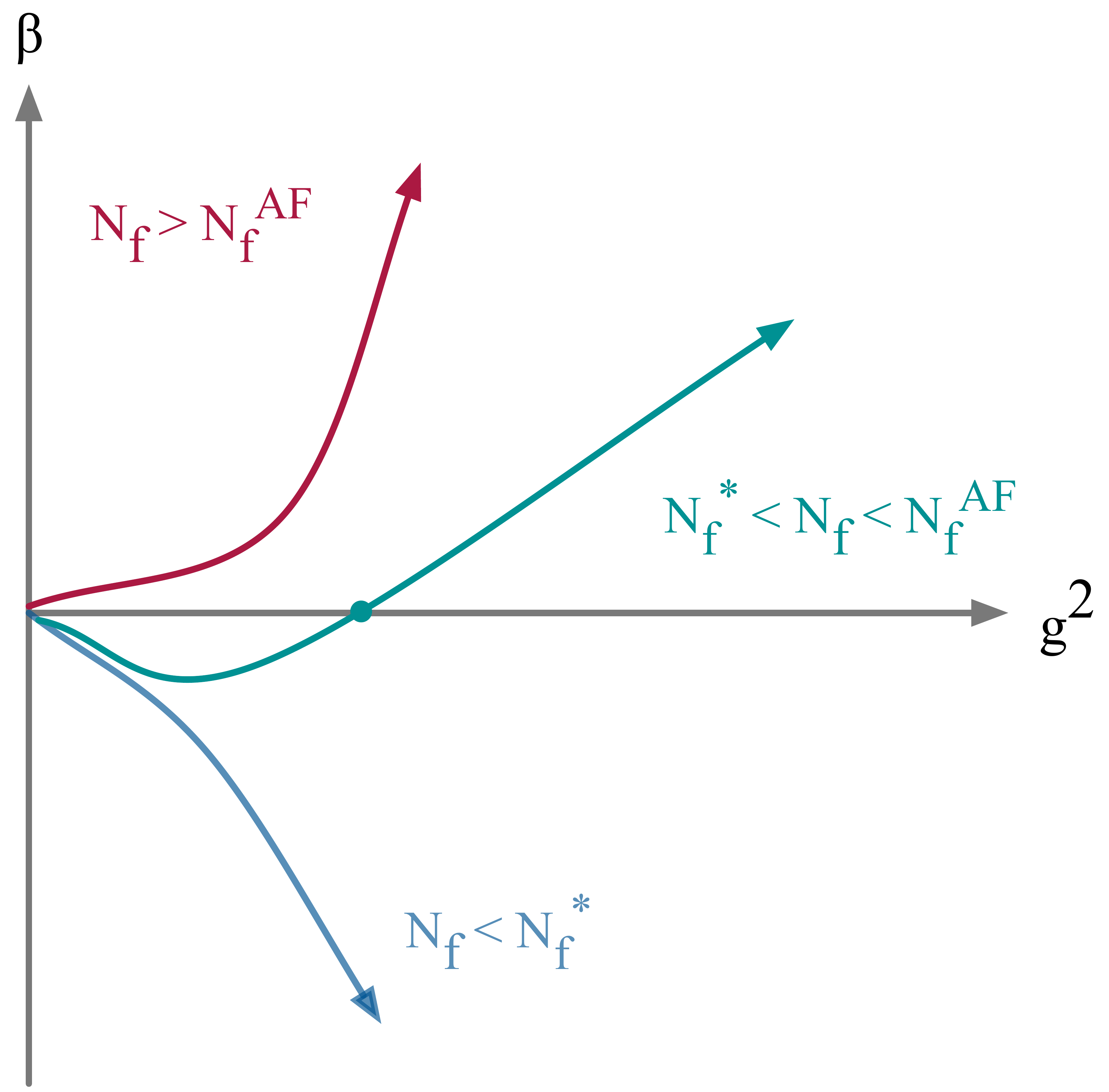}\caption{\label{fig:beta-functions}Three possible behaviors for the beta function.
If $N_{f}<N_{f}^{*},$ the theory is asymptotically free but has no
non-trivial fixed point. If $N_{f}>N_{f}^{AF}$, the gauge theory
is no longer asymptotically free. For $N_{f}^{*}<N_{f}<N_{f}^{AF},$
there is a conformal window with a non-trivial infrared fixed point.}
\end{figure}

\begin{figure}
\includegraphics[scale=0.3]{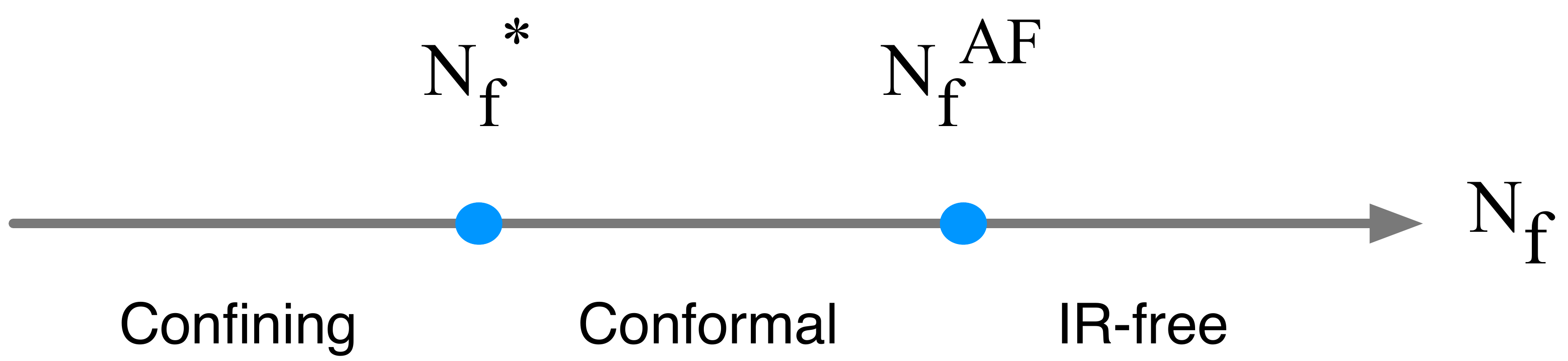}\caption{\label{fig:conformal-window-Nf} Phase structure of a theory with
a conformal window as a function of the number of flavors $N_{f}$.}
\end{figure}

Three possible behaviors may be obtained from perturbation theory:
\begin{lyxlist}{00.00.0000}
\item [{1)}] $b_{1}<0$ : the theory is no longer asymptotically free at
high energies, and $g=0$ is an infrared fixed point, as is the case
in QED. For a non-Abelian gauge theory, this is typically the case
when the matter content of the theory becomes too large. Only the
gauge fields themselves give a negative contribution to $b_{1}$,
so a sufficient number of additional matter fields can drive $b_{1}$
negative. 
\item [{2)}] $b_{1}>0$ and $b_{2}>0$: in this case the coupling runs
out of the region of perturbative computability in the IR. This is
the QCD-like case, where the theory is asymptotically free at high
energies (the UV) and confining at low energies (the IR).
\item [{3)}] $b_{1}>0$ and $b_{2}<0$: Perturbation theory predicts a
non-trivial fixed point, given by
\begin{equation}
g_{BZ}^{2}=-16\pi^{2}\frac{b_{1}}{b_{2}}.
\end{equation}
\end{lyxlist}
These three behaviors are shown in \ref{fig:beta-functions}. 
The corresponding phase structure as a function of the number of flavors is shown in figure \ref{fig:conformal-window-Nf}.
As was first pointed out by Banks and Zaks \cite{Banks:1981nn}, this
fixed point may or may not be located in the region where perturbation
theory is reliable. However, when it is located in that region, the
prediction of an infrared-stable fixed point is reliable. Note however
that a change in the renormalization prescription, which redefines
the coupling constant, will change the beta function. This in turn
will change the location of a non-trivial fixed point. Given that
such theories are possible, we would like to understand their properties.
Such theories are not generally candidates for describing the real
world. Theories with non-trivial infrared fixed points have correlation
functions that decay algebraically at large distances, in a manner
familiar from critical behavior at second-order phase transitions.
Requiring that models not have an infrared fixed point is a condition
that likely requires non-perturbative understanding that lattice gauge
theories can supply. The case of a non-trivial infrared fixed point,
case 3), is intermediate between the familiar case 2) and non-asymptotically
free theories. Typically, for a given gauge group and choice of additional
representation $R$, there is some number of flavors $N_{f}^{AF}$
such that asymptotic freedom is lost, and some smaller value $N_{f}^{*}$
where an infrared fixed point appears. Theories with $N_{f}^{*}<N_{f}<N_{f}^{AF}$
are said to lie in the conformal window.

There is a close relation between the existence of the conformal window
and possible beyond the standard model physics associate with the
Higgs field. In the standard model, the Higgs field is a fundamental
scalar field, in the sense of not being a composite; in fact it is
the only fundamental scalar field in the standard model. Because the
bare Higgs mass receives corrections which are quadratic in any ultraviolet
cutoff, there is a problem of {}``naturalness'' or {}``fine-tuning''
in achieving a sufficiently small Higgs mass. This problem would be
circumvented if the Higgs were a bound state of some new fermions
rather than a fundamental scalar. Theories of this type that have
a non-fundamental Higgs are called technicolor models, and the gauge
group of the new fermions is the technicolor group; see \cite{Sannino:2009za}
for a recent review. In the standard model, the Higgs is responsible
not just for the gauge boson masses, but also for the fermion masses.
The class of models needed for successful phenomenology are called
extended technicolor models, and seem to require a gauge theory very
close to the conformal window. Such theories are often referred to
as {}``walking'' because the running coupling constant moves slowly
if the beta function is small, as it would be near a zero.

There has been much work done on the location of the conformal window
but definitive results have not yet been achieved. Much of the analytic
work uses the chiral order parameter $\bar{\psi}\psi$ as the order
parameter for the transition from a confining phase, where $\bar{\psi}\psi$
is expected to be non-zero, to a conformal phase which has no dynamically
generated scale. One method is based on truncated Schwinger-Dyson
equations \cite{Appelquist:1988yc,Cohen:1988sq,Miransky:1996pd,Appelquist:1996dq,Appelquist:1998rb,Dietrich:2006cm}
while another approach uses renormalization group methods \cite{Gies:2005as}.
In many supersymmetric theories, the boundary of the conformal window
can be determined, as discussed in \cite{Intriligator:1995au}. There
are also estimates for non-supersymmetric models based on ideas from
supersymmetry \cite{Ryttov:2007sr}. Introductory reviews of lattice
gauge theory research in this area can be found in \cite{Fleming:2008gy,DeGrand:2009mt,DeGrand:2010ba,DelDebbio:2011rc}.
Recent research includes the study of model with fermions in the fundamental
representation \cite{Hasenfratz:2009ea,Fodor:2009wk,Hasenfratz:2010fi,Fodor:2011tu,Appelquist:2011dp},
the adjoint representation \cite{DelDebbio:2008zf,Hietanen:2009az,DelDebbio:2009fd,Bursa:2009we,DelDebbio:2010hx,Fukano:2010yv,DeGrand:2011qd},
and the symmetric representation \cite{Fodor:2009ar,DeGrand:2009hu,Kogut:2010cz,DeGrand:2010na,Kogut:2011ty}.

\subsection{Conformality and duality in Abelian models}

There is an interesting class of lattice models that share some of
the features of gauge theories with conformal windows: $Z(N)$ spin
systems in two dimensions and $Z(N)$ gauge theories in four dimensions.
Our understanding of these models originates in the two-dimensional
XY model, which is a spin model with a global $U(1)$ symmetry. The
order parameters of the model are the operators
\begin{equation}
S_{p}\left(x\right)=e^{ip\phi\left(x\right)}
\end{equation}
where $\phi\left(x\right)$ takes on values between $0$ and $2\pi$.
The XY model provides the principle example of a phase transition
driven by topological excitations, the Berezinsky-Kosterlitz-Thouless
(BKT) transition \cite{Berezinsky:1970fr,Kosterlitz:1973xp}. Low-temperature
arguments indicate that the XY model has a low-temperature gapless
phase represented by a line of critical points, along which critical
indices vary continuously with the temperature. On the other hand,
high-temperature expansions indicate a gapped phase at high temperatures.
The phase transition that separates the two phases is driven by vortices,
configurations of spins that have a non-trivial winding number. In
a continuum notation, these are configurations that have
\begin{equation}
\oint\nabla\phi\cdot dx=2\pi n
\end{equation}
 where $n$ is a non-zero integer. Vortices act as a two-dimensional
classical Coulomb gas, interacting via a long-ranged logarithmic interaction.
In the low-temperature phase, vortex pairs bind tightly to form bound
states of zero vorticity and have no effect on the large-distance
behavior. At the critical temperature, vortices unbind to form a classical
Coulomb plasma, giving rise to a mass gap, the Debye mass of the plasma.
From a naive continuum field theory point of view, the XY model has
only massless spin-wave excitation, with an Euclidean action given
by
\begin{equation}
S=\int d^{2}x\frac{J^{2}}{2}\left(\partial\phi\right)^{2}
\end{equation}
where $J^{2}$ is a coupling constant inversely proportional to the
temperature. It is convenient to rescale the field $\phi$ by $\phi\rightarrow\phi/J$
so that the kinetic term of the action has conventional normalization.
Vortices are created and destroyed by operators of the form
\begin{equation}
V_{p}\left(x\right)=e^{i2\pi pJ\tilde{\phi}\left(x\right)}
\end{equation}
 where $\phi$ and $\tilde{\phi}$ are related by
\begin{equation}
i\partial_{\mu}\phi=\epsilon_{\mu\nu}\partial^{\nu}\tilde{\phi}
\end{equation}
The effect of vortices gives rise to an effective action
\begin{equation}
S_{eff}=\int d^{2}x\left[\frac{J^{2}}{2}\left(\partial\phi\right)^{2}-\sum_{p=1}^{\infty}2y_{p}\Lambda^{2}\cos\left(2\pi pJ\tilde{\phi}\right)\right]
\end{equation}
where $\Lambda$ is a cutoff on the order of the lattice spacing and
$y_{p}$ is a dimensionless activity for a vortex of winding number
$\pm p$ \cite{Jose:1977gm,Ogilvie:1981mj}. This generalized sine-Gordon
model gives rise to Coulomb gas of vortices when the partition function
is expanded in the activities. Generally speaking, it is only necessary
to consider the $p=\pm1$ vortices. This model can be perturbed to
a model with $Z(p)$ symmetry by the addition of an operator $-h_{p}\Lambda^{2}\cos\left(pJ^{-1}\phi\right)$
which explicitly breaks the $U(1)$ symmetry down to $Z(p)$:
\begin{equation}
S_{eff}=\int d^{2}x\left[\frac{J^{2}}{2}\left(\partial\phi\right)^{2}-2y_{1}\Lambda^{2}\cos\left(2\pi J\tilde{\phi}\right)-2h_{p}\cos\left(p\phi/J\right)\right].
\end{equation}
This model has a duality under the interchange
\begin{eqnarray*}
\phi & \leftrightarrow & \tilde{\phi}\\
h_{p} & \leftrightarrow & y_{1}\\
\frac{p}{J} & \leftrightarrow & 2\pi J
\end{eqnarray*}
which is a generalization of Kramers-Wannier duality for the Ising
model \cite{Kramers:1941kn,Kramers:1941zz}. 

\begin{figure}
\includegraphics[scale=0.25]{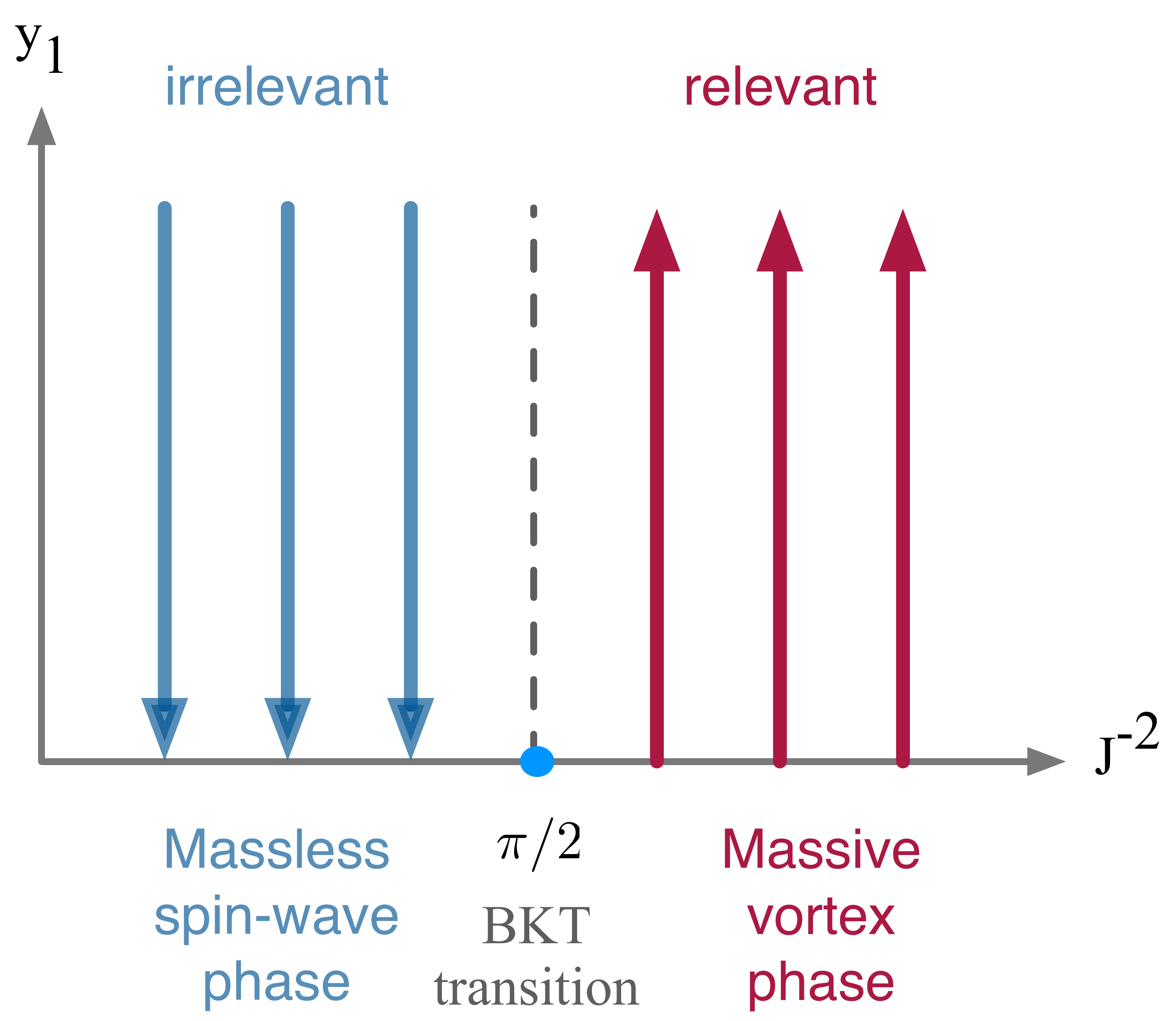}\caption{\label{fig:simple-SG-rg-flow}Relevance of $V_{1}$ for the sine-Gordon
model.}
\end{figure}

\begin{figure}
\includegraphics[scale=0.25]{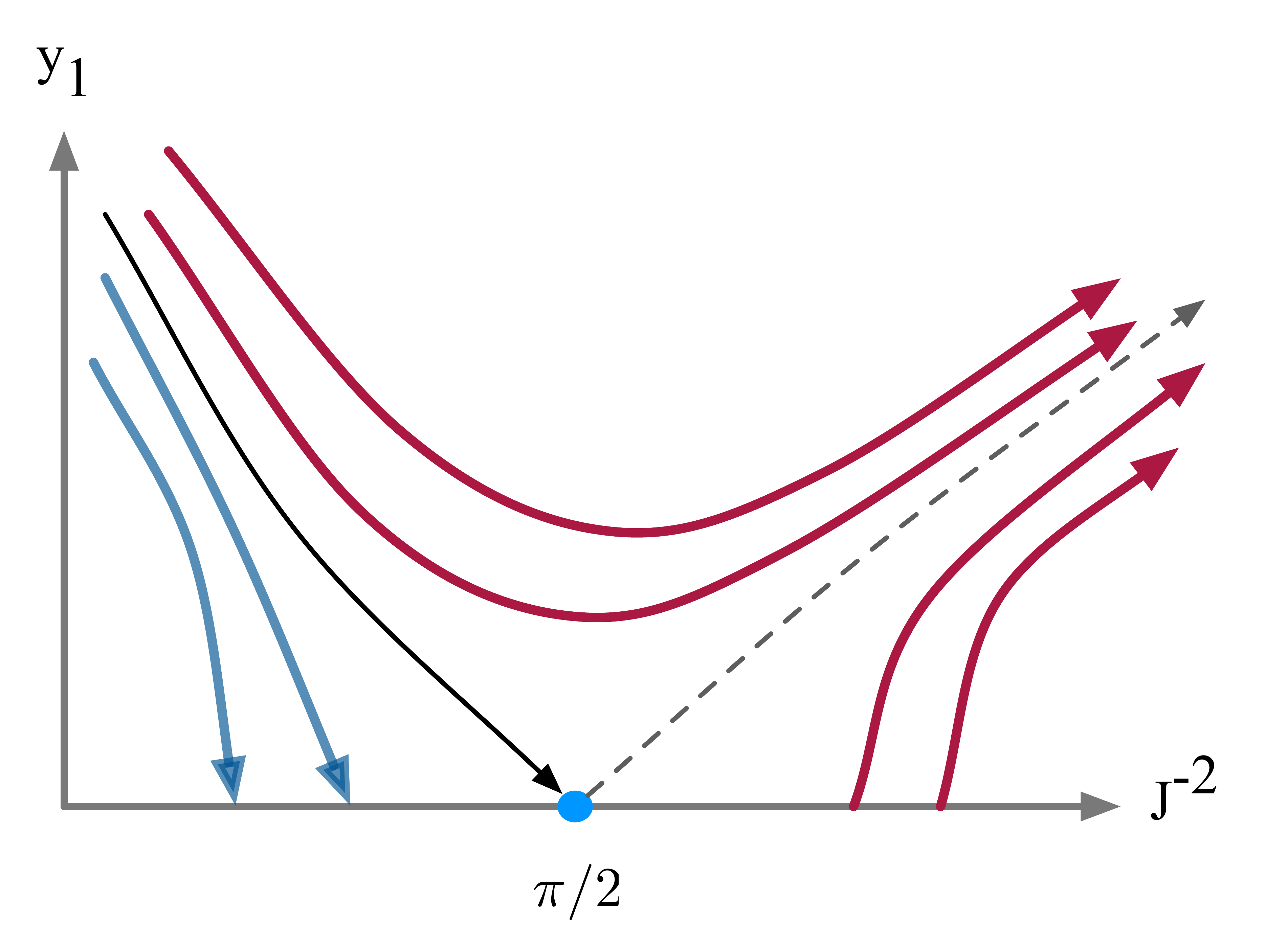}
\caption{\label{fig:RG-flow-sine-Gordon}Actual sine-Gordon flow.}
\end{figure}
\begin{figure}
\includegraphics[scale=0.25]{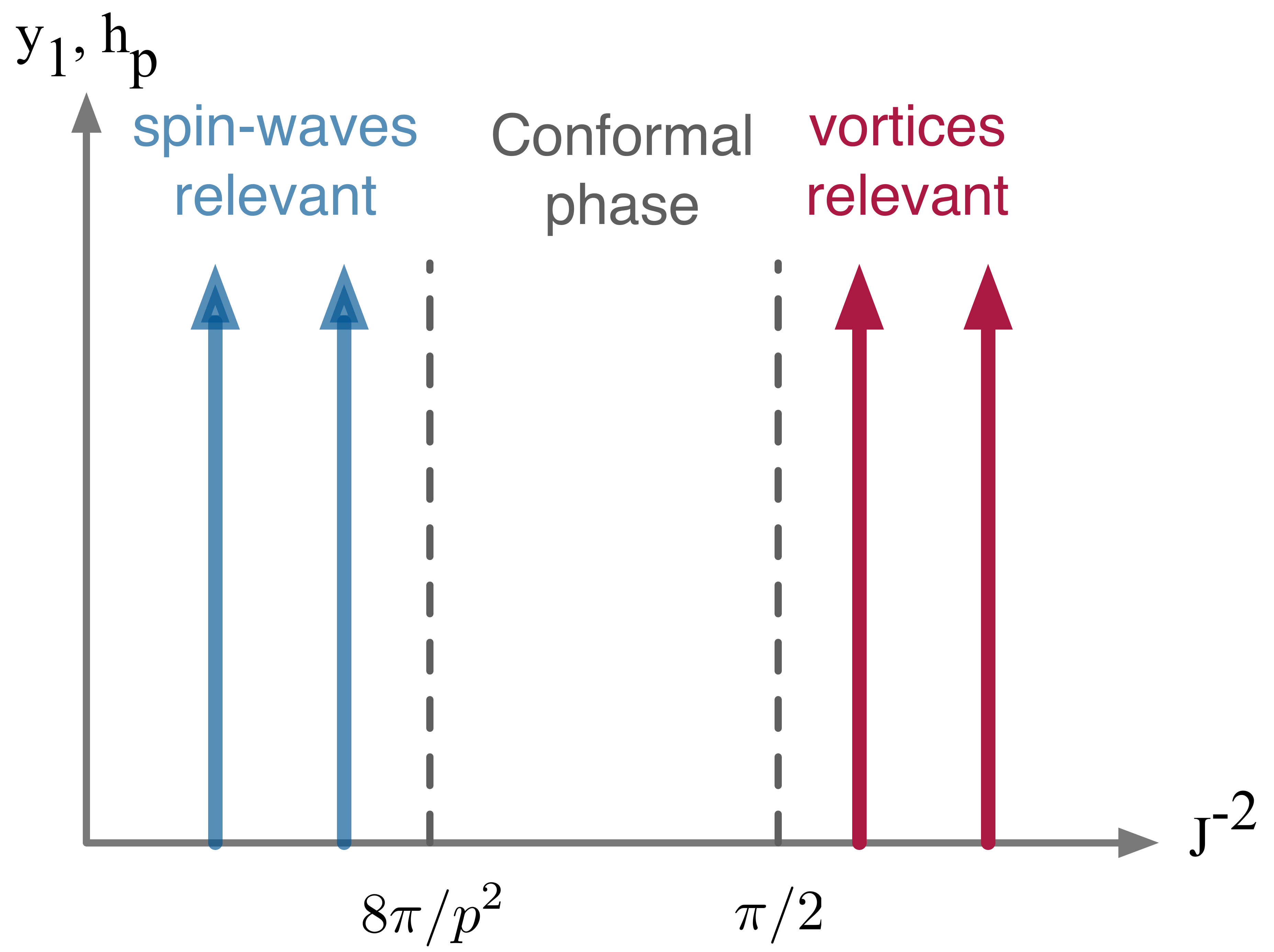}\caption{\label{fig:Z(p)-Relevance}Relevance of $S_{p}$ and $V_{1}$ operators
in a $Z(p)$ model for $p>4$.}
\end{figure}

A simple understanding of the renormalization-group behavior of the
XY model is provided by analyzing the multiplicative renormalization
of the spin-wave and vortex operators associated with normal ordering
\cite{Coleman:1974bu,Amit:1979ab}. For a massless free field, the
relation between the bare field $\exp\left(i\beta\phi\right)$ and
the normal-ordered field $N\left[\exp\left(i\beta\phi\right)\right]$
is 
\begin{equation}
\Lambda^{2}e^{i\beta\phi}=\mu^{2}\left(\frac{\Lambda}{\mu}\right)^{2-\frac{\beta^{2}}{4\pi}}\left[e^{i\beta\phi}\right]
\end{equation}
where $\mu$ is arbitrary. An intereaction term of this type will
be relevant if $\beta^{2}<8\pi$ and irrelevant if $\beta^{2}>8\pi$\@.
When applied to the XY model, this implies that vortices are relevant
only if $J^{-2}>\pi/2$, in the high-temperature phase. This leads
to the simple picture shown in Fig. \ref{fig:simple-SG-rg-flow}.
Taking into account the renormalization of the vortex activity $y_{1}$
leads to a more complete picture of the renormalization group flow
for the XY model, as shown in Fig. \ref{fig:RG-flow-sine-Gordon}.
Applying the same simple condition to the $Z(p)$ model indicates
that the spin-wave symmetry breaking term $S_{p}+S_{-p}$ is relevant
if $J^{-2}<8\pi/p^{2}$, corresponding to low temperatures. If $p>4$,
there is a gap between the regions where neither spin-waves induced
by $h_{p}\ne0$ nor vortices are relevant, as shown in Fig. \ref{fig:Z(p)-Relevance}.
In this intermediate conformal window, correlation functions decay
algebraically; there is no mass gap. This phase structure is only
part of a larger picture for more general $Z(p)$ lattice models.
As we have seen in Sections \ref{sec:Symmetries-of-gauge-theories}
and \ref{sec:Phases-on-R3xS1}, systems with $Z(p)$ symmetries can
break spontaneously to a non-trivial subgroup of $Z(p)$. This also
occurs in $d=2$ $Z(p)$ spin systems \cite{Alcaraz:1980sa,Alcaraz:1980bb,Dorey:1998fx};
for example $Z(6)$ can break spontaneously to $Z(2)$ or $Z(3)$.
For $p=2$ and $p=3$, the regions of vortex and spin-wave relevance
overlap, and must be handled as special cases. 

These results extend to gauge theories in several different ways.
The analysis of two dimensional systems by Jose \emph{et al.} \cite{Jose:1977gm}
can be extended to Abelian lattice gauge theories. In four-dimensional
Abelian lattice models, the analogs of the point-like vortices of
the two-dimensional spin models are closed magnetic monopole loops
\cite{Banks:1977cc}. Similar to two-dimensional spin systems, four-dimensional
$Z(p)$ lattice gauge theories have a duality between monopole loops
and loops of charged particles. For $p>4$, there is a gapless intermediate
phase where neither the electric nor the magnetic operators are relevant
\cite{Ukawa:1979yv}. In three-dimensional Abelian lattice models,
the topological excitations are point-like, as is the case in Polyakov's
treatment of the continuum three-dimensional $SU(2)$ gauge theory
broken to $U(1)$. This is not accidental, because there is a clear
relation between Abelian duality on the lattice and in the continuum.
At finite temperature, \emph{i.e.}, in a $R^{2}\times S^{1}$ geometry,
Polyakov's model can be analyzed in a manner analogous to $d=2$ spin
systems. Three-dimensional monopole solutions made periodic on $S^{1}$
play the role of vortices, and early work identified the phase transition
to be in the XY-model, or $U(1)$, universality class \cite{Agasian:1997wv}.
As in $3+1$ dimensions, adoint Polyakov lines associated with $W^{\pm}$
gauge bosons ensure that the symmetry group of the low-energy theory
remains $Z(2)$ \cite{Dunne:2000vp}.

Another interesting connection between the behavior of $d=2$ spin
models and $d=4$ gauge theories is found in the suggestion by Kaplan
et al. that there may be two nontrivial fixed points in gauge theories
with a conformal window \cite{Kaplan:2009kr}. In this scenario, there
is a UV fixed point at $g^{2}=0$, followed by a non-trivial infrared
fixed point, followed by a non-trivial ultraviolet fixed point at
a larger value of the $g^{2}$. As $N_{f}$ decreases, the two non-trivial
fixed points annihilate, which is to say they move off into the complex
plane. This behavior would give rise to BKT scaling behavior in the
region near where the fixed point merger occurs.

\subsection{Relevance of topological objects on $R^{3}\times S^{1}$}

Unsal and Poppitz have proposed an alternative criterion for locating
the boundary between confining and conformal gauge theories \cite{Poppitz:2009uq,Poppitz:2009tw}.
Their work is based on the relevance or irrelevance of the topological
objects relevant for confinement on $R^{3}\times S^{1}$, analogous
to the criterion for the relevance of vortices in $d=2$ Abelian spin
systems. For all center-symmetric gauge theories on $R^{3}\times S^{1}$,
the mass scale induced by topological excitations can be written in
the form 
\begin{equation}
m\left(L\right)\sim\frac{1}{L}\exp\left[-q\frac{8\pi^{2}}{g^{2}\left(L\right)N}\right]
\end{equation}
where $q$ is $\mathcal{O}\left(1\right)$ and depends on the details
of the theory. With the use of the renormalization group, we can write
this as
\begin{equation}
m\left(L\right)\sim\Lambda\left(\Lambda L\right)^{q\bar{b}_{1}-1}
\end{equation}
where $\bar{b}_{1}\equiv b_{1}/N$/ Requiring that the theory remain
in the center-symmetric, confined phase as $L$ becomes large puts
a constraint on the theory. In the limit $L\rightarrow\infty$, the
topological excitations are relevant if $q\bar{b}_{1}>1$ and the
theory is confined for large $L$. If $q\bar{b}_{1}<1$ , the theory
is not confined. While conceptually simple, the details of the calculations
for different models are somewhat technical. The parameter $q$ is
an integer determined essentially by how many monopoles, each of action
$8\pi^{2}/g^{2}N$, are required in the process that generates the
mass gap. For $N_{f}^{D}$ flavors of Dirac fermions in the fundamental
representation, the conformal window is estimated to lie in the range
\begin{equation}
\frac{5}{2}N\le N_{f}^{D}\le\frac{11}{2}N
\end{equation}
while for $N_{f}$ flavors of Weyl fermions in the adjoint representation
the range is
\begin{equation}
4\le N_{f}^{W}\le\frac{11}{2}
\end{equation}
 where $N_{f}^{W}=2N_{f}^{D}$ if Dirac fermions are used. Estimates
of the conformal window for other representations can also be made
\cite{Poppitz:2009uq}. The estimates are comparable to estimates
made using other methods, but the connection between this method and
others is at present unclear.

\section{\label{sec:Large-N}Gauge theories in the large-N limit}

Many theories simplify in the large-$N$ limit, where $N$ is the
dimension of some internal symmetry group such as $SU(N)$ taken to
be large. Although physical QCD has $N=3,$ there is reason to believe
the large-$N$ limit provides a good description of many features
\cite{'tHooft:1973jz,Witten:1979kh}. Lattice simulations at finite
values of $N$ have been able to provide accurate extrapolations to
the large-$N$ limit for some quantities \cite{Teper:2005hx,Narayanan:2007fb}.
For example, lattice simulations of $SU(N)$ gauge theories in $\left(2+1\right)$
dimensions have provided evidence for $1/N$ corrections in string
tension scaling laws \cite{Bringoltz:2008nd}. The AdS/CFT correspondence,
which relates gauge theories to gravity duals
 in the large-$N$ limit \cite{Maldacena:1997re},
provides another reason for interest in large-$N$ gauge theories;
see \cite{Peeters:2007ab,CasalderreySolana:2011us} for reviews.

The large-$N$ limit began as an approximation in statistical mechanics
\cite{Stanley:1968gx}, but was quickly applied to models with scalars
$\phi^{a}$ \cite{Coleman:1974jh} or fermions $\psi^{a}$ \cite{Gross:1974jv}
in the vector representation of the groups $O\left(N\right)$ or $SU\left(N\right)$
. Such models are exactly solvable in the large-$N$ limit. Field
theories with particles in the adjoint representation, such as gauge
bosons, are much more difficult.

The original approach of 't Hooft to the large-$N$ limit in $SU(N)$
Yang-Mills theories was based around a detailed analysis of classes
of Feynman diagrams using an ingenious double line notation \cite{'tHooft:1973jz}.
This notation is based on the group properties of $SU(N)$ propagators:
for fundamental representation fermions, we have
\begin{equation}
\left\langle \psi^{a}\left(x\right)\bar{\psi}_{b}\left(y\right)\right\rangle =\delta_{b}^{a}S\left(x-y\right)
\end{equation}
 but for gauge bosons, which are in the adjoint representation, we
have
\begin{equation}
\left\langle A_{\mu b}^{a}\left(x\right)A_{\nu d}^{c}\left(y\right)\right\rangle =\left[\delta_{d}^{a}\delta_{b}^{c}-\frac{1}{N}\delta_{b}^{a}\delta_{d}^{c}\right]D_{\mu\nu}\left(x-y\right)
\end{equation}
In this notation, quarks in $N$, the fundamental representation of
$SU(N)$, and antiquarks in $\bar{N}$ are represented by single lines
with arrows. Gauge bosons in the adjoint representation are represented
by a double line with one arrow in each direction; other representations
may be represented by other combinations of arrows.

The Euclidean Lagrangian for an $SU(N)$ gauge theory can be written
as
\begin{equation}
L_{E}=\frac{1}{2g^{2}}Tr_{F}\left[F_{\mu\nu}^{2}\right]=\frac{N}{2\lambda}Tr_{F}\left[F_{\mu\nu}^{2}\right]
\end{equation}
 we have introduced $\lambda\equiv g^{2}N$, which will be taken to
be finite in a limit where $g\rightarrow0$ as $N\rightarrow\infty$.
It is obvious that every gauge boson propagator will carry a factor
of $1/N$, and the cubic and quartic vertices will carry a factor
of $N$. Fermion fields can be rescaled by $\psi\rightarrow N^{1/2}\psi$
so that similar properties hold.

Any vacuum diagram can be regarded as a polygon, where every closed
loop can be regarded as a face. A vacuum diagram can then be characterized
by its number of faces $F$, number of vertices $V$ and number of
edges $E$. Each face generates a factor of $N$ from a trace. Every
edge is a propagator, and carries with a factor of $N^{-1}$ while
every vertex carries a factor of $N$. Thus each vacuum graph carries
a factor of
\begin{equation}
N^{F-E+V}
\end{equation}
However, $F-E+V$ is the Euler number $\chi_{E}$ of the surface,
a topological invariant. It may be calculated from the number of handles
$H$ of a surface and the number of holes $B$ (for boundary) as
\begin{equation}
\chi_{E}=2-2H-B.
\end{equation}
Thus the maximum power of $N$ associated with a vacuum diagram is
$N^{2}$, for diagrams with $H=B=0$. Because each fundamental representation
loop gives rise to a hole, fundamental representation particles begin
contributing at order $N$. This behavior is what we see in the perturbative
calculation of the effective potential for the Polyakov effective
potential. The effects of quarks are suppressed by a factor $N_{f}/N$
relative to gluons. It should be noted that there is another interesting
large-$N$ limit, the Veneziano limit, in which $N_{f}$ is taken
to infinity at the same time as $N$, with the ratio $x=N_{f}/N$
fixed \cite{Veneziano:1976wm} . This generates a large-$N$ expansion
where particles in the fundamental representation are not suppressed.

A key feature is large-$N$ factorization. Suppose, for example, that
$O_{1}$ and $O_{2}$ are operators transforming according to the
adjoint representation of the group. Then $Tr\, O_{i}/N$ has a smooth
limit as $N\rightarrow\infty$. Large-$N$ factorization states that
in the large-$N$ limit
\begin{equation}
\left\langle \frac{1}{N}TrO_{1}\frac{1}{N}TrO_{2}\right\rangle \rightarrow\left\langle \frac{1}{N}TrO_{1}\right\rangle \left\langle \frac{1}{N}TrO_{2}\right\rangle +O\left(1/N^{2}\right)
\end{equation}
 up to terms of $O(1/N^{2})$. 
For Wilson loops, large-$N$ factorization leads to a closed, exact equation 
in the $N\rightarrow\infty$ limit, 
the Migdal-Makeenko equation \cite{Makeenko:1979pb}.
It was quickly realized that many aspects
are better captured by postulating the existence of a large-$N$ saddle
point of some large-$N$ effective action analogous to the gap equation
in vector-like models. This saddle point, really a gauge orbit in
the space of fields, is often referred to as the master field \cite{0521318270}.
While it is possible to find the master field for some model systems
in lower dimensions with fields in the adjoint representation \cite{Brezin:1977sv},
the problem of determining the master field for four-dimensional gauge
theories has not been solved.


In recent years, much of the work on gauge theories in the large-$N$
limit has employed gauge-string duality to explore strong-coupling
behavior using generalizations of the AdS/CFT correspondence \cite{Maldacena:1997re}.
For a review emphasizing the relation of gauge/string duality to
finite temperature QCD, see \cite{CasalderreySolana:2011us}.
Attempts to model the phase structure of QCD based on the
AdS/QCD approach \cite{Erlich:2005qh} may be found in 
\cite{Kajantie:2006hv,BallonBayona:2007vp,Gursoy:2008bu,Gursoy:2009jd}.

\subsection{Eguchi-Kawai models}

A direct attempt at constructing a large-$N$ reduction for lattice
gauge theories was first made in 1982 by Eguchi and Kawai \cite{Eguchi:1982nm}.
They defined a lattice gauge theory on a single lattice site and showed
that Wilson loops in an infinite volume theory could also be obtained
from their reduced model in the large-$N$ limit if two conditions
were met: large-$N$ factorization and center symmetry must both hold
in all directions. This latter restriction implies that the reduction
can only hold in the confined phase.

The original Eguchi-Kawai (EK) model may be derived by observing that
the master field must be translation invariant. For a lattice gauge
theory, this means that the value of the link variable at every site
must be a gauge transform of the field at the origin
\begin{equation}
U_{\mu}\left(x\right)\rightarrow D^{\dagger}(x)\tilde{U}_{\mu}D\left(x+\mu\right)
\end{equation}
 where $\tilde{U}_{\mu}$ denotes the link variables at the origin.
The matrix $D(x)$ can be taken to be a diagonal matrix of the form
\begin{equation}
D(x)=\exp\left(-ix\cdot P\right)
\end{equation}
where $P_{\mu}$ is a diagonal Hermitian matrix of the form $P_{\mu}=diag\left(p_{1\mu},...,p_{N\mu}\right)$.
Gauge-invariant quantities immediately collapse. For example, a rectangular
$R\times T$ Wilson loop in the $\mu-\nu$ plane becomes
\begin{equation}
\left\langle Tr\left(\tilde{U}_{\mu}\right)^{T}\left(\tilde{U}_{\nu}\right)^{R}\left(\tilde{U}_{\mu}^{\dagger}\right)^{T}\left(\tilde{U}_{\nu}^{\dagger}\right)^{R}\right\rangle 
\end{equation}
 and the Wilson action collapses to
\begin{equation}
S_{EK}=\mathcal{VT}\sum_{\mu>\nu}\frac{\beta}{N}Re\, Tr\left[\tilde{U}_{\mu}\tilde{U}_{\nu}\tilde{U}_{\mu}^{\dagger}\tilde{U}_{\nu}^{\dagger}\right]
\end{equation}
 where $\mathcal{VT}$ is the volume of space-time. Note that global
center symmetry transformations of the form $\tilde{U}_{\mu}\rightarrow z_{\mu}\tilde{U}_{\mu}$
is a symmetry of $S_{EK}$.

In the infinite volume gauge theory, the expectation value of non-closed
paths such as
\begin{equation}
\left\langle Tr\left(\tilde{U}_{\mu}\right)^{T}\right\rangle 
\end{equation}
is zero as a consequence of gauge-invariance and Elitzur's theorem.
In the Eguchi-Kawai model, this expectation value will vanish if center
symmetry is unbroken, but need not vanish if center symmetry is unbroken.
A detailed comparison of the Schwinger-Dyson equations for both theories
confirms that they are equivalent only if center symmetry is unbroken.
The similarity with the case of finite temperature immediately suggests
that center symmetry will be broken in the large-$N$ limit, and the
equivalence will fail. This turns out to be the case \cite{Bhanot:1982sh,Kazakov:1982gh,Okawa:1982ic}.
Several attempts have been made to modify the original EK model to
maintain center symmetry. The first was the Quenched EK (QEK) model,
in which the eigenvalues of the link variables are treated as quenched
variables, forcing the link variables to maintain center symmetry
\cite{Bhanot:1982sh}. While early lattice simulation did indicate
that the QEK model was viable, but recent lattice simulations provides
strong evidence for the breakdown of the reduction \cite{Bringoltz:2008av}.
Another variant of the original model that originally seemed promising
is the Twisted EK (TEK) model \cite{GonzalezArroyo:1982ub}, but more
recent extensive lattice simulations indicate center symmetry breaking
in this model as well \cite{Teper:2006sp,Azeyanagi:2007su,Bietenholz:2007xh}.
An interesting alternative to single-site reduction is to simulate
large-$N$ theories on lattices of size $N_{s}^{4}$, and keep $N_{s}$
sufficiently large that the deconfinement transition is avoided \cite{Kiskis:2002gr,Narayanan:2007fb}.
As long as center symmetry is maintained, such models will be equivalent
to infinite-volume theories in the limit of infinite $N$. This behavior,
a hallmark of a successful large-$N$ reduction, is often generally
described as volume independence.

\subsection{Planar equivalence}

The most prominent example of planar equivalence is the equivalence,
at large $N$, of the $\mathcal{N}=1$ super-Yang-Mills model to an
$SU(N)$ gauge theory with fermions in either the symmetric or the
antisymmetric two-index representation. In other words, in the large-$N$
limit, $SU\left(N\right)$ gauge theories with fermions in the adjoint
representation (Adj) are equivalent to the same gauge theory with
fermions in either the symmetric (S) or antisymmetric (A) representation.
The history of the subject has an interesting arc; see \cite{Armoni:2007vb}
for an introductory discussion. There is a very general formalism
for describing many forms of large-$N$ equivalences \cite{Kovtun:2004bz},
with which a detailed discussion of many forms of planar equivalence
can be carried out \cite{Kovtun:2005kh,Unsal:2006pj,Kovtun:2007py}.
For our purposes, the equivalence of $SU(N)_{Adj,A,S}$ can be understood
from the equivalence of the loop equations for the different theories
\cite{Kovtun:2003hr}. Supersymmetric gauge theories have Majorana
fermions in the adjoint representation which are superpartners of
the gauge bosons. While the effects of fundamental representation
fermions are suppressed in the large-$N$ limit, the effects of adjoint
representation fermions are not, because the number of components
in the representation grows as $N^{2}$ in the large-$N$ limit. In
general, the dimensionality $d_{R}$ of a given representation $R$
is easily determined from its character $\chi_{R}(U)\equiv Tr_{R}U$
via $d_{R}=\chi_{R}(I)$. For example, we have for the adjoint representation
\begin{equation}
\chi_{Adj}\left(U\right)=\chi_{F}\left(U\right)\chi_{\bar{F}}\left(U\right)-1
\end{equation}
so $d_{Adj}=N^2-1$.
The characters of the symmetric and antisymmetric
representations $\chi_{S/A}(U)$
are formed from the product of two characters in the fundamental
representation
and given by
\begin{equation}
\chi_{S/A}(U)=\frac{1}{2}\left[\chi_{F}^{2}(U)\pm\chi_{F}(U)\right]
\end{equation}
with dimensions $N(N\pm1)/2$. These representations are sometimes
referred to as bifermion representations. Because the number of fields
grows as $N^{2}$, we expect the effects of internal loops of $S/A$
fermions to survive in the large-$N$ limit in the same way that adjoint
fermion loops do. 

Let us imagine integrating out the fermions in an $SU(N)$ gauge theory
with either adjoint, symmetric or antisymmetric fermions \cite{Armoni:2004ub}.
In all cases, the logarithm of the fermion determinant can be expressed
as a sum of Wilson loops in $R^{4}$; in a geometry like $R^{3}\times S^{1}$
or $T^{4}$, the sum includes Polyakov loops as well as Wilson loops.
In the large-$N$ limit, factorization tells us that on $R^{4}$ we
have
\begin{equation}
2\left\langle \chi_{S/A}(W)\right\rangle \sim\left\langle \chi_{F}(W)\right\rangle \left\langle \chi_{F}(W)\right\rangle \sim\left\langle \chi_{F}(W)\right\rangle \left\langle \chi_{F}^{*}(W)\right\rangle \sim\left\langle \chi_{Adj}(W)\right\rangle 
\end{equation}
where the expectation value is over the gauge fields. The apparent
mis-match of a factor of two disappears in the functional determinants
because the fermions in the symmetric or antisymmetric representations
are Dirac fermions, while the adjoint fermions are Majorana fermions
with half the degrees of freedom. A proof of the equivalence at the
level of Feynman diagrams can also be given using 't Hooft's double
line notation \cite{Armoni:2003gp}. In the confined phase on $T^{4}$
or a similar geometry, all the expectation value for the Polyakov
loops are zero
\begin{equation}
2\left\langle \chi_{S/A}(P)\right\rangle \sim\left\langle \chi_{F}(P)\right\rangle \left\langle \chi_{F}(P)\right\rangle \sim\left\langle \chi_{F}(P)\right\rangle \left\langle \chi_{F}^{*}(P)\right\rangle \sim\left\langle \chi_{Adj}(P)\right\rangle =0
\end{equation}
and any of the reduced gauge theories that is confining is completely
equivalent to the theories on $R^{4}$, realizing volume independence.
However, this equivalence fails in non-confining phases, because the
$N$-ality of the symmetric and antisymmetric representations is two,
but for the adjoint representation it is zero. Thus we return once
again to the issue of maintaining confinement in small volumes.

\subsection{Confinement for finite volume}

As we have seen in section \ref{sec:Phases-on-R3xS1}, suitable modifications
of the action can lead to a confining phase on $R^{3}\times S^{1}$
for small circumference $L$. It is natural to look at these methods
for the possible construction of useful reduced large-$N$ models.
However, the confining phase is only one of many possible phases as
$N$ becomes large \cite{Myers:2009df,Meisinger:2009ne}. In principle,
it is possible to modify the large-$N$ gauge action on $T^{4}$ by
adding double-trace deformations
\begin{equation}
S\rightarrow S+\sum_{k=1}^{\left[N/2\right]}\sum_{\mu}a_{k}\int d^{4}x\, Tr_{A}P_{\mu}\left(x\right)^{k}.
\end{equation}
where $P_{\mu}\left(x\right)$ is the Polyakov loop in the direction
$\mu$ based at the spacetime point $x$. There is a harmless redundantcy
induced by the integration over $x$, because all points a given loop
path are integrated over. For a single-site lattice action, the corresponding
modification is
\begin{equation}
S\rightarrow S+\sum_{k=1}^{\left[N/2\right]}\sum_{\mu}a_{k}Tr_{A}U_{\mu}^{k}.
\end{equation}
because the single-site link variable is the Polyakov loop. If the
coefficients $a_{k}$ are sufficiently large, then the system will
remain in the confined phase, with $\left\langle Tr_{A}P^{k}\right\rangle =0$
. In the confined phase, the precise values of the coefficients don't
matter; we would have a self-consistent finite-volume reduction of
the pure gauge theory \cite{Unsal:2008ch}.

Adjoint fermions with periodic boundary conditions offer another possible
route to the construction of a large-$N$ finite-volume reduction
\cite{Bedaque:2009md,Bringoltz:2009mi,Hanada:2009kz,Bringoltz:2009kb,Hollowood:2009sy,Hanada:2009hd,Bringoltz:2009fj,Poppitz:2009fm,Hietanen:2009ex,Azeyanagi:2010ne,Catterall:2010gx,Hietanen:2010fx}.
This model is also of direct phenomenological interest. In physical
QCD with $N=3$, we normally think of quarks in the fundamental representation
as playing no role in the large-$N$ limit. However, antiquarks in
physical QCD are in the $\bar{3}$ representation, which is an antisymmetric
representation. In this sense, physical QCD has two different large-$N$
limits. The large-$N$ limit, an $SU(N)$ gauge theory with $N_{f}$
Dirac fermions in the adjoint representation is equivalent to the
theory with $2N_{f}$ Dirac fermions in the antisymmetric representation,
which is QCD with $2N_{f}$ flavors when $N=3$. Analytic calculations
of the lattice form of the one-loop effective potential \cite{Bedaque:2009md,Bringoltz:2009mi,Hollowood:2009sy,Bringoltz:2009fj,Poppitz:2009fm}
indicate a rich phase structure as the fermion mass is varied, similar
to that found in the continuum on $R^{3}\times S^{1}$ \cite{Myers:2009df, Hollowood:2009sy}. 
However,
it should be kept in mind that chiral symmetry breaking effects, which
affect Polyakov loop terms in the effective potential, are generally
not included in these calculations; at present, these effects can
only be estimated using phenomenological models \cite{Nishimura:2009me}.
Thus lattice simulations are essential not only to determine the properties
of the confined phase, but also to confirm that it exists. The key
issue is finding a confined phase in lattice simulations that survives
the extrapolation to the continuum limit where the bare lattice coupling
$g^{2}$ goes to zero. Of course, the success of the large-$N$ reduction
may depend on the lattice fermion implementation. In the single-site
model, analytic calculations indicate that center symmetry is spontaneously
broken with naive fermions but unbroken with overlap fermions \cite{Hietanen:2009ex}.
Lattice simulations of $SU(3)$ with 2 flavors of staggered fermions
in the adjoint representation on $R^{3}\times S^{1}$ show that a
confined phase does exist for sufficiently small fermion mass \cite{Cossu:2009sq}.
Simulations of the single-site theory using Wilson fermions also show
a region in the parameter space of fermion mass and lattice coupling
where center symmetry is unbroken and an extrapolation to the continuum
limit appears plausible \cite{Bringoltz:2009kb,Catterall:2010gx};see
also \cite{Azeyanagi:2010ne}. Simulations with overlap fermons \cite{Hietanen:2010fx}
also indicated a confined region. The most recent lattice results
\cite{Bringoltz:2011by} of a single-site model with two flavors of
Wilson fermions show a center-symmetric region for $N$ as large as
$53$ and 't Hooft coupling $g^{2}N$ as small at $0.005$, so the
prospects for a successful large-$N$ reduction appear bright.

\section{\label{sec:Conclusions}Conclusions}

We have entered a new phase in our understanding of the phase structure
of gauge theories. The phase structure of gauge theories at finite
temperature was known to be rich. By itself, however, temperature
alone does not give us a large window where continuum analytical results
and lattice simulations are both useful. We have now a class of theories
that can be successfully studied using both analytical methods and
lattice simulations, built on our understanding of finite temperature
physics. On $R^{3}\times S^{1}$, the use of double-trace deformations
or periodic adjoint fermions allows us to study confinement in a region
where semiclassical methods are valid, and check our results with
lattice simulations. At the same time, we have access to many new
phases, most of which are partially-confining phases. Nor are we restricted
to an $R^{3}\times S^{1}$ geometry; other geometries are available,
and largely unexplored. By extending finite-temperature physics to
this larger class of models, we have seen how instantons and non-Abelian
monopoles can play a role in confinement, realizing some long-held
ideas about the nature of the confining phase. At the same time, we
have seen how topological effects can survive in the large-$N$ limit.
We also have a new, promising class of single-site large-$N$ models
to explore.

There is more to be done. Most of the lattice simulations to date
have only checked analytic predictions of overall phase structure,
and predictions for, \emph{e.g.}, the behavior of string tensions
have not been checked in simulations. Although semiclassical techniques
have a natural range of validity on $R^{3}\times S^{1}$ given by
$N\Lambda L\ll1$, we do not know the general validity of the concepts
useful for small $L$. We know that a double-trace deformation connects
the small-$L$ confining phase to the large-$L$ confining phase without
an intervening phase transition, but we do not know how physical quantities
change along a path that connects the two. We do not yet have a complete
understanding of string-tension scaling laws, even in the small-$L$
region, and there are many other questions that remain to be asked
and answered. 
\begin{acknowledgments}
The author would like to thank his collaborators Peter Meisinger,
Joyce Myers, and Hiro Nishimura for sharing their insights into this
subject, Mithat Unsal for many stimulating discussions, and his colleagues
Mark Alford and Francesc Ferrer for their helpful advice. The author
gratefully acknowledges the support of this work by the U.S. Dept.
of Energy under Grant 91ER40628.
\end{acknowledgments}
\bibliography{Phases_review}

\end{document}